\documentclass[reprint, amsmath, amssymb, pra, longbibliography, nofootinbib, floatfix]{revtex4-2}

\usepackage[colorlinks, citecolor=blue, linkcolor=blue, urlcolor=blue, hypertexnames=false]{hyperref}
\usepackage[T1]{fontenc}
\usepackage{graphicx}
\usepackage{amsmath, amssymb, amsthm}
\usepackage{bm}
\usepackage{booktabs}
\usepackage{multirow}
\usepackage{xcolor}
\usepackage{float}
\makeatletter
\let\newfloat\newfloat@ltx
\makeatother
\usepackage{algorithm}
\usepackage{algpseudocode}

\newtheorem{theorem}{Theorem}
\newtheorem{proposition}{Proposition}
\newtheorem{corollary}{Corollary}

\newcommand{\W}{\mathrm{W}_1}
\newcommand{\MMD}{\mathrm{MMD}}
\newcommand{\CFI}{\mathrm{CFI}}
\newcommand{\TV}{\mathrm{TV}}
\newcommand{\Tr}{\mathrm{Tr}}
\newcommand{\RKHS}{\mathcal{H}_k}
\newcommand{\simplex}{\Delta_N}

\begin{document}

\title{Certifying quantum generative models beyond the Fisher trace}

	\thanks{This manuscript has been authored by UT-Battelle, LLC, under Contract No.\ DE-AC05-00OR22725 with the U.S.\ Department of Energy. The United States Government retains and the publisher, by accepting the article for publication, acknowledges that the United States Government retains a non-exclusive, paid-up, irrevocable, world-wide license to publish or reproduce the published form of this manuscript, or allow others to do so, for the United States Government purposes. The Department of Energy will provide public access to these results of federally sponsored research in accordance with the DOE Public Access Plan.}

\author{Vicente Leyton-Ortega}
\email{leytonorteva@ornl.gov}
\affiliation{Quantum Computational Science Group, Quantum Information Science Section, Computational Sciences and Engineering Division, Oak Ridge National Laboratory, Oak Ridge, Tennessee 37831, USA}

\author{Kathleen Hamilton}
\email{hamiltonke@ornl.gov}
\affiliation{Quantum Computational Science Group, Quantum Information Science Section, Computational Sciences and Engineering Division, Oak Ridge National Laboratory, Oak Ridge, Tennessee 37831, USA}
\date{\today}

\begin{abstract}
	A diagnostic that certifies a trained quantum generative model is trusted once it separates good from defective models on a labeled benchmark, which it can do by recovering how the benchmark was built.
	We determine which diagnostics can certify the real-amplitude quantum circuit Born machine, in noiseless simulation.
	For half-angle rotation circuits, the classical and quantum Fisher information coincide almost everywhere, so the Fisher trace equals the parameter count for every model and carries no information about fit quality.
	The Fisher eigenvalue rank counts the support of the converged distribution and scores at chance on full-support targets.
	Certification must therefore compare distributions, and we combine the Wasserstein distance and the maximum mean discrepancy in an escalation rule with a finite-sample false-pass bound.
	For two exactly solvable defect families, we compute in closed form the window in which the discrepancy rejects generators that the transport distance admits; it opens only at certification-grade tolerances and closes as the qubit number grows.
	At the tolerances a near-term experiment can resolve, the discrepancy changed no verdict in our fixed-threshold test.
	We prove that every normalized kernel admits a parity defect with exponentially small discrepancy, so a certificate must name the correlation orders it targets.
	Together with a sharp tolerance ratio on the Hamming cost, these results tell a practitioner in advance which checks carry information about the fit, which defects each one misses, and when the second metric is worth its shots.
\end{abstract}


\maketitle

%% ============================================================
\section{Introduction}\label{sec:intro}

Quantum generative models learn a target probability distribution by optimizing the parameters of a variational quantum circuit, with applications ranging from quantum-chemistry data generation to financial modeling and combinatorial sampling.
The main families are quantum generative adversarial networks (QGANs)~\cite{Zoufal2019,Stein2022}, which have been used to learn and load distributions onto quantum hardware, quantum circuit Born machines (QCBMs)~\cite{Liu2018,Benedetti2019}, and quantum generative diffusion models~\cite{Chen2024}.

In earlier work we used the QCBM to benchmark the training of shallow circuits~\cite{Benedetti2019} and built an adversarial framework, Re-QGAN, around a real-amplitude generator circuit~\cite{LeytonOrtega_ReQGAN}.
As such models move from simulation toward deployment, one has to decide whether a trained generator has learned its target before its samples are used downstream, and this certification problem is the subject of the present work.

In practice, a trained generator is certified by scoring its output distribution against the target with a distributional metric, most commonly the Wasserstein-1 distance~\cite{Zoufal2019} or the maximum mean discrepancy (MMD)~\cite{Gretton2012}.
Quantities derived from the Fisher information of the circuit, introduced as measures of model capacity~\cite{Abbas2021,Meyer2021,Haug2021}, are natural companions to such scores.
A diagnostic of either kind is trusted once it separates good from defective generators on a labeled benchmark.

A benchmark, however, is built by varying a known control, typically a perturbation of a good target, and its generators are produced by a particular optimizer.
Hence, a diagnostic can score well by recovering how the benchmark was built or how the generator was trained, while saying little about whether the output distribution matches the target.

Clearly, a diagnostic should be tested against this possibility before it is used to certify.
Earlier studies have compared several metrics on trained models~\cite{Gili2024}, explored alternative training losses~\cite{Coyle2020}, and shown that kernel discrepancies weight high-order correlations weakly~\cite{Rudolph2023,Kurkin2025}.
Yet, these studies examine the metrics themselves, and whether a diagnostic that scores well on a labeled benchmark measures the quality of the output distribution or the design of the benchmark remains open, as does the question of when a second metric adds protection to a first.

In this paper, we ask which diagnostics can certify a trained generator and what each of them misses.
We carry out this examination for the QCBM built on the universal real-amplitude state-preparation circuit of M\"ott\"onen \emph{et al.}~\cite{Mottonen2004,Mottonen2005}, in noiseless simulation at three to eight qubits.
We first show that for this circuit, and for every circuit that prepares real amplitudes with half-angle rotations, the classical and quantum Fisher information coincide almost everywhere, as follows from results of Facchi \emph{et al.}~\cite{Facchi2010} and Liu \emph{et al.}~\cite{Liu2020}.
The trace of the Fisher matrix is therefore fixed by the number of parameters and carries no information about the fit. 

We then turn to the eigenvalue rank of the same matrix, whose capacity interpretation is due to Larocca \emph{et al.}~\cite{Larocca2023} and Delgado \emph{et al.}~\cite{Delgado2023}.
The rank separates our benchmark classes perfectly. We show that it does so because it counts the support of the output distribution, which on that benchmark coincides with the label. On targets with full support it scores at chance, and we retain it only as a flag for verdicts that are sensitive along few parameter directions.
In turn, the pooled Wasserstein and MMD scores change with the convergence of the optimizer and with the reference distribution against which they are computed. They become perfect once the runs labeled good are restricted to converged fits.

Certification thus has to rest on the two distributional metrics, and we analyze an escalation rule that consults the discrepancy only when the transport distance passes narrowly. We prove a finite-sample bound on its false-pass probability that improves quadratically on the single-metric rule for defects close to the threshold.
For two families of bitstring defects on which both metrics have closed forms, we then compute exactly the range of qubit numbers and tolerances in which the discrepancy rejects generators that the transport threshold admits. This window opens only at strict tolerances and narrows as the qubit number grows.

Two further results hold beyond these families.
On the Hamming cost, the discrepancy vote is redundant once its tolerance exceeds a threshold quadratic in the transport tolerance, so a practitioner knows in advance when the second metric is worth its shots.
For any normalized kernel, moreover, some parity defect has a discrepancy exponentially small in the qubit number, so a certificate must specify the correlation orders it targets.
Our guarantees are stated for noiseless sampling.
At the tolerances a near-term experiment can resolve, the second metric changed no verdict in a pre-registered test on independent targets, in agreement with the redundancy result, and its protection is located at strict tolerances and small qubit numbers, where the closed-form window places it.

The Fisher geometry of a real-amplitude circuit therefore reports the circuit and the benchmark design, and it should not be read as evidence of certification. A distributional test can certify, and its blind spots are known before any measurement: the defects a transport tolerance admits, the tolerances at which the discrepancy catches them, and the correlation orders that no kernel resolves. The controls we use to separate these cases, a restriction to converged fits, a severity split and a full-support benchmark, apply to any diagnostic validated on a constructed benchmark.

The paper is organized as follows.
Section~\ref{sec:setup} fixes the metrics and tolerances, the ansatz, the benchmarks and the decision rules. Section~\ref{sec:cap} shows that the Fisher trace and rank do not certify, and Sec.~\ref{sec:pooled} states the conventions under which the distributional scores do. Section~\ref{sec:protocol_sec} defines the escalation rule, bounds its false-pass probability and computes its window, and Sec.~\ref{sec:validation} tests these statements. Section~\ref{sec:discussion} discusses noise and the limitations, and Sec.~\ref{sec:conclusion} concludes. Appendix~\ref{app:proofs_main} collects the proofs of the main results, and the Supplemental Material appended to this article gives the design details, the controls, the remaining proofs and the reproduction protocol.

\section{Setting}\label{sec:setup}\label{sec:design}

We consider quantum generative models whose output is a probability distribution $q \in \simplex$, the probability simplex over the outcome space $\mathcal{X} = \{0,1\}^n$ with $N = 2^n$ elements.
The distribution is obtained by measuring an $n$-qubit parameterized quantum circuit (PQC) in the computational basis, whereas the target distribution $p \in \simplex$ is either known analytically or estimated from samples.
The certification question is then: given $p$ and $q$, is $q$ an acceptable approximation of $p$?

\subsection{Metrics and tolerances}\label{sec:metrics}\label{sec:w1}\label{sec:mmd}\label{sec:qfi}

The transport channel is the Wasserstein-1 distance $\W(p, q; C)$, the minimal cost $\langle C, \Gamma \rangle$ over couplings $\Gamma \geq 0$ whose row and column marginals are $p$ and $q$, for a ground-cost matrix $C \in \mathbb{R}^{N \times N}_{\geq 0}$. We use it mainly through its Kantorovich dual,
\begin{equation}\label{eq:w1_dual}
	\W(p, q; C) = \max_{f} \sum_{x \in \mathcal{X}} f(x)\bigl(p(x) - q(x)\bigr),
\end{equation}
where the maximum is over $C$-Lipschitz potentials, $f(x) - f(y) \leq C(x,y)$ for all $x, y$, and we write $\TV(p, q) = \tfrac12\|p - q\|_1$ for the total-variation distance.
We use three ground costs: the Hamming cost $C_1(x,y) = d_H(x,y)$; a weighted variant $C_2(x,y) = \sum_i w_i |x_i - y_i|$ with qubit weights $w_i > 0$; and a circuit-structure cost $C_3(x,y) = d_H(x,y) + \alpha_W \sum_i |s_i(x) - s_i(y)|$ with nearest-neighbor parities $s_i(x) = x_i \oplus x_{i+1}$.
All three are metrics, and their maximum $D_{\max} = \max_{x,y} C(x,y)$ sets the scale of $\W$ and enters the sample-complexity bounds.

The discrepancy channel is the maximum mean discrepancy (MMD)~\cite{Gretton2012} of a positive-definite kernel $k$ on $\mathcal{X}$, the distance between the kernel mean embeddings $\mu_p = \mathbb{E}_{x \sim p}[k(x, \cdot)]$ of the two distributions in the reproducing kernel Hilbert space (RKHS) $\RKHS$,
\begin{equation}\label{eq:mmd}
	\MMD_k^2(p, q) = \| \mu_p - \mu_q \|_{\RKHS}^2 = (p - q)^{\top} K\, (p - q),
\end{equation}
with $K_{xy} = k(x,y)$ the kernel matrix. Hence, the squared discrepancy is a quadratic form in $p - q$, so it grows quadratically along a defect direction, whereas $\W$ grows linearly; Sec.~\ref{sec:dominance} turns this difference into the tolerance-mismatch window.
We use the radial basis function (RBF) kernel on Hamming distance, $k(x,y) = \exp(-\gamma\, d_H(x,y))$ with $\gamma = 1/n$. It is characteristic, so that $\MMD_k(p,q) = 0$ only if $p = q$~\cite{Gretton2012}.

Throughout this paper, the tolerance $\tau_{\MMD}$ acts on the squared statistic: the loose benchmark-comparison tolerance is $\MMD_k^2 \leq 10^{-2}$ and the certification-grade tolerance is $\MMD_k^2 \leq 10^{-4}$, equivalently $\MMD_k \leq 0.1$ and $\MMD_k \leq 0.01$. The labels ``loose'' and ``certification-grade'' refer to $\tau_{\MMD}$ alone; the transport threshold is $\tau_{\W} = 0.05$ in every analysis. We compare each statistic with its tolerance; a permutation test would test the sharp null $p = q$, which certification does not ask (Sec.~\ref{sm:permutation} of the Supplemental Material appended to this article, SM below).

The spectral channel is built from the classical Fisher information matrix (CFIM) of the output distribution $p_{\bm{\theta}}$ of a PQC with parameters $\bm{\theta} \in \mathbb{R}^d$,
\begin{equation}\label{eq:fim}
	\bigl(F_{\mathrm{cl}}\bigr)_{jk}(\bm{\theta}) = \sum_{x \in \mathcal{X}} \frac{1}{p_{\bm{\theta}}(x)} \frac{\partial p_{\bm{\theta}}}{\partial \theta_j}(x) \frac{\partial p_{\bm{\theta}}}{\partial \theta_k}(x),
\end{equation}
and the diagnostic is its trace, $\CFI_w(\bm{\theta}) = \Tr F_{\mathrm{cl}}(\bm{\theta})$.
The CFIM is dominated in the L\"owner (positive-semidefinite, PSD) order by the quantum Fisher information matrix~\cite{Safranek2018,Rath2021}. We compute the Jacobians exactly by the parameter-shift rule and clamp the inverse probability at $p_{\mathrm{floor}} = 10^{-10}$ (SM Sec.~\ref{sm:fisher_estimator}).

\subsection{Ansatz and training}\label{sec:mottonen}\label{sec:qcbm}

The generator is the quantum circuit Born machine (QCBM) built on the real-amplitude state-preparation circuit of M\"ott\"onen \emph{et al.}~\cite{Mottonen2004,Mottonen2005}. Its output is the Born distribution $q_{\bm{\theta}}(x) = |\langle x | U(\bm{\theta}) | 0^{\otimes n} \rangle|^2$.
The circuit is recursive: a length-$k$ block consists of two length-$(k-1)$ blocks joined by a CNOT between its first and last qubit, and stage $k = 1, \dots, n$ applies such a block to the first $k$ qubits. Stage $k$ carries $2^{k-1}$ rotations and $2^{k-1} - 1$ CNOTs, so the circuit has $d = 2^n - 1$ rotation parameters and $2^n - n - 1$ CNOT gates. The ansatz is universal for real-amplitude states with the minimal parameter count.

Two properties of this construction drive Sec.~\ref{sec:cap}. The gate set $\{R_Y, \mathrm{CNOT}\}$ and the initial state are real, so the prepared state is real for every $\bm{\theta}$; and each rotation $R_Y(\theta_k) = \exp(-i\theta_k Y/2)$ enters with the half-angle convention and the Hermitian-unitary generator $Y$.

We train by minimizing $\W(p, q_{\bm{\theta}}; C)$ with the covariance matrix adaptation evolution strategy (CMA-ES)~\cite{Hansen2016}, a derivative-free optimizer, under a budget of $5{,}000$ evaluations at $n = 3$ rising to $50{,}000$ at $n = 8$ (SM Sec.~\ref{sm:budgets}). Like any optimizer it meets the flatness of the cost function~\cite{Arrasmith2021}, and it does not converge within the budget on every run.

\subsection{Benchmarks and scoring conventions}\label{sec:benchmarks}

Benchmark B1 is a GHZ-type synthetic target. The clean target is the two-peak Born distribution $p_0 = 0.6\,\delta_{x, 0^{n}} + 0.4\,\delta_{x, 1^{n}}$ of the state $\sqrt{0.6}\,|0\rangle^{\otimes n} + \sqrt{0.4}\,|1\rangle^{\otimes n}$, and the perturbed targets are the convex blends
\begin{equation}\label{eq:b1_blend}
	p_\epsilon = (1 - \epsilon)\, p_0 + \epsilon\, \nu, \qquad \epsilon \in \{0.00, 0.02, 0.05, 0.10, 0.20\},
\end{equation}
where $\nu$ is a random background distribution drawn once per seed (i.i.d.\ uniform weights on $\{0,1\}^n$, normalized). The blend spreads support over all $2^n$ outcomes, and $\epsilon$ supplies the quality labels.

Benchmark B2 is a structured target whose Shannon entropy grows with $n$ while staying well below the uniform maximum. At $n = 3$ it is the output distribution of our trained Re-QGAN generator~\cite{LeytonOrtega_ReQGAN}, $1{,}000$ generated images each encoded in three bits; at $n \geq 4$ it is a synthetic four-peak distribution on a random background. Benchmark B3 is a synthetic near-uniform target whose entropy lies within $3 \times 10^{-4}$ nats of $n\ln 2$, a stress test for targets with broad support. SM Sec.~\ref{sm:entropy} gives the entropies, and SM Sec.~\ref{app:reproduction} the constructions.

\begin{figure*}[t]
	\centering
	\includegraphics[width=\textwidth]{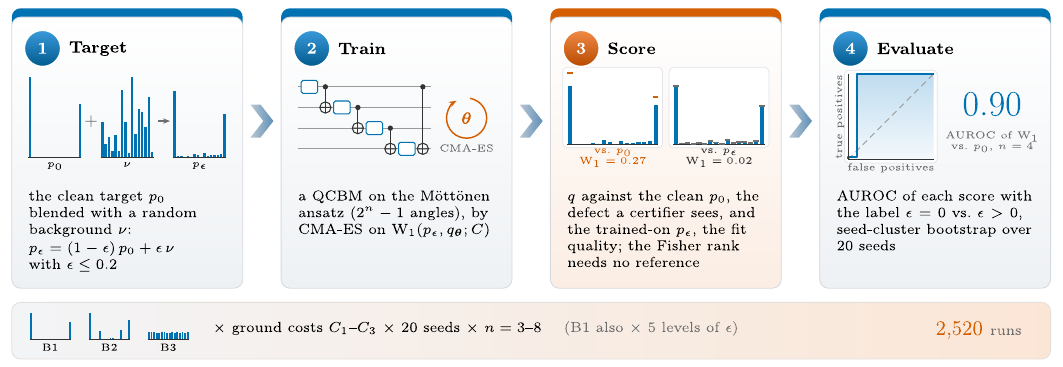}
	\caption{Campaign design: one run in four steps, with data from the campaign at $n = 4$ and cost $C_1$. (1) The training target $p_\epsilon$ blends the clean target $p_0$ with a random background $\nu$ (B1, $\epsilon = 0.2$, seed $0$). (2) The QCBM is trained by CMA-ES against $p_\epsilon$. (3) Its output $q$ (bars) is scored against both references (ticks): against the trained-on $p_\epsilon$ the run meets the tolerance $\tau_{\W} = 0.05$, while against the clean $p_0$ (vermilion) it carries the injected defect. The Fisher rank uses no reference. (4) Each score enters an AUROC with the perturbation level as the label; the curve is the clean-reference $\W$ over the $100$ B1 runs at $n = 4$. Bottom: the targets of the three benchmarks and the size of the campaign.}\label{fig:design}
\end{figure*}

The campaign (Fig.~\ref{fig:design}) comprises $2{,}520$ noiseless statevector runs at $n = 3$--$8$, $420$ per qubit count: for B1, three costs times five levels times $20$ seeds, and for B2 and B3, three costs times $20$ seeds each. A further $420$-run batch at $n = 9$ is excluded from every analysis, because CMA-ES fits neither the unperturbed nor the perturbed targets there within any budget we tried.

Two reference distributions can be used to score a run.
A score against the \emph{trained-on} target $p_\epsilon$ is a training residual and measures fit quality; a score against the \emph{clean} target $p_0$ measures defect magnitude. In particular, a perturbed run that is fitted perfectly has a training residual of zero while carrying the largest defect the benchmark injects, and a certifier in deployment sees the clean score.

The area under the receiver operating characteristic (AUROC) is the probability that a diagnostic scores a randomly chosen perturbed run ($\epsilon > 0$) as worse than a randomly chosen unperturbed one ($\epsilon = 0$). The $20$ seeds of a cell are shared across levels and costs, so every interval is a seed-cluster bootstrap, and two scores are compared through the paired difference of their AUROCs on the same draws. The stress test, the full-support benchmark and the fixed-threshold test were specified before execution in dated documents (SM Secs.~\ref{sm:statistics} and~\ref{sm:prereg}).

\subsection{Decision rules}

We frame certification as a hypothesis test in which the certifier decides PASS or FAIL.
A \emph{false pass} (type II error) occurs when a defective generator is declared certified; a \emph{false fail} (type I error) occurs when a good generator is rejected.
Given thresholds $\tau_{\W}$ and $\tau_{\MMD}$, the $\W$-only baseline $D_1$ passes iff $\W \leq \tau_{\W}$, and the escalation rule $D_4$ uses $\W$ as the primary channel and consults the discrepancy only inside the escalation window $\tau_{\W} - \eta < \W \leq \tau_{\W}$, where it passes only if $\MMD_k^2 \leq \tau_{\MMD}$ as well. Its acceptance set is therefore
\begin{equation}\label{eq:acceptance}
	\begin{split}
	&\{\W \leq \tau_{\W} - \eta\}\\
	&\cup \{\tau_{\W} - \eta < \W \leq \tau_{\W},\ \MMD_k^2 \leq \tau_{\MMD}\}.
	\end{split}
\end{equation}
The escalation keeps the sample cost of $D_1$ in the clear case and applies the AND of the two votes where the transport vote passes narrowly.

\section{The Fisher trace and rank as certification channels}\label{sec:cap}\label{sec:failures}

In this section we show that the two Fisher quantities of the M\"ott\"onen QCBM carry no certification signal: the trace is pinned to the parameter count, and the rank counts the support of the converged distribution.
The ingredients of the trace identity are standard, since Facchi \emph{et al.}~\cite{Facchi2010} established the classical--quantum Fisher equality for real-amplitude pure states, the domination of the classical by the quantum Fisher matrix is due to Braunstein and Caves~\cite{BraunsteinCaves1994}, and the unit quantum-Fisher diagonal of half-angle involutive generators is a textbook result~\cite{Helstrom1976,Holevo1982,Liu2020}.

\subsection{Pinning of the Fisher trace}\label{sec:cap_general}\label{sec:cap_mottonen}

\begin{proposition}[Structural Fisher-trace identity]\label{prop:cap_general}
	Let $|\psi(\bm{\theta})\rangle = U(\bm{\theta})|0\rangle^{\otimes n}$ be the output state of an $n$-qubit parameterized quantum circuit with parameters $\bm{\theta} \in \mathbb{R}^d$, and let $p_{\bm{\theta}}(x) = |\langle x | \psi(\bm{\theta}) \rangle|^2$ be the computational-basis probability distribution.
	Let $F_{\mathrm{cl}}(\bm{\theta})$ be the classical Fisher information matrix of $p_{\bm{\theta}}$, Eq.~\eqref{eq:fim} with the sum running over outcomes with $p_{\bm{\theta}}(x) > 0$, and $F_{\mathrm{Q}}(\bm{\theta})$ the quantum Fisher information matrix of $|\psi(\bm{\theta})\rangle$.
	Suppose the circuit satisfies the following two conditions:
	\begin{enumerate}
		\item[$({\rm cond}_1)$] (Real amplitudes.) The state $|\psi(\bm{\theta})\rangle$ has real amplitudes in the computational basis for every $\bm{\theta}$.
		\item[$({\rm cond}_2)$] (Half-angle rotation generators.) Each parameter $\theta_k$ enters $U(\bm{\theta})$ exactly once, as a half-angle rotation $R_{G_k}(\theta_k) = \exp(-i\theta_k G_k / 2)$, where $G_k$ is a Hermitian unitary acting on some subset of qubits, i.e.\ an involution $G_k^2 = I$ (canonically a Pauli string, as for the single-qubit $Y$ rotations of the M\"ott\"onen ansatz).
	\end{enumerate}
	Then the following two statements hold for the Fisher matrices of the circuit.
	\begin{enumerate}
		\item[\textup{(a)}] For every $\bm{\theta} \in \mathbb{R}^d$, $\Tr F_{\mathrm{Q}}(\bm{\theta}) = d$ and
		\begin{equation}\label{eq:cap_general}
			\Tr F_{\mathrm{cl}}(\bm{\theta}) \;\leq\; d.
		\end{equation}
		\item[\textup{(b)}] Let $\psi_{\bm{\theta}}(x) = \langle x|\psi(\bm{\theta})\rangle$ and let $\mathcal{Z} = \{\bm{\theta} : \psi_{\bm{\theta}}(x) = 0 \text{ and } \partial_k\psi_{\bm{\theta}}(x) \neq 0 \text{ for some } x, k\}$ be the zero-amplitude set. For every $\bm{\theta} \notin \mathcal{Z}$,
		\begin{equation}\label{eq:cap_equality}
			F_{\mathrm{cl}}(\bm{\theta}) = F_{\mathrm{Q}}(\bm{\theta}), \qquad \Tr F_{\mathrm{cl}}(\bm{\theta}) = d,
		\end{equation}
		and for $\bm{\theta} \in \mathcal{Z}$ the inequality of Eq.~\eqref{eq:cap_general} is strict.
	\end{enumerate}
	If, in addition, the parameterization is minimal, that is, if $({\rm cond}_3)$ the parameter count satisfies $d = 2^n - 1$, matching the dimension of the real-amplitude pure-state sphere $\mathbb{S}^{2^n - 1}_{\mathbb{R}} \subset \mathbb{R}^{2^n}$ (modulo global sign), then $d$ in Eqs.~\eqref{eq:cap_general} and~\eqref{eq:cap_equality} reads $2^n - 1$.
\end{proposition}

Appendix~\ref{app:cap_proof} gives the proof: reality forces the phase overlap $\langle\psi|\partial_k\psi\rangle$ to vanish, the half-angle involutive generator fixes every diagonal entry of the quantum Fisher matrix at one, and the classical and quantum matrices differ only by the zero-amplitude terms.

Part~(b) turns the inequality of Eq.~\eqref{eq:cap_general} into an identity on the operational parameter space, because $\mathcal{Z}$ has measure zero: each amplitude is a real-analytic function of $\bm{\theta}$, and the zero set of a real-analytic function that is not identically zero has Lebesgue measure zero.
The M\"ott\"onen ansatz satisfies $({\rm cond}_1)$--$({\rm cond}_3)$ by the two properties of Sec.~\ref{sec:mottonen} and its parameter count, so its Fisher trace obeys
\begin{equation}\label{eq:qfi_cap_mottonen}
	\CFI_w(\bm{\theta}) \;\leq\; d \;=\; 2^n - 1,
\end{equation}
with equality for every $\bm{\theta} \notin \mathcal{Z}$.
The Fisher trace of the QCBM is thus fixed by the circuit architecture alone, irrespective of the target distribution and of how well the circuit fits it.
Condition $({\rm cond}_3)$ enters only to identify $d$ with $2^n - 1$, so the identity holds for any real-amplitude half-angle ansatz, including the layered hardware-efficient ansatz of $L{+}1$ $R_Y$ layers ($d = n(L+1)$); we found $\Tr F_{\mathrm{cl}}/d = 1.000$ at random parameters for $n = 3$--$8$ and $L = 1$--$5$. The value $d$ belongs to the circuit angles; in the node angles of the amplitude tree the trace is pinned at $n$ instead (SM Sec.~\ref{sm:coordinates}).

In the campaign, with the probability floor removed, $\Tr F_{\mathrm{cl}}(\bm{\theta}) = d$ to within $10^{-8}$ in every run at every qubit count. With the floor in place, the clamped trace lies within $0.03\%$ to $1.1\%$ of the cap in the mean, and every sub-cap run carries an outcome below the floor, whose deficit we predict in closed form to a relative residual of $1.2 \times 10^{-8}$ (SM Sec.~\ref{sec:cap_saturation}). Observing the identity from shots requires $S \approx 3 \times 10^4$ at $n = 3$ and $10^5$--$3 \times 10^5$ beyond (SM Sec.~\ref{sec:finite_shot}).

\subsection{The Fisher rank as a support count}\label{sec:rank_results}

Proposition~\ref{prop:cap_general} pins the trace of $F_{\mathrm{cl}}(\bm{\theta})$ but not its spectrum. The capacity interpretation of its rank is due to Larocca \emph{et al.}~\cite{Larocca2023} and, for the QCBM specifically, to Delgado \emph{et al.}~\cite{Delgado2023}; here we ask whether it certifies.
We use the numerical rank with eigenvalue threshold $10^{-6}$ relative to the largest eigenvalue, $\mathrm{rank}_{10^{-6}}(F_{\mathrm{cl}})$, at the converged parameters of every run.

On B1 it separates the unperturbed from the perturbed runs perfectly through $n = 7$ at costs $C_1$ and $C_2$, reaches $0.999$ at $C_3$, and falls to $0.959$--$0.966$ at $n = 8$, while the pooled $\W$ and MMD AUROCs fall to $0.70$--$0.78$ over $n = 5$--$7$ (SM Table~\ref{tab:auroc}). The separation survives doubling the unperturbed class to $40$ seeds and thresholds of $10^{-5}$ and $10^{-7}$, but its cuts were identified after inspecting the campaign (SM Secs.~\ref{sm:gap} and~\ref{sm:sweep}).

\begin{figure}[t]
	\centering
	\includegraphics[width=\columnwidth]{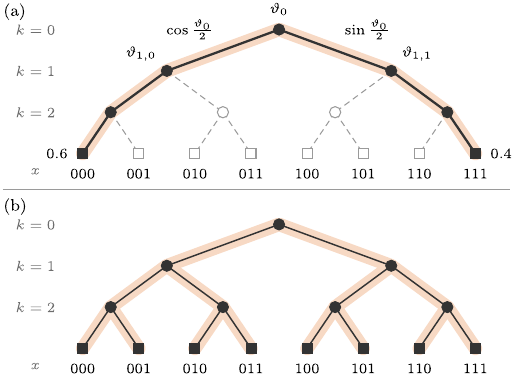}
	\caption{Amplitude tree of the M\"ott\"onen ansatz at $n = 3$. The node at depth $k$ on the path of the prefix $x_{<k}$ carries the angle $\vartheta_{k,x_{<k}}$ and multiplies the amplitude by $\cos(\vartheta_{k,x_{<k}}/2)$ on the left branch ($x_k = 0$) and by $\sin(\vartheta_{k,x_{<k}}/2)$ on the right branch ($x_k = 1$); the leaves are the outcomes $x$. (a)~Exact fit to the two-peak target $0.6\,\delta_{x,000} + 0.4\,\delta_{x,111}$. The $2n - 1 = 5$ live nodes on the two shaded root-to-leaf paths are the parameters on which the amplitudes depend, and the pruned subtrees (open symbols, dashed edges) carry no mass, so $F_{\mathrm{Q}}$ has rank $5$. (b)~Full-support target. All $d = 2^n - 1 = 7$ nodes are live and the rank is full.}\label{fig:amplitude_tree}
\end{figure}

\begin{table}[t]
	\caption{What the rank counts, on B1 at cost $C_1$. $2n-1$ is the number of live nodes of the amplitude tree (SM Sec.~\ref{app:locus}) for an exact fit to the two-peak target, the rank of the quantum Fisher matrix there; ``at $2n-1$'' is the fraction of converged unperturbed runs (training residual below $0.005$; below $0.02$ at $n = 8$, where none reaches $0.005$) whose $\mathrm{rank}_{10^{-6}}$ equals it, with the number of such runs in parentheses. The classes are gap-separated (largest unperturbed rank below smallest perturbed rank) through $n = 7$.}\label{tab:rank_structure}
	\begin{ruledtabular}
		\begin{tabular}{cccccccc}
			$n$ & $d$ & $2n-1$ & at $2n-1$ & \multicolumn{2}{c}{unperturbed} & \multicolumn{2}{c}{perturbed} \\
			 &  &  &  & median & max & min & full rank \\
			\hline
			3 & 7   & 5  & $1.00\ (20)$ & 5   & 5   & 7   & $1.00$ \\
			4 & 15  & 7  & $1.00\ (18)$ & 7   & 7   & 15  & $1.00$ \\
			5 & 31  & 9  & $0.86\ (14)$ & 9   & 27  & 30  & $0.97$ \\
			6 & 63  & 11 & $0.00\ (15)$ & 19.5 & 40  & 54  & $0.66$ \\
			7 & 127 & 13 & $0.00\ (15)$ & 59.5 & 91  & 101 & $0.26$ \\
			8 & 255 & 15 & $0.00\ (13)$ & 141 & 219 & 156 & $0.00$ \\
		\end{tabular}
	\end{ruledtabular}
\end{table}

What the rank counts follows from the amplitude tree of Fig.~\ref{fig:amplitude_tree}: every amplitude factorizes along a binary tree into $n$ half-angle factors, one per node on the path of its outcome (SM Sec.~\ref{app:locus}).
An exact fit to the two-peak target prunes every subtree that carries no mass, and the amplitude Jacobian vanishes on the angles inside a pruned subtree. The parameters that survive are the live nodes on the two root-to-leaf paths, $2n - 1$ of them, which is the rank of the quantum Fisher matrix $F_{\mathrm{Q}}$ at that point. The clamped estimator reproduces the rank of $F_{\mathrm{Q}}$ in $96.5\%$ of the runs and is off by at most two in the rest (SM Sec.~\ref{app:rank_clamp}).

The converged unperturbed runs realize the live-node count exactly at small $n$ (Table~\ref{tab:rank_structure}): every run with residual below $0.005$ has rank $5$ at $n = 3$ and $7$ at $n = 4$, $86\%$ have rank $9$ at $n = 5$, and no unperturbed run at any scale attains full rank. From $n = 6$ the count is exceeded, because CMA-ES leaves background mass above the eigenvalue threshold on outcomes that an exact fit would zero.

Perturbed targets have full support, and their fits are at or near full rank, as are the converged fits of B2 and B3. The rank is therefore the support of the output distribution at the resolution of the eigenvalue threshold, and on B1 it coincides with the label. A count of the tree nodes that carry mass above $10^{-6}$ predicts the rank across all $2{,}520$ runs with Pearson $r = 0.9996$ (SM Sec.~\ref{app:rank_nodes}).

Three controls confirm that the rank reads the support and is blind to the defect.
First, a median split on $\epsilon$ among the perturbed runs, which asks each diagnostic to grade defect severity, reduces the mean rank AUROC at $n = 3$--$6$ to $0.54$--$0.56$, while $\W$ and MMD keep $0.89$--$0.95$ against the clean reference.

Second, on the pre-registered stress test, whose families keep the support fixed across levels, the rank AUROC is exactly $0.50$ on a correlation shift of the uniform target at every tested $n$, and $0.55$ falling to $0.37$ (CI $[0.23, 0.53]$) on a collapse of the two-peak target toward one mode (SM Sec.~\ref{app:reproduction} defines both families).
Third, on the full-support benchmark of Sec.~\ref{sec:b3p} the rank AUROC lies between $0.494$ and $0.525$ at every $n = 3$--$8$.

Retraining B1 with Powell's method and with Adam shows that the count the rank reaches is set by the training objective: Adam on $\W$ reproduces the CMA-ES result, whereas Adam on an MMD objective leaves every unperturbed fit at full rank at $n = 3$ (SM Sec.~\ref{sm:sweep}).

The rank does carry one piece of information about a certified verdict. At the certification-grade operating point the rank-deficient share of the certified runs grows from $67$ of $399$ at $n = 3$ to $101$ of $132$ at $n = 8$, and it includes every certified fit to the unperturbed B1 target. These runs resist random parameter drift: at $\|\delta\bm\theta\| = 0.1$ and $n = 4$--$6$ they break in $24$--$50\%$ of random directions against $73$--$80\%$ for full-rank certified runs, but along their top Fisher eigendirection they break in $100\%$ of cases at $n = 6$. The rank thus flags anisotropic robustness (SM Sec.~\ref{sec:rank_fragility}).

\section{Distributional scores and their conventions}\label{sec:pooled}\label{sec:auroc}\label{sec:response}

With both Fisher quantities removed, the pooled distributional scores remain. Their AUROC on B1 moves with the convergence of the optimizer and with the reference distribution.

\subsection{Convergence and reference}

The pooled scores do not decay monotonically with $n$: the clean-reference $\W(C_1)$ AUROC falls to $0.712$ at $n = 5$ and recovers to $0.920$ at $n = 8$, and it is depressed over $n = 5$--$7$ by an amount the seed-cluster intervals resolve (SM Table~\ref{tab:auroc}).
The depression is a labeling effect. The label calls every $\epsilon = 0$ run ``good'', CMA-ES does not always fit the unperturbed target (the unperturbed residuals are $0.058$--$0.170$ at $n \geq 4$), and a run labeled good while distributionally wrong penalizes a metric for correctly flagging it.

\begin{figure}[t]
	\centering
	\includegraphics[width=\columnwidth]{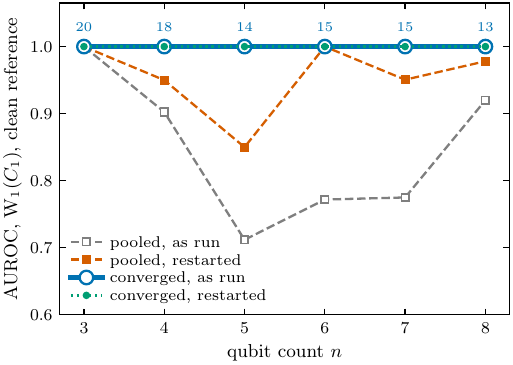}
	\caption{Convergence control on the B1 $\W(C_1)$ clean-reference AUROC under both campaign variants. Pooled (dashed) it moves by up to $0.23$ between the campaign as run and the restarted one; conditioned on converged negatives (residual $\leq 0.02$) it is $1.000$ everywhere and identical under both. Small numerals give the surviving negatives, which bound the precision of each conditioned cell.}\label{fig:conditioned_auroc}
\end{figure}

When we restrict the negative class to converged unperturbed runs, dropping the $\epsilon = 0$ runs whose training residual exceeds $\tau_{\mathrm{fit}} = 0.02$, every $(n, \text{cost})$ cell reaches $\mathrm{AUROC} = 1.000$ for both metrics, at every scale from $n = 3$ to $n = 8$ (Fig.~\ref{fig:conditioned_auroc}).
The same holds for a cut on the other metric, $\MMD_k^2 \leq 10^{-4}$, and for a cut on the total-variation residual.

The cut spends negatives, leaving $20, 18, 14, 15, 15$ and $13$ per $C_1$ cell at $n = 3$ through $8$, and it lies in an observed gap, so the degenerate intervals describe the surviving sample. The optimizer's own stopping rule cannot supply the cut, because CMA-ES exhausts its budget in $20$ of $20$ unperturbed runs at $n = 5$.
A post hoc warm restart of CMA-ES, whose coverage differs between the label classes ($99\%$ of the eligible unperturbed B1 runs against $48\%$ of the perturbed ones), raises the pooled AUROC by $+0.05$ to $+0.34$ and leaves the conditioned AUROC at $1.000$ everywhere (SM Sec.~\ref{app:restart}).

In turn, the reference distribution moves the pooled score independently of convergence.
At $n = 8$ the trained-on and clean scores separate for MMD, $0.735$ against $0.896$ at $C_1$, and the paired seed-cluster difference is $+0.161\,[0.084, 0.243]$; it excludes zero at every cost for MMD and at $C_1$ for $\W$. At $n \leq 7$ the two conventions differ by at most $0.06$ outside $n = 3$.
At $n = 8$ the perturbed targets are the harder ones to fit, so scoring against the trained-on target rewards the wrong runs.

The remaining differences are small: no ground cost is best at every scale, $\W$ scores at or above MMD against the clean reference in all $18$ B1 cells by up to $0.07$, and neither the debiased Sinkhorn divergence~\cite{Feydy2019} nor a change of kernel bandwidth alters the picture (SM Sec.~\ref{sm:conditioning}).

Hence, we use the distributional scores as certification channels only with the scoring convention stated as part of the claim: the negative class is conditioned on convergence, scores are taken against the clean reference, and any post hoc re-optimization is reported with its coverage by label class.

\subsection{A full-support labeled benchmark}\label{sec:b3p}

Every AUROC above is computed on B1, whose perturbation label coincides with the support of the target.
To separate the two, we built a labeled benchmark whose clean target already has full support: the near-uniform B3 target blended with the same random background as B1, $p_\epsilon = (1 - \epsilon)\,p_{\mathrm{B3}} + \epsilon\,\nu$, at the five levels of Eq.~\eqref{eq:b1_blend}, cost $C_1$, $n = 3$--$8$ and $20$ seeds per level.

Three predictions were fixed before launch in a dated document: (P1) the rank AUROC lies in $[0.35, 0.65]$ at every $n$; (P2) the convergence cut lifts the clean-reference $\W$ and MMD AUROCs to at least $0.90$ in every cell with at least $10$ surviving negatives; (P3) converged clean fits are full rank.
The runs at $n = 7$ and $8$ were launched after the outcome at $n \leq 6$ was known, under a dated addendum, written before their launch, that restates P2 for these scales as a residual criterion (SM Sec.~\ref{sm:b3p_full}).

\begin{table}[t]
	\caption{Perturbed full-support benchmark (B3 blends, cost $C_1$, $100$ runs per $n$), clean-reference AUROC on the $\epsilon = 0$ vs $\epsilon > 0$ labels. Pooled: the rank and $\W$. Conditioned: $\W$ after the $\W$ cut (residual $\leq 0.02$) or the MMD cut ($\MMD_k^2 \leq 10^{-4}$), surviving negatives in parentheses, dashed below two. Residual: median of the unperturbed fits; defect: clean-reference $\W$ of the injected defect at $\epsilon = 0.02$; last two columns: per-level $\W$ AUROC. SM Table~\ref{tab:b3p} adds MMD and the intervals.}\label{tab:b3p_short}
	\footnotesize
	\begin{ruledtabular}
		\begin{tabular}{cccccccc}
			$n$ & rank & $\W$ & $\W$ cut & MMD cut & resid. & defect & $\epsilon{=}0.02$\,/\,$0.20$ \\
			\hline
			3 & $0.500$ & $0.981$ & $0.981\ (20)$ & $0.981\ (20)$ & $0.000$ & $0.0069$ & $0.930$\,/\,$1.000$ \\
			4 & $0.500$ & $0.991$ & $0.991\ (20)$ & $0.991\ (20)$ & $0.002$ & $0.0063$ & $0.965$\,/\,$1.000$ \\
			5 & $0.500$ & $0.890$ & $0.996\ (16)$ & $0.935\ (18)$ & $0.004$ & $0.0062$ & $0.823$\,/\,$1.000$ \\
			6 & $0.494$ & $0.811$ & $0.937\ (13)$ & $0.822\ (19)$ & $0.014$ & $0.0061$ & $0.598$\,/\,$1.000$ \\
			7 & $0.525$ & $0.696$ & --\ (1) & $0.738\ (17)$ & $0.043$ & $0.0057$ & $0.495$\,/\,$0.978$ \\
			8 & $0.494$ & $0.734$ & --\ (0) & $0.734\ (20)$ & $0.046$ & $0.0055$ & $0.453$\,/\,$0.975$ \\
		\end{tabular}
	\end{ruledtabular}
\end{table}

P1 and P3 hold at every $n$ (Table~\ref{tab:b3p_short}): the rank AUROC lies between $0.494$ and $0.525$, and $597$ of the $600$ fits are full rank.
P2 as first stated fails. Under the $\W$ cut it fails at $n = 6$, where the conditioned MMD AUROC is $0.863$ ($\W$: $0.937$); under the MMD cut, which the prediction also admits, it fails at $n = 6$--$8$, where $17$ to $20$ negatives survive and the $\W$ AUROC is $0.822$, $0.738$ and $0.734$.

At $n = 7$ and $8$ the clean target does not converge within the budget, and at most one unperturbed fit survives the $\W$ cut, so P2 is untestable there under that cut. The restated criterion names no metric, and we report it without a verdict: at $\epsilon = 0.20$ the per-level AUROC is $0.978$ and $0.975$ for $\W$ and $0.777$ and $0.807$ for MMD (SM Sec.~\ref{sm:b3p_full}).

P2 fails because the injected defect on the near-uniform target is small: at $\epsilon = 0.02$ its clean-reference $\W$ is $0.005$--$0.007$, against $0.016$--$0.058$ on B1, and from $n = 6$ on it lies below the unperturbed residual, where the per-level AUROC falls to chance. Across the $24$ level cells the per-level AUROC rises with the ratio of the defect to the median residual (Spearman rank correlation $0.86$ for $\W$ and $0.93$ for MMD).

The distributional channels therefore resolve a defect on a full-support target to the extent that it exceeds the fit residual. In contrast, the perfect conditioned scores on B1 are a property of the sparse benchmark, on which the smallest injected defect exceeds the largest converged residual by a factor of at least $2.8$ at every $n$ and cost.

\section{The escalation rule and its scope}\label{sec:protocol_sec}\label{sec:protocol}

Certification therefore rests on the two distributional channels. An AUROC of one guarantees that some threshold separates the classes, whereas a fixed pair $(\tau_{\W}, \tau_{\MMD})$ may still fail to, so every claim below is made at a stated operating point. We bound the false-pass probability of the escalation rule $D_4$ and compute, for two exactly solvable defect families, the window of tolerances and qubit numbers in which its discrepancy vote rejects what the transport vote admits.

\begin{figure}[t]
	\centering
	\includegraphics[width=\columnwidth]{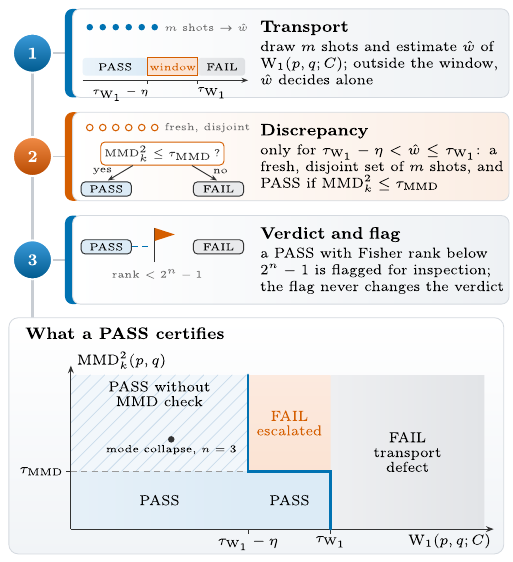}
	\caption{Escalation protocol $D_4$. (1) Stage 1 estimates $\hat w$ from $m$ shots of the generator; an estimate in the clear-PASS zone (blue) or the clear-FAIL zone (grey) decides the verdict alone. (2) Inside the one-sided escalation window $\tau_{\W} - \eta < \hat w \leq \tau_{\W}$ (vermilion), a fresh, disjoint set of $m$ shots decides through the discrepancy vote. (3) A PASS whose Fisher rank is below $2^n - 1$ is flagged for inspection (SM Sec.~\ref{sec:rank_fragility}). Bottom: the acceptance set in the $(\W, \MMD_k^2)$ plane, schematic and not to scale. The blue line bounds the accepted pairs. The discrepancy tolerance is enforced only inside the window, so a pair to its left passes without an MMD evaluation (hatched), as the mode-collapse example at $n = 3$ does.}\label{fig:protocol_schematic}
\end{figure}

Figure~\ref{fig:protocol_schematic} shows the decision flow of $D_4$ and the set of pairs it accepts, and SM Algorithm~\ref{alg:protocol} states the rule in full.
A \emph{transport-defective} pair has $\W(p,q;C) > \tau_{\W}$, the sense of the false-pass bound, and a transport-good pair has $\W \leq \tau_{\W}$.
A PASS of $D_4$ certifies the transport tolerance everywhere and the discrepancy tolerance only inside the window [Eq.~\eqref{eq:acceptance}], so a pair with $\W \leq \tau_{\W} - \eta$ and $\MMD_k^2 > \tau_{\MMD}$ is admitted without a discrepancy evaluation (on the mode-collapse family of Sec.~\ref{sec:dominance} at $n = 3$, $\W = 0.03$ gives $\MMD_k^2 = 1.5 \times 10^{-4}$, which $D_4$ admits at $\eta = 0.01$).
The tolerance-mismatch region of Secs.~\ref{sec:dominance}--\ref{sec:matched} is transport-good by construction, so rejecting it enforces the discrepancy tolerance on the window and corrects no transport false pass.

The window is one-sided because a run on the FAIL side is rejected by the AND rule irrespective of the MMD vote, and letting an MMD=PASS override a $\W$=FAIL would reintroduce the false passes the rule exists to exclude.

With measured shots the protocol samples in two adaptive stages: the $\W$ stage draws its shots unconditionally, and the MMD stage draws a fresh, disjoint set only when the window is entered. The sampling scheme thus enforces the independence the false-pass bound assumes, at an expected cost of $m$ plus $m$ times the escalation frequency; SM Sec.~\ref{sec:dependent_alpha} states the bound for same-data estimation, as in the exact-statevector campaign.

\subsection{False-pass bound}\label{sec:combined}\label{sec:concentration}

Both metrics concentrate at Hoeffding rates when estimated from $m$ samples (the shot count $S$ of the experiments below); SM Sec.~\ref{app:concentration} states the two bounds as SM Propositions~\ref{prop:w1} and~\ref{prop:mmd}. With the target known exactly and the generator sampled, the empirical transport distance deviates from its mean by more than $t$ with probability at most $2\exp(-2mt^2/D_{\max}^2)$, and its upward bias is at most $D_{\max}\sum_x\sqrt{q(x)}/(2\sqrt{m})$. For our ground costs $D_{\max} = O(n)$, so the fluctuation complexity is $O(n^2/t^2)$.

In contrast, the unbiased MMD $U$-statistic deviates from its population value by more than $t$ in either direction with probability at most $\exp(-mt^2/16K_{\max}^2)$ per direction (SM Proposition~\ref{prop:mmd}; for two-sample estimation this is Theorem~10 of Ref.~\cite{Gretton2012}), so with $K_{\max} = 1$ the discrepancy needs $m \geq 16\ln(1/\delta)/t^2$ shots, independent of $n$.

We apply $D_4$ to the empirical estimate $\hat{w}$ and assume that the two statistics use disjoint sample sets, so that their estimation errors are independent.
We write $\delta_{\W}(t) = 2\exp(-2mt^2/D_{\max}^2)$ and $\delta_{\MMD}(t) = \exp(-mt^2/16K_{\max}^2)$ for the tail probabilities of SM Propositions~\ref{prop:w1} and~\ref{prop:mmd} at margin $t$, with $\delta_{\W}(t) := 1$ when the formula exceeds one, and we define the margin zone $\mathcal{M}_\eta = \{(p,q) : |\W(p,q) - \tau_{\W}| < \eta\}$ (the same $\eta$ sets the window and the zone).

Since $\W(p, \cdot\,; C)$ is convex, $\bar{W}_m \geq \W(p, q; C)$ by Jensen's inequality, so a false pass requires a downward deviation from the mean at least as large as that from the true value, and no bias correction is needed.
For $(p,q) \notin \mathcal{M}_\eta$ the escalation adds no protection to the transport vote, and the false-pass probability obeys
\begin{equation}\label{eq:combined_clear}
	\Pr(D_4 \text{ false-pass}) \leq \delta_{\W}(\eta),
\end{equation}
which is identical to the bound for the $\W$-only rule $D_1$.
Inside the margin zone, in contrast, the escalation improves the bound: if $(p,q)$ is \emph{defective} with defect size $\Delta = \W(p,q;C) - \tau_{\W} > 0$ and \emph{jointly detectable}, in the sense that $\MMD_k^2(p,q) \geq \tau_{\MMD} + \xi$ for an MMD margin $\xi > 0$, then
\begin{equation}\label{eq:combined_bound}
	\Pr(D_4 \text{ false-pass}) \leq \underbrace{\delta_{\W}(\eta)}_{\text{escalation miss}} + \underbrace{\delta_{\W}(\Delta)\,\delta_{\MMD}(\xi)}_{\text{joint misvote}},
\end{equation}
as we prove in SM Sec.~\ref{sm:bounds}.
The escalation-miss term appears because $D_4$ triggers on the empirical $\hat{w}$: a defective pair whose estimate falls at $\hat{w} \leq \tau_{\W} - \eta$ is certified without a discrepancy vote. For defects with margin $\Delta \geq t$ the second term is the quadratic product $\delta_{\W}(t)\,\delta_{\MMD}(\xi)$. The joint-detectability hypothesis is necessary, and the constants are loose at realistic shot budgets (Sec.~\ref{sec:finite_shot_falsepass}).

The bound translates into an explicit quadratic improvement on false pass over single-metric certification.
We fix a target false-pass level $\delta$ for defects with margin $\Delta \geq t$, allocate $\delta_{\W}(t) = \delta_{\MMD}(\xi) = \sqrt{\delta/2}$ to the joint-misvote term, and size the escalation window so that $\delta_{\W}(\eta) = \delta/2$, which yields
\begin{equation}\label{eq:window_ratio}
	\eta = t\,\sqrt{\frac{2\ln(4/\delta)}{\ln(8/\delta)}}.
\end{equation}
With this choice Eq.~\eqref{eq:combined_bound} gives $\Pr(D_4 \text{ false-pass}) \leq \delta/2 + \delta/2 = \delta$, while the $\W$-only rule $D_1$ at the same per-metric sample size achieves only $\delta_{\W}(t) = \sqrt{\delta/2}$.
For $\delta = 0.05$ the window ratio of Eq.~\eqref{eq:window_ratio} is $\eta \approx 1.31\,t$, and the comparison reads $D_4 \leq 0.05$ vs.\ $D_1 \leq 0.158$.
This compares upper bounds at equal sample size per metric. SM Sec.~\ref{sm:baselines} restates the comparison at equal total budget, compares $D_4$ with the shifted rule $D_1(\tau_{\W} - \eta)$, bounds the false-fail probability, and identifies the composition as the intersection--union test of Berger and Hsu~\cite{BergerHsu1996}.

\subsection{The tolerance-mismatch window on the bitstring space}\label{sec:dominance}

On a finite outcome space $\W$ and the MMD of a characteristic kernel induce the same topology~\cite{Sriperumbudur2010,Binkowski2018}, and the rejection region of any AND rule contains that of the transport rule. We therefore compute the size of the difference at a stated operating point, exactly, for two defect families on which both metrics admit closed forms.

\begin{theorem}[Tolerance-mismatch window on $\{0,1\}^n$]\label{thm:dominance}
	Let $n \geq 2$, let $k(x,y) = \exp(-\gamma\, d_H(x,y))$ be the RBF--Hamming kernel with bandwidth $\gamma > 0$, and fix a mode $x_0 \in \{0,1\}^n$.
	Take the uniform target $p = \mathrm{Unif}(\{0,1\}^n)$ and the mode-contaminated generator family
	\begin{equation}\label{eq:mixture_witness}
		q_\lambda = (1-\lambda)\,\mathrm{Unif}(\{0,1\}^n) + \lambda\,\delta_{x_0}, \qquad \lambda \in (0, 1],
	\end{equation}
	which places an excess weight $\lambda$ on the single mode $x_0$ and interpolates between the target at $\lambda \to 0$ and complete collapse onto $x_0$ at $\lambda = 1$, the sense in which we call the family mode collapse. Then the following identities are exact for every $\lambda$, $n \geq 2$, and $\gamma > 0$:
	\begin{enumerate}
		\item[\textup{(a)}] $\W(p, q_\lambda; C_1) = \lambda\, n/2$ under the Hamming cost;
		\item[\textup{(b)}] $\TV(p, q_\lambda) = \lambda\,(1 - 2^{-n})$;
		\item[\textup{(c)}] $\MMD_k^2(p, q_\lambda) = \lambda^2 \Phi_n$, with $\Phi_n = 1 - c^n$ and $c = (1 + e^{-\gamma})/2$.
	\end{enumerate}
	Fix a $\W$ threshold $\tau_{\W} > 0$ and an escalation window $\eta \in (0, \tau_{\W})$. For every
	\begin{equation}\label{eq:witness_interval}
		\lambda \in \Bigl(\tfrac{2(\tau_{\W} - \eta)}{n},\ \tfrac{2\tau_{\W}}{n}\Bigr],
	\end{equation}
	the witness has $\W(p, q_\lambda; C_1) \in (\tau_{\W} - \eta,\ \tau_{\W}]$ and lies in the bi-metric rejection region but not in the $\W$-only one, provided
	\begin{equation}\label{eq:dominance_condition}
		\tau_{\MMD} \;\leq\; \left(\frac{2(\tau_{\W} - \eta)}{n}\right)^{\!2}\Phi_n
	\end{equation}
	holds, with $\mathcal{R}_{\W} = \{q : \W(p,q;C_1) > \tau_{\W}\}$ and $\mathcal{R}_{\textup{bi}} = \{q : D_4(q) = \textup{FAIL}\}$; the tolerance-mismatch region $\mathcal{R}_{\textup{bi}} \setminus \mathcal{R}_{\W}$ is then non-empty. The condition holds for every window $\eta \leq \eta_{\max}(n)$, with
	\begin{equation}\label{eq:eta_max}
		\eta_{\max}(n) = \tau_{\W} - \frac{n}{2}\sqrt{\frac{\tau_{\MMD}}{\Phi_n}}.
	\end{equation}
\end{theorem}

We give the proof, which evaluates the transport plan, the total variation and the mean embedding of the witness in closed form, in SM Sec.~\ref{sm:bounds}.

At the certification-grade tolerance $\tau_{\MMD} = 10^{-4}$, with $\tau_{\W} = 0.05$ and $\gamma = 1/n$, Eq.~\eqref{eq:eta_max} gives $\eta_{\max}(n) = 0.025,\ 0.017,\ 0.009,\ 0.001$ at $n = 3, 4, 5, 6$, an open window of relative width $0.51,\ 0.35,\ 0.19,\ 0.03$ times $\tau_{\W}$. The separation is robust for $n \leq 5$ and narrows at $n = 6$ to the knife-edge $\lambda = 2\tau_{\W}/n$, the limit $\eta \to 0$. For $n \geq 7$ the window closes, $\eta_{\max}(n) = -0.007,\ -0.015$ at $n = 7, 8$, because the right-hand side of Eq.~\eqref{eq:dominance_condition} at $\eta = 0$ drops below $\tau_{\MMD} = 10^{-4}$ (to $7.8 \times 10^{-5}$ and $6.0 \times 10^{-5}$).

We call the window \emph{open} at $n$ when $\eta_{\max}(n) > 0$: every escalation window $\eta \leq \eta_{\max}(n)$ then rejects all in-window members of the family, and $\eta_{\max}(n)/\tau_{\W}$ is the relative width of the window. A non-empty mismatch interval alone requires less, and SM Sec.~\ref{sm:matched} keeps the two notions apart.

The mismatch originates in a linear-versus-quadratic asymmetry: the $\W$ penalty is linear in the contamination while the MMD penalty is quadratic, so the witness is detected only while $\tau_{\W} \gtrsim (n/2)\sqrt{\tau_{\MMD}}$.

\subsection{Two defect families}\label{sec:blindspots}

We call the mode-collapse construction of Theorem~\ref{thm:dominance} Family~1.

For Family~2, the correlation shifts, we let $p$ be uniform on $\{0,1\}^n$ and take
\begin{equation}
	q(x) = 2^{-n}\bigl[1 + \beta\,(-1)^{x_0 \oplus x_1}\bigr], \qquad \beta \in (0, 1),
\end{equation}
which matches every single-qubit marginal of $p$ but introduces a parity correlation $\beta$ between qubits $0$ and $1$; it is normalized because $\sum_x (-1)^{x_0 \oplus x_1} = 0$.
Its transport distance equals its total-variation weight, $\W(p, q; C_1) = \TV(p, q) = \beta/2$: flipping $x_0$ carries the excess on the strings with $x_0 = x_1$ to the deficit at unit cost, and the potential $\tfrac12 (-1)^{x_0 \oplus x_1}$ attains the bound (SM Sec.~\ref{sm:bounds}).
The RBF--Hamming MMD, in turn, reads exactly
\begin{equation}\label{eq:family2_mmd}
	\MMD_k^2(p, q) = \beta^2 A_n, \qquad A_n = b^2 c^{\,n-2},
\end{equation}
with $b = (1 - e^{-\gamma})/2$ and $c$ as in Theorem~\ref{thm:dominance}.
The transport response is linear in $\beta$, whereas the discrepancy response is quadratic, with a bracket $A_n$ that decays as $n^{-2}$ at $\gamma = 1/n$ (SM Eq.~\eqref{eq:family2_asymptotic}).

Because both metrics are again exact, the window admits the same closed form as Eq.~\eqref{eq:eta_max}, now reading $\eta_{\max}(n) = \tau_{\W} - \tfrac12\sqrt{\tau_{\MMD}/A_n}$. It closes far sooner than the mode-collapse window: at $(\tau_{\W}, \tau_{\MMD}, \gamma) = (0.05, 10^{-4}, 1/n)$ it is open only at $n = 3$, with relative width $0.24$, and already negative at $n = 4$.

The correlation signal is the smaller per unit transport distance, by the factor $\Phi_n/(n^2A_n) = 2.4$--$2.5$ at $n = 3$--$8$ (limit $2.6$), the square of the threshold ratio of Sec.~\ref{sec:matched}.

\begin{figure}[t]
	\centering
	\includegraphics[width=\columnwidth]{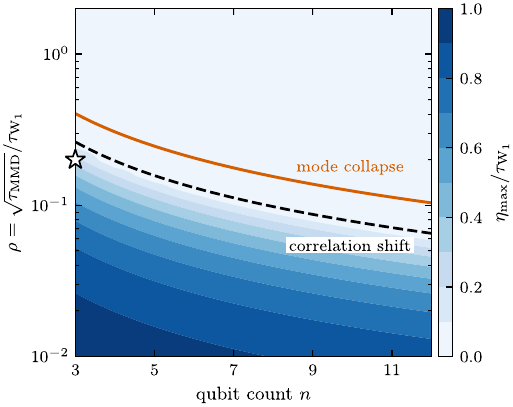}
	\caption{Escalation window in the $(n, \rho)$ plane, $\rho = \sqrt{\tau_{\MMD}}/\tau_{\W}$, the only combination through which the operating point enters either closed form. Shading: the deployment window, the minimum over the two families. Lines: closure boundaries for mode collapse (solid) and correlation shift (dashed). The star marks the certification-grade point, where the deployment window is open at $n = 3$ alone.}\label{fig:window_phase}
\end{figure}

Both windows depend on the operating point through a single dimensionless ratio, so the scope claim holds for every pair $(\tau_{\W}, \tau_{\MMD})$ with the same ratio. Writing $\rho = \sqrt{\tau_{\MMD}}/\tau_{\W}$, the two closed forms collapse to
\begin{equation}\label{eq:window_ratio_form}
	\frac{\eta_{\max}}{\tau_{\W}} = 1 - \frac{n}{2}\frac{\rho}{\sqrt{\Phi_n}}
	\qquad\text{and}\qquad
	1 - \frac{1}{2}\frac{\rho}{\sqrt{A_n}},
\end{equation}
so each family closes above a critical ratio, $\rho^{*}_1(n) = 2\sqrt{\Phi_n}/n$ and $\rho^{*}_2(n) = 2\sqrt{A_n}$, however $\tau_{\W}$ and $\tau_{\MMD}$ are chosen separately. Figure~\ref{fig:window_phase} maps that plane, and the certification-grade point of Sec.~\ref{sec:blindspot_demo} sits at $\rho = 0.2$.

A practitioner certifying a generator does not know which defect is present, so the escalation window must be sized against the minimum over the families considered plausible. At $\rho = 0.2$ that minimum is $+0.24$ at $n = 3$ and negative from $n = 4$ onward, the correlation-shift family binding at every scale. The mode-collapse figures, open through $n = 6$, therefore describe the best case. Reopening the window at larger $n$ requires lowering $\rho$, that is, tightening $\tau_{\MMD}$ faster than $\tau_{\W}^2$.

Adding the parity feature of the suspected pair to the kernel restores an $n$-independent signal $\beta^2$; against an unknown pair the certified shot cost of such a repair grows as $O(n^4/\beta^4)$ (SM Sec.~\ref{sm:kernel_repair}).

\subsection{Correlation order and the redundancy boundary}\label{sec:parity}

The two families are instances of a structure that the RBF--Hamming kernel makes exact.
We write $\chi_S(x) = (-1)^{\sum_{i \in S} x_i}$ for the parity of a subset $S \subseteq \{0, \ldots, n-1\}$, so that the $2^n$ parities are orthonormal under the uniform distribution $u$.
The kernel factorizes over qubits, $e^{-\gamma d_H(x,y)} = \prod_i [\,c + b\,(-1)^{x_i \oplus y_i}]$ with $c$ and $b$ as above, and expanding the product gives
\begin{align}\label{eq:kernel_fourier}
	k(x,y) &= \sum_{S} c^{\,n - |S|}\, b^{\,|S|}\, \chi_S(x)\chi_S(y), \nonumber\\
	\MMD_k^2(p,q) &= \sum_{S \neq \varnothing} c^{\,n - |S|}\, b^{\,|S|} \bigl(\mathbb{E}_p\chi_S - \mathbb{E}_q\chi_S\bigr)^2 .
\end{align}
The weight of a parity-moment mismatch falls geometrically in its order $|S|$.
Theorem~\ref{thm:dominance} and Family~2 read off from it: mode collapse excites every parity, $\lambda^2\sum_{S \neq \varnothing} c^{n-|S|}b^{|S|} = \lambda^2(1 - c^n) = \lambda^2\Phi_n$, and the correlation shift excites a single parity of order two, $\beta^2 c^{n-2}b^2 = \beta^2 A_n$.
The expansion also bounds, for any kernel, how evenly that sensitivity can be spread.

\begin{theorem}[Parity defects]\label{thm:parity}
	Let $u$ be uniform on $\{0,1\}^n$ and, for $\varnothing \neq S \subseteq \{0, \ldots, n-1\}$ and $0 < \beta < 1$, let $q_{S,\beta}(x) = 2^{-n}[1 + \beta\chi_S(x)]$.
	The distribution $q_{S,\beta}$ has full support, matches every marginal of $u$ on fewer than $|S|$ qubits, and is the Born distribution of a circuit of $n$ single-qubit rotations and $|S| - 1$ CNOT gates.
	\begin{enumerate}
		\item[\textup{(a)}] $\TV(u, q_{S,\beta}) = \W(u, q_{S,\beta}; C_1) = \beta/2$ for every $S$.
		\item[\textup{(b)}] For the RBF--Hamming kernel, $\MMD_k^2(u, q_{S,\beta}) = \beta^2 c^{\,n - |S|} b^{\,|S|}$, which at $\gamma = 1/n$ and fixed $r = |S|$ equals $e^{-1/2}\beta^2(2n)^{-r}[1 + O(n^{-1})]$.
		\item[\textup{(c)}] For every positive-semidefinite kernel with $k(x,x) = 1$,
		\begin{equation}\label{eq:minimax}
			\min_{S \neq \varnothing} \MMD_k^2(u, q_{S,\beta}) \leq \frac{\beta^2}{2^n - 1},
		\end{equation}
		with equality for the centered kernel $k_*(x,y) = (2^n\,\mathbf{1}[x = y] - 1)/(2^n - 1)$, so that the supremum of the left-hand side over normalized kernels equals the right-hand side.
	\end{enumerate}
\end{theorem}

\begin{figure}[t]
	\centering
	\includegraphics[width=\columnwidth]{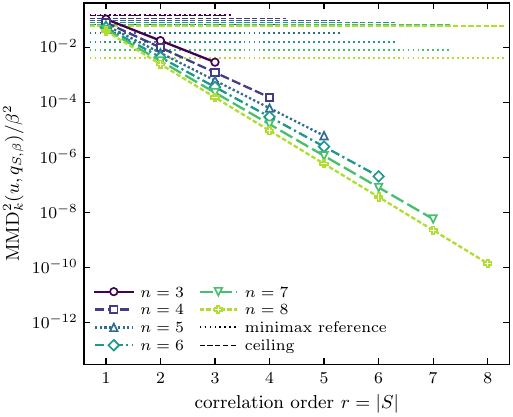}
	\caption{Squared discrepancy of the parity defect $q_{S,\beta}$ against the uniform target at fixed transport distance $\beta/2$, by correlation order $r = |S|$, for the RBF--Hamming kernel at $\gamma = 1/n$ (Theorem~\ref{thm:parity}(b)). Dotted: the minimax reference $\beta^2/(2^n - 1)$ of Eq.~\eqref{eq:minimax}, the largest worst-parity signal any normalized kernel attains; it bounds the minimum over $S$ from above. Dashed: the Hamming-cost ceiling $2(1 - e^{-\gamma})\W^2$ of Eq.~\eqref{eq:redundancy}.}\label{fig:parity}
\end{figure}

The proof is in Appendix~\ref{app:parity_proofs}; part~(c) rests on the identity $\sum_S a_S = 1$ for the parity weights $a_S = 2^{-2n}\chi_S^{\!\top}K\chi_S$ of any unit-diagonal kernel, so that a kernel has one unit of sensitivity to distribute over $2^n - 1$ correlation directions.
Figure~\ref{fig:parity} shows the hierarchy of part~(b): at the same transport distance $\beta/2$, a single-qubit bias costs the discrepancy $\Theta(\beta^2/n)$, a two-qubit correlation $\Theta(\beta^2/n^2)$, an $r$-qubit correlation $\Theta(\beta^2/n^r)$, and the full parity $\beta^2 b^n$.

Part~(c) states that no bandwidth, and no other normalized kernel, gives a scale-independent signal on all of these alternatives: the centered kernel $k_*$ flattens the hierarchy, responding equally to every nonempty parity, but only at the level $\beta^2/(2^n - 1)$, which for $\beta = 0.1$ lies below the certification-grade tolerance from $n = 7$ on.
Hence, a certification verdict must name the correlation orders it is meant to cover.

Part~(c) concerns the population signal and fixes the kernel before the defect direction is chosen, the quantifier order certification needs. Kurkin \emph{et al.}~\cite{Kurkin2025} show the converse order, a pair at total variation one that defeats every stationary characteristic kernel of a class; SM Sec.~\ref{sm:kurkin} compares the two statements, neither of which contains the other.

In the other direction, the kernel bounds the discrepancy by the transport distance, which says exactly when the second channel can contribute nothing.

\begin{theorem}[Population redundancy boundary]\label{thm:redundancy}
	For the Hamming cost $C_1$ and the RBF--Hamming kernel, every pair $p, q \in \simplex$ satisfies
	\begin{equation}\label{eq:redundancy}
		\MMD_k^2(p,q) \leq 2\,(1 - e^{-\gamma})\,\W(p,q;C_1)^2,
	\end{equation}
	with equality for $p = \delta_x$ and $q = (1 - w)\delta_x + w\delta_y$ at Hamming-adjacent $x, y$ and $0 \leq w \leq 1$.
	Consequently, for $0 < \tau_{\W} \leq 1$, the discrepancy vote on the population values is redundant on the whole transport acceptance region, so that $D_4$ and $D_1(\tau_{\W})$ evaluated on the same pair $(p, q)$, exact or plug-in, agree for every window $\eta > 0$, if and only if $\tau_{\MMD} \geq 2(1 - e^{-\gamma})\,\tau_{\W}^2$.
\end{theorem}

The inequality is an instance of the bound $\MMD_k(p,q) \leq \W(p,q;\rho_k)$ of Theorem~21(i) of Sriperumbudur \emph{et al.}~\cite{Sriperumbudur2010} with the kernel-induced ground metric; the theorem adds the sharpness, since the adjacent pair attains the constant, and the decision statement. The proof is in Appendix~\ref{app:parity_proofs}.

At $\tau_{\W} = 0.05$ and $\gamma = 1/n$ the boundary is $1.4 \times 10^{-3}$, $1.1 \times 10^{-3}$, $9.1 \times 10^{-4}$, $7.7 \times 10^{-4}$, $6.7 \times 10^{-4}$ and $5.9 \times 10^{-4}$ at $n = 3$--$8$. The loose tolerance $\tau_{\MMD} = 10^{-2}$ makes the discrepancy redundant for every pair of distributions at every scale we study, the intermediate $10^{-3}$ from $n = 5$ on, and the certification-grade $10^{-4}$ at none. This is the population form of the finding of SM Sec.~\ref{sec:cross_run} and Sec.~\ref{sec:finite_shot_falsepass} that the escalation changes no verdict at the loose tolerance.
The theorem concerns population values; the finite-shot agreement at the loose tolerance is measured in Sec.~\ref{sec:b3p_validation}, and SM Sec.~\ref{sm:redundancy} gives the finite-shot margin and the adjacent-pair window, which contains the window of every defect family.

\subsection{Matched tolerances}\label{sec:matched}

On each family the escalation acts as a transport-only rule with an effective threshold. Inside the window of family $i$ it is the shifted rule $D_1(\tau_{\W} - \eta)$, and outside every window it is $D_1(\tau_{\W})$ (SM Sec.~\ref{sm:matched}). The discrepancy tolerance corresponds on each family to the transport distance at which that family's discrepancy crosses it, $\tau^{*}_1 = \tfrac{n}{2}\sqrt{\tau_{\MMD}/\Phi_n}$ and $\tau^{*}_2 = \tfrac12\sqrt{\tau_{\MMD}/A_n}$. With the certification-grade tolerances at $n = 3$ these are $0.025$ and $0.038$; a window $\eta = 0.005$ places both below $\tau_{\W} - \eta = 0.045$, so the escalation acts on either family as the single rule $D_1(0.045)$.

The ratio $\tau^{*}_2/\tau^{*}_1 = \sqrt{\Phi_n/(n^2 A_n)}$ equals $1.54$ to $1.58$ at $n = 3$--$8$ under the Hamming cost, with limit $2\sqrt{e^{1/2} - 1} \approx 1.61$, and does not depend on the calibration. At $\rho = 0.2$ the discrepancy threshold is tighter than the transport threshold on both families at $n = 3$, tighter on mode collapse and looser on correlation shift at $n = 4$--$6$, and looser on both from $n = 7$. A certifier who wants a given transport tolerance enforced on both families must therefore name the family.

\section{Tests of the escalation rule}\label{sec:validation}

We test the statements of Sec.~\ref{sec:protocol_sec} on sampled defective pairs, on the held-out full-support benchmark, and on trained generators; SM Sec.~\ref{sm:campaign} decomposes the verdicts of the exact-statevector campaign.
Throughout, $\tau_{\W} = 0.05$ and $\tau_{\MMD}$ is $10^{-2}$ (loose) or $10^{-4}$ (certification grade) on the squared discrepancy; $D_1$ is the transport-only rule and $D_4$ the escalation, $\eta$ is the escalation window, $\Delta$ the transport excess of a defective pair over $\tau_{\W}$, and $S$ the number of shots per metric (SM Table~\ref{tab:symbols} lists every symbol).

\subsection{Finite-shot false-pass rates}\label{sec:finite_shot_falsepass}

The campaign is exact-statevector, so it never exercises the independence mechanism that makes the bound of Eq.~\eqref{eq:combined_bound} quadratic.
For each defective pair we therefore draw two disjoint sets of $S$ shots from the generator, evaluate $\W$ on one and the MMD $U$-statistic on the other, and run both $D_4$ and $D_1$, repeating $2{,}000$ times per configuration at $S = 10^2$ through $10^5$ with the window $\eta = \sqrt{2} \times 0.01$. The defective pairs are mode-collapse witnesses of Theorem~\ref{thm:dominance} placed a distance $\Delta \in \{0.005, 0.02\}$ above $\tau_{\W}$ and nine trained generators whose clean-reference $\W$ exceeds $\tau_{\W}$ (SM Sec.~\ref{sm:finite_shot}).

At the loose tolerance the escalation contributes essentially nothing: $D_4$ and $D_1$ return identical false-pass rates in $55$ of $60$ configurations, and the five exceptions occur at $S = 100$ only.
In contrast, at the certification-grade tolerance the escalation lowers the false-pass rate: among the $29$ configurations in which $D_1$ falsely certifies at a nonzero rate, $D_4$ drives the rate to zero in $2{,}000$ repetitions in $10$, and where both are nonzero the reduction is a median factor of $2.0$ (range $1.2$ to $55$), against the factor of $3.2$ of the upper bounds.

The realized rates stay below the bound of Eq.~\eqref{eq:combined_bound} in all $120$ configurations; the bound is loose because its Hoeffding tails use only the range of the statistic, and an empirical-Bernstein tail at the measured variance is tighter by a factor of $2.6 \times 10^{3}$ to $2.0 \times 10^{5}$.
The escalation keeps the false-fail rate of $D_1(\tau_{\W})$ to within a factor of two, whereas the shifted rule $D_1(\tau_{\W} - \eta)$ obtains its lower false-pass rate at the cost of a false-fail rate forty times larger inside the window. Of the $776$ unperturbed campaign fits that the transport threshold admits, $70$ lie in the window; the escalation keeps $62$ of them and rejects $8$, where the shifted rule would reject all $70$.

\subsection{Fixed thresholds on the full-support benchmark}\label{sec:b3p_validation}

We also test the protocol on targets never used to set a threshold. The $600$ fits of the full-support benchmark of Sec.~\ref{sec:b3p} are certified against the clean B3 target at the thresholds, kernel, window and budgets of the finite-shot test, with $200$ repetitions per fit. We compare $D_4$ with $D_1(\tau_{\W})$ and with a single transport threshold $\tau^{*}$ recalibrated on held-out seeds to the false-fail rate and the shot cost of $D_4$, the comparison least favorable to the escalation; the matching rule and the held-out calibration were fixed in advance (SM Sec.~\ref{sm:fixed_threshold}).

At the loose tolerance no fit is a mismatch pair at any $n$, as Theorem~\ref{thm:redundancy} requires of the population values, and $D_4$ returns the rates of $D_1(\tau_{\W})$ to the last digit in every cell, with no disagreement on any of the $41{,}771$ in-window draws taken on the fixed panel of $600$ fits (one-sided $95\%$ upper limit $7 \times 10^{-5}$ per in-window draw).
At certification grade the mismatch fits number $22$, $11$, $5$ and $1$ at $n = 3$--$6$ and none beyond, and at $S = 10^5$ the escalation rejects more of them than the recalibrated rule at $n = 4$--$6$ ($65$--$97\%$ against $0$--$60\%$) and fewer at $n = 3$ ($43\%$ against $55\%$).

On the transport-defective fits the escalation's false-pass rate equals that of the recalibrated threshold $\tau^{*}$, to the third decimal, in six of the nine cells with false passes; the recalibrated rule is lower in the two cells at $n = 5$, and the escalation is lower in one cell, at $n = 3$ and $S = 10^4$, by $0.003$.
Moreover, at small budgets the upward bias of $\hat{w}$ rejects nearly every draw of the good fits under every rule (false-fail rate above $0.999$ at $n = 5$ and $6$ with $S = 10^3$, and at $n \geq 7$ with $S \leq 10^4$), because their $\W$ sits near the threshold.

\subsection{Trained generators in the tolerance-mismatch region}\label{sec:blindspot_demo}

Theorem~\ref{thm:dominance} and the Family 2 construction of Sec.~\ref{sec:blindspots} establish the mismatch regions as differences of acceptance regions. To confirm that trained generators can occupy them, we instantiate both constructions directly at $n = 3$ through $6$. On the uniform reference $p = u$ we build the defective target of each family, placing its parameter at the midpoint of the escalation window so that $\W(u, q) \leq \tau_{\W}$ by construction. We then train the M\"ott\"onen QCBM to fit that target under the $C_1$ cost, twelve seeds per cell, and certify the trained output $\hat{q}$ against the clean reference $u$.

Both families admit an exact window, so the construction fixes the thresholds.
At the certification-grade operating point the mode-collapse window is open through $n = 6$ and the correlation-shift window only at $n = 3$ (Secs.~\ref{sec:dominance} and~\ref{sec:blindspots}). Where the window is too narrow to train into reliably, we tighten $\tau_{\MMD}$ to a quarter of the value at which the window closes, which pins $\eta_{\max}$ to $\tau_{\W}/2$ exactly, and keep the cell. Table~\ref{tab:blindspot_demo} marks those cells, and SM Sec.~\ref{app:blindspot_detail} gives their thresholds.

\begin{table}[t]
	\caption{Trained generators in the two mismatch regions ($12$ seeds per cell), certified against the clean uniform reference at $\tau_{\W} = 0.05$. Cells marked $\dagger$ use the tightened $\tau_{\MMD}$ given in SM Sec.~\ref{app:blindspot_detail}. Seeds absent from the $D_1$ PASS column overshot $\tau_{\W}$ and are rejected by $\W$ alone, so they lie outside the mismatch region.}\label{tab:blindspot_demo}
	\begin{ruledtabular}
		\begin{tabular}{ccccc}
			Family & $n$ & $D_1$ PASS & $D_4$ FAIL & mean $\MMD_k^2(u,\hat q)$ \\
			\hline
			\multirow{4}{*}{mode collapse}     & 3 & 12/12 & 12/12 & $2.3 \times 10^{-4}$ \\
			                                   & 4 & 11/12 & 11/11 & $1.7 \times 10^{-4}$ \\
			                                   & 5 &  7/12 &  6/7  & $1.4 \times 10^{-4}$ \\
			                                   & 6 & 10/12 & 10/10 & $7.3 \times 10^{-5}$\,$\dagger$ \\
			\hline
			\multirow{4}{*}{correlation shift} & 3 & 12/12 & 12/12 & $1.3 \times 10^{-4}$ \\
			                                   & 4 & 12/12 & 12/12 & $5.4 \times 10^{-5}$\,$\dagger$ \\
			                                   & 5 &  9/12 &  9/9  & $6.3 \times 10^{-5}$\,$\dagger$ \\
			                                   & 6 &  5/12 &  5/5  & $4.9 \times 10^{-5}$\,$\dagger$ \\
		\end{tabular}
	\end{ruledtabular}
\end{table}

Table~\ref{tab:blindspot_demo} reports the outcome: across the eight cells $78$ of the $96$ trained generators land inside the $\W$-only acceptance region, and the escalation rejects $77$ of those $78$. The single escape is a mode-collapse run at $n = 5$ whose trained output drifted to $\MMD_k^2 = 9.6 \times 10^{-5}$, just under the $10^{-4}$ tolerance, in the untightened cell with the narrowest window ($\eta_{\max}/\tau_{\W} = 0.19$). The eighteen generators that overshot $\tau_{\W}$ cluster at $n = 5$ for mode collapse, the same narrow-window cell, and at $n = 6$ for correlation shift.

We also find mismatch pairs in the campaign itself at certification grade. Among the $1{,}052$ campaign runs that $\W$-only certification admits against the clean reference, $56$ carry $\MMD_k^2 > 10^{-4}$, and the escalation rejects the $37$ of them that lie in its window. Thirty-five of the $56$ are B1 runs at $\epsilon = 0.05$, whose injected defect lies inside the transport tolerance; the other $21$, unperturbed B2 and B3 fits and thus the route of ordinary training toward a legitimate target, are $2.7\%$ of the $776$ admitted unperturbed fits, and the escalation rejects the $8$ of them in its window. At $\tau_{\MMD} = 10^{-3}$ and $10^{-2}$ the population is empty at every scale, as Theorem~\ref{thm:redundancy} requires at $10^{-2}$ (SM Sec.~\ref{sm:campaign}).

\section{Discussion and limitations}\label{sec:discussion}\label{sec:rank_diagnostic}

The headline score of a diagnostic on a constructed benchmark does not show whether it measured the distributional property or recovered the control variable.
Sections~\ref{sec:cap} and~\ref{sec:pooled} give three instances of this failure: the trace is pinned by construction, the rank counts the support of the converged distribution, and the pooled scores move with the scoring convention.
Five inexpensive controls expose them: a severity split among the perturbed runs, a convergence restriction of the negative class, a stress test across independently constructed defect families, a replication under further optimizers and objectives, and a labeled benchmark whose clean target has full support. We recommend them whenever a diagnostic is scored by AUROC against a constructed benchmark.

The $\W$/MMD escalation is one choice among the distributional channels that remain; total variation, the Sinkhorn divergence and spectral refinements of the rank are admissible in its place. We chose the pair because both metrics admit closed forms on the defect families of Sec.~\ref{sec:blindspots} and their composition admits the finite-sample bound of Eq.~\eqref{eq:combined_bound}. The escalation is useful only where the generator is approximately correct but may harbor a defect one metric cannot see; Theorem~\ref{thm:redundancy} makes the redundancy exact on the Hamming cost.

The rank does not enter the decision. It flags certified verdicts with anisotropic robustness, and we use it to trigger inspection where structured drift is plausible; adding it as a decision channel would change the certification question from ``does the output distribution match the target?'' to ``does it match, and is the converged parameter regime robust?''.

The Hamming cost $C_1$ is the cost induced by the order-1 quantum-Wasserstein distance of De Palma, Marvian, Trevisan, and Lloyd, which on computational-basis-diagonal states reduces to the classical $\W$ with $C_1$~\cite[Proposition~6]{DePalma2021}, a reduction restated as Eq.~(67) of Ref.~\cite{Beatty2024}; $C_2$ and $C_3$ are classical refinements that are not restrictions of it. The induced cost is not the best-scoring one at every scale, since $C_2$ scores higher by $0.24$ at $n = 5$, so which cost certifies best at a given scale remains an empirical question.

The identity rests on a pure, real state, and noise removes it. A depolarizing rate of $10^{-3}$, applied to the stored Jacobians, drops the trace to $0.033$--$0.14$ of the parameter count on the sparse B1 fits and leaves the near-uniform B3 fits at $0.998$, so under noise the trace regains a dependence on the entropy of the target, which is unrelated to fit quality (SM Sec.~\ref{sec:noise}).

Moreover, the naive certification AUROC inverts under depolarizing and readout error, because noise pushes outputs toward uniform and penalizes concentrated targets more than diffuse ones. Settling the correct reference under noise and extending the bound of Eq.~\eqref{eq:combined_bound} to device noise requires circuit-level simulation, and our false-pass guarantee holds for noiseless sampling only.

The campaign covers $n = 3$--$8$, bounded by the exact-statevector requirement and the trainability frontier at $n = 9$. The closure of the mode-collapse window at $n \geq 7$ near that frontier is accidental: Eq.~\eqref{eq:window_ratio_form} involves no property of the training, and the window is open at any $n$ for sufficiently small $\rho$, at a shot cost above any budget we exercise (SM Sec.~\ref{sec:shot_budget}).

The empirical specialization of Proposition~\ref{prop:cap_general} covers the M\"ott\"onen ansatz trained by CMA-ES, with the rank control replicated under Powell and Adam at $n \leq 5$. The theoretical results are model-agnostic on the appropriate ansatz class, whereas the disagreement rates are specific to this model; QGANs and diffusion models may occupy the mismatch regions differently.

At certification grade, four of the eight trained-generator cells required a $\tau_{\MMD}$ tightened below $10^{-4}$, and the ground-cost mismatch has no trained-generator demonstration. In addition, on the fixed-threshold test a recalibrated transport threshold matches the false-pass rate of the escalation in most cells. Theorem~\ref{thm:parity}(b) and Theorem~\ref{thm:redundancy} are exact for the Hamming cost and the RBF--Hamming kernel, and the redundancy constant changes with the cost of an adjacent flip.

We scope the convergence-conditioned claims to B1 and to defects that exceed the fit residual, since the synthetic B2 targets at $n \geq 4$ and the synthetic B3 target limit the external validity of the conclusions that rest on them. The zero-amplitude locus $\mathcal{Z}$ is explicit for the M\"ott\"onen ansatz, but its relation to the parameters the optimizer reaches is open.

\section{Conclusion}\label{sec:conclusion}

To conclude, we have shown that the Fisher information of a real-amplitude quantum circuit Born machine cannot certify the trained model, and that certification by distributional metrics has blind spots that can be computed before any measurement.

For any circuit that prepares real amplitudes through half-angle rotation generators, classical and quantum Fisher information coincide away from the zero-amplitude set, so the trace is pinned to the number of parameters and carries no signal about the fit. The Fisher rank separates the benchmark classes because it reads the support of the converged distribution: on the two-peak target an exact fit leaves $2n - 1$ live nodes of the amplitude tree. It falls to chance on a severity split and on full-support defect families.

We found that the pooled distributional scores respond to optimizer convergence and to the reference against which they are taken. Conditioned on converged negatives they are perfect on the sparse benchmark, and on a full-support benchmark they resolve a defect only when it exceeds the residual the optimizer leaves.

For the two distributional channels, we have shown that the escalation rule admits a finite-sample false-pass bound with a quadratic improvement on the single-metric rule for defects near the threshold. The closed-form window $\eta_{\max}(n)$ of two exactly evaluable defect families states at which scales and tolerance ratios the escalation is worth its extra shots, and one discrepancy tolerance corresponds to transport tolerances that differ between the families by a factor of about $1.6$. On the Hamming cost the discrepancy is redundant for every pair of distributions once $\tau_{\MMD} \geq 2(1 - e^{-\gamma})\tau_{\W}^2$, and for any fixed normalized kernel some parity defect of transport size $\beta/2$ has squared discrepancy at most $\beta^2/(2^n - 1)$, so a certificate is meaningful only for the correlation orders it specifies.

In our tests the discrepancy vote changed no verdict at the loose tolerance, in any in-window draw of the fixed-threshold test. In contrast, at certification grade $2.7\%$ of the admitted unperturbed campaign fits are mismatch pairs, and the escalation rejects the $1.0\%$ that lie inside its window.

Our guarantees hold for noiseless sampling, and device noise removes the trace identity (Sec.~\ref{sec:discussion}). Future investigations could focus on the choice of ground cost per scale, on certification under device noise, and on a variance-adaptive concentration bound for the discrepancy channel, which the measured variance suggests would become non-vacuous at the shot budgets of a hardware experiment.

For groups that train and deploy quantum generative models, these results turn certification into a procedure with stated guarantees. Fisher-based capacity scores of a real-amplitude circuit describe the circuit and are not evidence that a model has learned its target. A transport tolerance is paired with a discrepancy tolerance where the closed-form window is open, and the certificate names the correlation orders it covers.

%% ============================================================

\section*{Data availability}
The data that support the findings of this article are not publicly available. The data are available from the authors upon reasonable request.
Section~\ref{app:reproduction} of the Supplemental Material specifies every procedure and setting needed to reproduce them.
The data comprise the raw per-run results and merged tables of the 2{,}520-run campaign at $n = 3$--$8$, as run and warm-restarted, and of the excluded $420$-run batch at $n = 9$; the data of the finite-shot, shot-budget, mismatch-region, depolarizing, optimizer-replication and full-support-benchmark studies; and the dated pre-registration documents and their addenda.

\begin{acknowledgments}
	This material is based upon work supported by the U.S.\ Department of Energy, Office of Science, Accelerated Research in Quantum Computing, Fundamental Algorithmic Research toward Quantum Utility (FAR-Qu).
\end{acknowledgments}

\appendix
\section{Proofs}\label{app:proofs_main}

This appendix proves Proposition~\ref{prop:cap_general} and Theorems~\ref{thm:parity} and~\ref{thm:redundancy}; the concentration bounds, the false-pass bound and Theorem~\ref{thm:dominance} are proved in Sec.~\ref{sm:bounds} of the Supplemental Material.

\subsection{Proof of Proposition~\ref{prop:cap_general}}\label{app:cap_proof}

\begin{proof}[Proof of Proposition~\ref{prop:cap_general}]
	We prove the statement in five steps, combining the Braunstein--Caves matrix inequality with the half-angle structure of the generators.

	\textbf{Step 1 (Classical--quantum Fisher matrix bound).}
	For any pure parameterized state $|\psi(\bm{\theta})\rangle$ and any fixed positive operator-valued measure (POVM), in particular the projective measurement in the computational basis, the classical Fisher information matrix of the resulting outcome distribution is dominated by the quantum Fisher information matrix in the Löwner (PSD) partial order~\cite{BraunsteinCaves1994,Paris2009}:
	\begin{equation}\label{eq:cr_matrix}
		F_{\mathrm{cl}}(\bm{\theta}) \preceq F_{\mathrm{Q}}(\bm{\theta}).
	\end{equation}
	Taking traces preserves the inequality: $\Tr F_{\mathrm{cl}}(\bm{\theta}) \leq \Tr F_{\mathrm{Q}}(\bm{\theta})$.

	\textbf{Step 2 (Quantum FIM diagonal for pure states).}
	For a pure state $|\psi(\bm{\theta})\rangle$ depending smoothly on $\bm{\theta}$, the diagonal entries of $F_{\mathrm{Q}}$ are~\cite{Liu2020}
	\begin{equation}\label{eq:fq_pure}
		F_{\mathrm{Q}}^{kk}(\bm{\theta}) = 4\bigl(\|\partial_k \psi\|^2 - |\langle\psi|\partial_k\psi\rangle|^2\bigr).
	\end{equation}
	By condition $({\rm cond}_1)$, $|\psi(\bm{\theta})\rangle$ is a real vector for every $\bm{\theta}$, so its derivative $\partial_k|\psi\rangle$ is real as well, and the overlap $\langle\psi|\partial_k\psi\rangle$ is therefore real.
	Differentiating the normalization $\langle\psi|\psi\rangle = 1$ and using this reality gives
	\begin{equation}\label{eq:norm_derivative}
		\langle\psi|\partial_k\psi\rangle + \langle\partial_k\psi|\psi\rangle = 2\langle\psi|\partial_k\psi\rangle = 0,
	\end{equation}
	whence
	\begin{equation}\label{eq:phase_vanish}
		\langle\psi|\partial_k\psi\rangle = 0 \qquad \text{for all } \bm{\theta} \text{ and all } k.
	\end{equation}
	(For a general pure state this overlap is only purely imaginary; the reality condition $({\rm cond}_1)$ forces it to vanish outright.)
	This is the computational-basis specialization of the classical--quantum Fisher equality condition for real-amplitude pure states established by Facchi \emph{et al.}~\cite{Facchi2010}.
	Substituting Eq.~\eqref{eq:phase_vanish} into Eq.~\eqref{eq:fq_pure} gives
	\begin{equation}\label{eq:fq_simplified}
		F_{\mathrm{Q}}^{kk}(\bm{\theta}) = 4\|\partial_k \psi\|^2.
	\end{equation}

	\textbf{Step 3 (Tangent norm from a half-angle generator).}
	Condition $({\rm cond}_2)$ gives $\partial_k|\psi\rangle = -\tfrac{i}{2}\,\widetilde{G}_{(k)}|\psi\rangle$ with $\widetilde{G}_{(k)}$ unitary.
	Therefore
	\begin{equation}\label{eq:tangent_norm}
		\|\partial_k\psi\|^2 = \tfrac{1}{4}\|\widetilde{G}_{(k)}\,\psi\|^2 = \tfrac{1}{4}\,\|\psi\|^2 = \tfrac{1}{4},
	\end{equation}
	where the second equality uses the unitarity of $\widetilde{G}_{(k)}$ and the third the normalization of $|\psi\rangle$.

	\textbf{Step 4 (Combining).}
	Equations~\eqref{eq:fq_simplified} and~\eqref{eq:tangent_norm} give $F_{\mathrm{Q}}^{kk}(\bm{\theta}) = 1$ for every $k \in \{1,\ldots,d\}$ and every $\bm{\theta}$, so $\Tr F_{\mathrm{Q}}(\bm{\theta}) = d$ exactly.
	Combined with Step~1, this gives $\Tr F_{\mathrm{cl}}(\bm{\theta}) \leq \Tr F_{\mathrm{Q}}(\bm{\theta}) = d$, which is part~(a).

	\textbf{Step 5 (Equality off the zero-amplitude set).}
	For a real state, Eq.~\eqref{eq:phase_vanish} reduces the full quantum Fisher matrix to
	\begin{equation}\label{eq:fq_real_full}
		F_{\mathrm{Q}}^{jk} = 4\langle\partial_j\psi|\partial_k\psi\rangle = 4\sum_x \partial_j\psi_{\bm{\theta}}(x)\,\partial_k\psi_{\bm{\theta}}(x),
	\end{equation}
	the sum running over all outcomes.
	On an outcome with $p_{\bm{\theta}}(x) = \psi_{\bm{\theta}}(x)^2 > 0$ the classical summand is $\partial_j p\,\partial_k p / p = 4\,\partial_j\psi_{\bm{\theta}}(x)\,\partial_k\psi_{\bm{\theta}}(x)$, the same as the quantum one, so
	\begin{equation}\label{eq:cap_deficit}
		F_{\mathrm{cl}}(\bm{\theta}) = F_{\mathrm{Q}}(\bm{\theta}) - 4\!\!\sum_{x:\,\psi_{\bm{\theta}}(x) = 0}\!\!\bigl(\nabla_{\bm{\theta}}\psi_{\bm{\theta}}(x)\bigr)\bigl(\nabla_{\bm{\theta}}\psi_{\bm{\theta}}(x)\bigr)^{\!\top}.
	\end{equation}
	If $\bm{\theta} \notin \mathcal{Z}$, every outcome with $\psi_{\bm{\theta}}(x) = 0$ also has $\nabla_{\bm{\theta}}\psi_{\bm{\theta}}(x) = 0$, the correction vanishes, and $F_{\mathrm{cl}} = F_{\mathrm{Q}}$ with trace $d$ by Step~4. If $\bm{\theta} \in \mathcal{Z}$, the correction has trace $4\sum_{x:\,\psi_{\bm{\theta}}(x)=0}\|\nabla_{\bm{\theta}}\psi_{\bm{\theta}}(x)\|^2 > 0$ and the inequality is strict. This is part~(b).
	Steps 1--5 use only conditions $({\rm cond}_1)$ and $({\rm cond}_2)$; when $({\rm cond}_3)$ holds as well, $d = 2^n - 1$.
\end{proof}

\subsection{Proofs of Theorems~\ref{thm:parity} and~\ref{thm:redundancy}}\label{app:parity_proofs}

\begin{proof}[Proof of Theorem~\ref{thm:parity}]
	Write $N = 2^n$ and $d = q_{S,\beta} - u = \beta N^{-1}\chi_S$.

	(a) The excess of $q_{S,\beta}$ over $u$ sits on the $N/2$ strings with $\chi_S = +1$ and the deficit on their complements, each of total mass $\beta/2$, so $\TV = \beta/2$.
	Flipping any one bit of $S$ is a bijection from the excess set onto the deficit set at Hamming cost $1$, so the coupling that moves the excess $\beta N^{-1}$ of each excess string to its flip has cost $\beta/2$ and $\W \leq \beta/2$.
	Conversely, the potential $f = \chi_S/2$ changes by at most $1$ across any single bit flip, hence is $1$-Lipschitz for $C_1$, and $\langle f, q_{S,\beta} - u\rangle = \beta/2$, so $\W \geq \beta/2$ by Eq.~\eqref{eq:w1_dual}.

	(b) By Eq.~\eqref{eq:kernel_fourier} and the orthonormality of the parities, $\mathbb{E}_{q_{S,\beta}}\chi_{S'} - \mathbb{E}_u\chi_{S'} = \beta\,\mathbf{1}[S' = S]$, so the sum collapses to the single term $\beta^2 c^{n-|S|}b^{|S|}$; at $\gamma = 1/n$, $b = (2n)^{-1}[1 + O(n^{-1})]$ and $c^{n-r} = e^{-1/2}[1 + O(n^{-1})]$.
	The circuit prepares every qubit outside $S$ and all but one qubit of $S$ in $|+\rangle$, the remaining qubit $v \in S$ in $\sqrt{(1 + \beta)/2}\,|0\rangle + \sqrt{(1 - \beta)/2}\,|1\rangle$, and applies a CNOT from every other qubit of $S$ into $v$.
	The Born distribution is then uniform on every qubit but $v$, the parity $\chi_S$ equals $(-1)^{v_0}$ for the pre-CNOT value $v_0$ of the target, and $\mathbb{E}\chi_S = \beta$ (we verified the identity $|\langle x|U|0\rangle|^2 = q_{S,\beta}(x)$ for every $S$ at $n \leq 6$).

	(c) For a positive-semidefinite kernel matrix $K$ we define $a_S = N^{-2}\chi_S^{\!\top}K\chi_S \geq 0$, so that
	\begin{equation}\label{eq:parity_mmd_weight}
		\MMD_k^2(u, q_{S,\beta}) = d^{\top}Kd = \beta^2 a_S .
	\end{equation}
	Since $\sum_S \chi_S(x)\chi_S(y) = N\,\mathbf{1}[x = y]$, the weights sum to
	\begin{equation}\label{eq:parity_weight_sum}
		\sum_S a_S = N^{-2}\Tr\Bigl(K \sum_S \chi_S\chi_S^{\!\top}\Bigr) = N^{-1}\Tr K = 1
	\end{equation}
	when $k(x,x) = 1$, so the $N - 1$ nonempty parities share at most one unit of weight and the smallest receives at most $1/(N - 1)$.
	For $k_*$, $K_* = (N I - \mathbf{1}\mathbf{1}^{\top})/(N - 1)$ is positive semidefinite with unit diagonal, and $\chi_S^{\!\top}K_*\chi_S = N^2/(N - 1)$ for every $S \neq \varnothing$, so every nonempty parity receives exactly $1/(N - 1)$ and the bound is attained.
\end{proof}

\begin{proof}[Proof of Theorem~\ref{thm:redundancy}]
	Let $\varphi$ be the feature map of $k$. Then
	\begin{equation}\label{eq:feature_distance}
		\|\varphi(x) - \varphi(y)\|^2 = 2 - 2k(x,y) = 2\bigl(1 - e^{-\gamma d_H(x,y)}\bigr),
	\end{equation}
	and adjacent strings are $\sqrt{2(1 - e^{-\gamma})}$ apart.
	Along a shortest Hamming path from $x$ to $y$ the triangle inequality gives $\|\varphi(x) - \varphi(y)\| \leq d_H(x,y)\sqrt{2(1 - e^{-\gamma})}$.
	For an optimal coupling $\pi$ of $(p, q)$ we then have
	\begin{align}
		\MMD_k(p,q) &= \Bigl\|\sum_{x,y}\pi(x,y)[\varphi(x) - \varphi(y)]\Bigr\| \nonumber\\
		&\leq \sum_{x,y}\pi(x,y)\|\varphi(x) - \varphi(y)\| \nonumber\\
		&\leq \sqrt{2(1 - e^{-\gamma})}\,\W(p,q;C_1), \label{eq:redundancy_chain}
	\end{align}
	where the second line is the triangle inequality in the feature space and the third uses the path bound together with $\sum_{x,y}\pi(x,y)d_H(x,y) = \W(p,q;C_1)$. Squaring gives Eq.~\eqref{eq:redundancy}.
	For $p = \delta_x$ and $q = (1 - w)\delta_x + w\delta_y$ with $d_H(x,y) = 1$, $\W = w$ and $\MMD_k^2 = w^2\|\varphi(x) - \varphi(y)\|^2 = 2(1 - e^{-\gamma})w^2$, so the bound is attained.

	If $\tau_{\MMD} \geq 2(1 - e^{-\gamma})\tau_{\W}^2$, every pair with $\W \leq \tau_{\W}$ has $\MMD_k^2 \leq \tau_{\MMD}$ by Eq.~\eqref{eq:redundancy}, so the discrepancy vote passes whenever the transport vote does and $D_4 = D_1(\tau_{\W})$ for every $\eta$.
	If instead $\tau_{\MMD} < 2(1 - e^{-\gamma})\tau_{\W}^2$, the adjacent pair at $w = \tau_{\W} \leq 1$ has $\W = \tau_{\W}$, which lies in every window $(\tau_{\W} - \eta, \tau_{\W}]$, and $\MMD_k^2 > \tau_{\MMD}$, so $D_4$ rejects a pair that $D_1(\tau_{\W})$ admits.
\end{proof}

% The apsrev4-2 control entry (references.bib) is requested through the .aux file directly;
% \nocite would trigger a spurious natbib undefined-citation warning for this non-bibitem key.
\makeatletter\immediate\write\@auxout{\string\citation{apsrev42Control}}\makeatother
\bibliography{references}

%% ============================================================
%% Supplemental Material
%% ============================================================
\clearpage
\onecolumngrid
\begin{center}
	\textbf{\large Supplemental Material for ``Certifying quantum generative models beyond the Fisher trace''}
\end{center}
\twocolumngrid
\setcounter{section}{0}
\setcounter{subsection}{0}
\setcounter{equation}{0}
\setcounter{figure}{0}
\setcounter{table}{0}
\setcounter{proposition}{0}
\setcounter{corollary}{0}
\setcounter{algorithm}{0}
\renewcommand{\thesection}{S\arabic{section}}
\renewcommand{\theequation}{S\arabic{equation}}
\renewcommand{\thetable}{S\Roman{table}}
\renewcommand{\thefigure}{S\arabic{figure}}
\renewcommand{\theproposition}{S\arabic{proposition}}
\renewcommand{\thecorollary}{S\arabic{corollary}}
\renewcommand{\thealgorithm}{S\arabic{algorithm}}
\renewcommand{\appendixname}{}
\makeatletter
\@removefromreset{equation}{section}
\@removefromreset{table}{section}
\@removefromreset{figure}{section}
\makeatother

This Supplemental Material collects the design details, the controls and the proofs that support the main text. Equations, figures, tables and propositions carry the prefix S; unprefixed references point to the main text. Citations refer to the reference list of the main text.

\section{Design details}\label{sm:design}

This section collects the design details of Sec.~\ref{sec:setup}: the campaign size and cost, the benchmark entropies, the Fisher estimator, the statistical analysis, the pre-registration record, the symbols and observation models, the escalation algorithm and the reproduction protocol.

\subsection{Campaign size, budgets and cost}\label{sm:budgets}

Figure~\ref{fig:design} summarizes a run and the benchmark targets. The full campaign comprises 2{,}520 experiments (420 per qubit count, six qubit counts), structured around the three benchmarks B1/B2/B3 and three ground-cost functions C1/C2/C3 introduced above.
For B1, each cost contributes five perturbation levels times 20 seeds, giving 100 runs per cost and 300 runs per qubit count; for B2 and B3, each cost contributes 20 seeds per benchmark, giving 60 runs per benchmark and 120 runs per qubit count.
All experiments use noiseless statevector simulation (\texttt{default.qubit} backend in PennyLane~\cite{Bergholm2022}).

CMA-ES function-evaluation budgets scale with qubit count: 5{,}000 ($n = 3$), 8{,}000 ($n = 4$), 12{,}000 ($n = 5$), 20{,}000 ($n = 6$), 30{,}000 ($n = 7$), and 50{,}000 ($n = 8$); all six scales appear in Tables~\ref{tab:qfi}, \ref{tab:response}, and~\ref{tab:auroc}.
In the campaign, MMD is computed exactly from the statevector output distributions, with no shot noise; sampled estimators enter only in Secs.~\ref{sec:finite_shot_falsepass} and~\ref{sec:b3p_validation} and in Secs.~\ref{app:finite_shot} and~\ref{sec:shot_budget}.

All 2{,}520 runs completed with zero execution failures.\footnote{The campaign cost $1{,}043$ CPU-hours, distributed as $3$, $9$, $25$, $101$, $225$, and $680$ CPU-hours at $n = 3$ through $8$; a run averages $27$~s at $n = 3$ and $5{,}832$~s at $n = 8$. The excluded $n = 9$ batch alone cost a further $3{,}854$ CPU-hours.} A further 420-run batch at $n = 9$ is excluded from every analysis: at that scale CMA-ES fits neither the unperturbed nor the perturbed targets within any budget we tried, so the batch only locates a trainability frontier.

Every table and figure below reports the campaign as it was run, with two exceptions stated where they occur (Table~\ref{tab:depol}, and three of the nine campaign fits reused as finite-shot pairs in Sec.~\ref{sec:finite_shot_falsepass}); a later warm restart of a subset of runs is reported as a robustness check in Sec.~\ref{sec:auroc}.

\subsection{Benchmark entropies}\label{sm:entropy}

Table~\ref{tab:benchmark_entropy} gives the Shannon entropy of the two full-support benchmarks.

\begin{table}[t]
	\caption{Shannon entropy $H$ (nats) of the structured target B2 and the near-uniform target B3 across qubit counts, with the uniform maximum $n\ln 2$ for reference. B3 agrees with $n\ln 2$ to within rounding, so the last column gives its deficit $n\ln 2 - H_{\mathrm{B3}}$ in units of $10^{-3}$ nats.}\label{tab:benchmark_entropy}
	\begin{ruledtabular}
		\begin{tabular}{ccccc}
			$n$ & B2   & B3   & $n\ln 2$ & deficit ($10^{-3}$) \\
			\hline
			3   & 0.19 & 2.08 & 2.08     & 0.30 \\
			4   & 1.61 & 2.77 & 2.77     & 0.18 \\
			5   & 2.17 & 3.47 & 3.47     & 0.29 \\
			6   & 2.92 & 4.16 & 4.16     & 0.24 \\
			7   & 3.83 & 4.85 & 4.85     & 0.21 \\
			8   & 4.68 & 5.54 & 5.55     & 0.25 \\
		\end{tabular}
	\end{ruledtabular}
\end{table}

\subsection{Fisher estimator}\label{sm:fisher_estimator}

We compute the Fisher trace from the Jacobian $J = \partial p / \partial \bm{\theta}$, which the parameter-shift rule $\partial p / \partial \theta_k = \tfrac{1}{2}[\,p(\bm{\theta} + \tfrac{\pi}{2}\bm{e}_k) - p(\bm{\theta} - \tfrac{\pi}{2}\bm{e}_k)\,]$ gives exactly, with no step size and no $O(h^2)$ error. The rule applies because each variational angle enters a single half-angle generator [condition $({\rm cond}_2)$ of Sec.~\ref{sec:cap}]. The amplitude Jacobian of Sec.~\ref{sec:rank_results} uses the same shift on the real statevector, $\partial\psi / \partial\theta_k = [\,\psi(\bm{\theta} + \tfrac{\pi}{2}\bm{e}_k) - \psi(\bm{\theta} - \tfrac{\pi}{2}\bm{e}_k)\,]/(2\sqrt{2})$. The coefficient differs because here the shift acts on a half-angle rotation of the state, while the probability rule shifts an expectation value.

The inverse-probability weight in $F_{\mathrm{cl}} = J^\top \mathrm{diag}(1/p)\,J$ is clamped at a probability floor of $10^{-10}$ to regularize rare near-zero outcomes, a regularization distinct from Jacobian accuracy (Sec.~\ref{sec:cap_saturation}).
On the converged parameters the parameter-shift and a central-difference ($h = 10^{-4}$) Jacobian reproduce the trace and rank to numerical precision.

\subsection{Statistical analysis}\label{sm:statistics}

Statistical analysis uses bootstrap resampling for $95\%$ confidence intervals on all reported quantities, with $5{,}000$--$10{,}000$ resamples for means and $2{,}000$ seed-cluster resamples for AUROC cells.
The area under the receiver operating characteristic (AUROC) is the probability that a diagnostic scores a randomly chosen perturbed run as worse than a randomly chosen unperturbed one; we compute it with the perturbation level as a proxy quality label, $\epsilon = 0$ (good) against $\epsilon > 0$ (degraded).

The $20$ seeds of a cell are shared across the five perturbation levels, since the background $\nu$ of seed $i$ is the same at every $\epsilon$, and across the three ground costs, so runs are not independent across levels or costs. Every AUROC interval is therefore a seed-cluster bootstrap, which resamples the $20$ seed indices and carries all levels and costs of a seed together. The run-level stratified bootstrap, which treats the $100$ runs of a cell as independent, moves no endpoint by more than $0.04$ and changes the median interval width by $3\%$.

Two diagnostics or two costs are compared through the paired difference of their AUROCs on the same runs or the same seeds, with its own seed-cluster interval. A difference whose interval covers zero is reported as a non-detection, and the precision of every such comparison is set by its $20$ unperturbed runs.

\subsection{Pre-registration record}\label{sm:prereg}

Three experiments were specified before execution in dated documents that are available from the authors with the data (see Data availability in the main text). The first is the adversarial cross-family stress test of Sec.~\ref{sec:rank_results}: its hypotheses, its defect families (correlation shift and mode collapse), its design ($10$ seeds per family, qubit count, and level) and its pass threshold, a rank AUROC $\geq 0.90$, were fixed before the runs. The second, specified after the campaign analysis, is the perturbed full-support benchmark of Sec.~\ref{sec:b3p}, with three predictions fixed before launch and its extension to $n = 7, 8$ fixed in a dated addendum. The third is the fixed-threshold test of Sec.~\ref{sec:b3p_validation}, whose matching rule was written down before the run and whose held-out calibration was fixed in a dated addendum.

Three further controls, the convergence conditioning, the severity split, and the Powell and Adam replications, were specified after the campaign had been seen and before the analysis they support, and were run once without tuning. The eigenvalue-threshold and kernel-bandwidth sweeps are post hoc.

\subsection{Symbols and observation models}\label{sm:symbols}

Table~\ref{tab:symbols} collects the thresholds, margins and reference conventions of the decision analysis, and Table~\ref{tab:obsmodels} fixes which observation model each part of the paper uses.

\begin{table}[t]
	\caption{Symbols used in the decision analysis.}\label{tab:symbols}
	\footnotesize
	\begin{ruledtabular}
		\begin{tabular}{@{}l p{0.70\columnwidth}@{}}
			$\tau_{\W}$, $\tau_{\MMD}$ & transport and (squared) discrepancy tolerances; $0.05$ and $10^{-2}$ (loose) or $10^{-4}$ (certification grade) \\
			$\eta$ & escalation window on the PASS side of $\tau_{\W}$; also the half-width of the margin zone $\mathcal{M}_\eta$ \\
			$\Delta$ & defect size, $\W(p,q;C) - \tau_{\W} > 0$, of a transport-defective pair \\
			$t$ & concentration margin of the transport estimate (accuracy at which a bound is stated) \\
			$\xi$ & discrepancy margin, $\MMD_k^2(p,q) - \tau_{\MMD} > 0$ \\
			$m$, $S$ & samples per metric in a bound; shots per metric in an experiment \\
			$\rho$ & tolerance ratio $\sqrt{\tau_{\MMD}}/\tau_{\W}$; $0.2$ at certification grade \\
			$\Phi_n$, $A_n$ & kernel brackets of the two defect families, Theorem~\ref{thm:dominance}(c) and Eq.~\eqref{eq:family2_mmd} \\
			$\tau_{\mathrm{fit}}$, $p_{\mathrm{floor}}$ & convergence cut on the training residual ($0.02$); probability floor of the Fisher estimator ($10^{-10}$) \\
			trained-on, clean & reference distribution of a score: the target the run was optimized against, or the unperturbed target $p_0$ \\
		\end{tabular}
	\end{ruledtabular}
\end{table}

\begin{table}[t]
	\caption{Observation models. ``Exact'' means the statevector distribution; ``one-sample'' means the target is exact and the generator is sampled, with $\W$ and the MMD $U$-statistic on disjoint draws; ``two-sample'' means both are sampled.}\label{tab:obsmodels}
	\footnotesize
	\begin{ruledtabular}
		\begin{tabular}{@{}p{0.46\columnwidth} p{0.46\columnwidth}@{}}
			Campaign (Secs.~\ref{sec:cap}--\ref{sec:failures}, \ref{sec:cross_run}, \ref{sec:disagreement}--\ref{sec:rank_fragility}, \ref{sec:blindspot_demo}) & exact; both metrics on the same distributions \\
			Bounds, Eqs.~\eqref{eq:combined_clear}--\eqref{eq:window_ratio} & one-sample; two-sample constants in Sec.~\ref{app:budget} \\
			Finite-shot tests (Secs.~\ref{sec:finite_shot_falsepass}, \ref{sec:b3p_validation}; Secs.~\ref{sec:shot_budget}, \ref{app:finite_shot_exp}, \ref{app:b3p_detail}) & one-sample \\
			Theorems~\ref{thm:dominance}, \ref{thm:parity}, \ref{thm:redundancy} & population values \\
		\end{tabular}
	\end{ruledtabular}
\end{table}

\subsection{Threshold test instead of a permutation test}\label{sm:permutation}

We do not use a permutation test because its $p$-value tests the sharp null $p = q$. Certification, however, asks whether the discrepancy exceeds an operational tolerance, and a generator can differ from its target at any significance level while sitting well inside the tolerance. The false-pass guarantee of Sec.~\ref{sec:combined} therefore covers the threshold comparison.

\subsection{The escalation algorithm}\label{sm:algorithm}

\begin{algorithm}[t]
	\caption{Bi-metric escalation protocol ($D_4$)}\label{alg:protocol}
	\begin{algorithmic}[1]
		\Require Target $p$, generator output $q$ (when sampled, the MMD sample set is drawn fresh and disjoint from the $\W$ set, and only upon escalation), cost matrix $C$, kernel $k$, thresholds $(\tau_{\W}, \tau_{\MMD})$, escalation window $\eta$
		\Ensure Decision $\in \{\text{PASS}, \text{FAIL}\}$
		\State Compute $w \gets \W(p, q; C)$
		\If{$w \leq \tau_{\W} - \eta$} \Comment{Clear PASS: no escalation}
		\State \Return PASS
		\ElsIf{$w > \tau_{\W}$} \Comment{Clear FAIL: $\W$ vote already fails}
		\State \Return FAIL
		\EndIf
		\State Compute $\MMD_k^2(p, q)$ \Comment{Escalation window $\tau_{\W} - \eta < w \leq \tau_{\W}$}
		\State $v_{\MMD} \gets \mathbf{1}[\MMD_k^2(p, q) \leq \tau_{\MMD}]$
		\If{$v_{\MMD} = 1$} \Comment{$\W$ already PASS; conservative AND}
		\State \Return PASS
		\Else
		\State \Return FAIL
		\EndIf
	\end{algorithmic}
\end{algorithm}

The MMD verdict $v_{\MMD}$ thresholds the squared statistic against $\tau_{\MMD}$ (Sec.~\ref{sec:mmd}).

\subsection{Reproduction protocol}\label{app:reproduction}

This section specifies the numerical work in enough detail to reproduce it independently. Algorithm~\ref{alg:targets} builds the benchmark targets, Algorithm~\ref{alg:run} performs one campaign run, Algorithm~\ref{alg:analysis} computes an AUROC and its seed-cluster interval, and Algorithm~\ref{alg:finite_shot} runs the finite-shot test of Sec.~\ref{sec:finite_shot_falsepass}. Table~\ref{tab:repro_settings} collects the settings.

The circuits are simulated as exact statevectors with the \texttt{default.qubit} device of PennyLane~\cite{Bergholm2022}. CMA-ES is the \texttt{pycma} implementation~\cite{Hansen2016}, the transport linear program is solved by the HiGHS solver of SciPy~\cite{Virtanen2020}, and the Sinkhorn divergence is computed with the Python Optimal Transport library~\cite{Flamary2021}. Our environment used PennyLane 0.39, pycma 3.4, SciPy 1.17, POT 0.9 and NumPy 1.26. An outcome $x \in \{0,1\}^n$ has the index $\sum_i x_i 2^{\,n-1-i}$, so qubit $0$ is the most significant bit; this convention fixes where the peaks of B1 and B2 sit.

\begin{table}[t]
	\caption{Settings of the campaign and of the finite-shot test. Seeds are integers passed to the NumPy generators; the campaign seed is shared across the three costs and the five levels.}\label{tab:repro_settings}
	\footnotesize
	\begin{ruledtabular}
		\begin{tabular}{@{}l p{0.62\columnwidth}@{}}
			run seed & $s = 424242 + 1000\,o_b + i$, with $o_{\mathrm{B1}} = 1$, $o_{\mathrm{B2}} = 2$, $o_{\mathrm{B3}} = 3$ and seed index $i = 0, \dots, 19$ \\
			costs & $C_2$ weights $(1.0, 1.15, 0.85, 1.1, 0.9, 1.05, 0.95, 1.0)$ truncated to $n$ ($(1.0, 1.2, 0.8)$ at $n = 3$); $C_3$ parity weight $\alpha_W = 0.5$ \\
			CMA-ES & $\sigma_0 = 0.3$; bounds $[-\pi, \pi]$; population $4 + \lfloor 2\ln d \rfloor$; budget $5{,}000$, $8{,}000$, $12{,}000$, $20{,}000$, $30{,}000$, $50{,}000$ at $n = 3$--$8$; tolerances $10^{-12}$ (step), $10^{-8}$ (value), stagnation $100$ generations \\
			$\W$ solver & exact linear program; both marginals floored at $10^{-12}$ and renormalized \\
			kernel & $k(x,y) = e^{-d_H(x,y)/n}$; squared discrepancy computed exactly \\
			Sinkhorn & debiased, regularization $0.5$, at most $2{,}000$ iterations, against the trained-on target \\
			Fisher & parameter shift $\pm\pi/2$; probability floor $10^{-10}$; rank cutoff $10^{-6}\lambda_{\max}$ \\
			bootstrap & $2{,}000$ seed-cluster resamples, generator seed $1$ \\
			finite shot & $n = 3, 4, 5$; $S = 10^2, \dots, 10^5$; $2{,}000$ repetitions; $\eta = \sqrt{2} \times 0.01$; generator seed $20260806$ \\
		\end{tabular}
	\end{ruledtabular}
\end{table}

\begin{algorithm}[t]
	\caption{Benchmark targets}\label{alg:targets}
	\begin{algorithmic}[1]
		\Require qubit number $n$, benchmark $b$, level $\epsilon$, seed $s$; $N = 2^n$
		\If{$b = \mathrm{B1}$}
		\State $p_0 \gets 0.6\,\delta_{0} + 0.4\,\delta_{N-1}$
		\State $\nu_x \gets U_x / \sum_y U_y$ with $U_x \sim \mathrm{Unif}[0, 1)$ i.i.d., generator seeded with $s$
		\State \Return $(1 - \epsilon)\,p_0 + \epsilon\,\nu$
		\ElsIf{$b = \mathrm{B2}$ and $n = 3$}
		\State encode each of $1{,}000$ Re-QGAN images (784 pixels) as $x_j = \mathbf{1}[\text{mean of third } j > \text{image mean}]$, $j = 0, 1, 2$
		\State \Return the empirical distribution of the codes
		\ElsIf{$b = \mathrm{B2}$}
		\State $e_x \sim \mathrm{Exp}$ with mean $0.01$, i.i.d., seed $424242 + 2000 + n$
		\State $w \gets 0.35\,\delta_0 + 0.25\,\delta_{N-1} + 0.10\,\delta_{N/4} + 0.10\,\delta_{3N/4} + e$
		\State \Return $w / \sum_x w_x$
		\Else \Comment{B3; seed $424245$ at $n = 3$, else $424242 + 3000 + n$}
		\State draw $v^{(r)} \sim \mathcal{N}(0, I_N)$ for $r = 1, \dots, 4096$
		\State \Return $\frac{1}{4096}\sum_r v^{(r)} \odot v^{(r)} / \|v^{(r)}\|^2$
		\EndIf
	\end{algorithmic}
\end{algorithm}

\begin{algorithm}[t]
	\caption{One campaign run}\label{alg:run}
	\begin{algorithmic}[1]
		\Require $n$, benchmark $b$, cost $C$, level $\epsilon$, seed index $i$
		\State $s \gets 424242 + 1000\,o_b + i$; $p \gets$ \Call{Target}{$n, b, \epsilon, s$}; $p_0 \gets$ \Call{Target}{$n, b, 0, s$}
		\State gates $G \gets$ \Call{Block}{$0$} $\Vert$ \Call{Block}{$0, 1$} $\Vert \cdots \Vert$ \Call{Block}{$0, \dots, n-1$}
		\State $q_{\bm\theta}(x) \gets |\langle x|G(\bm\theta)|0\rangle^{\otimes n}|^2$, the $d = 2^n - 1$ angles assigned in gate order
		\State seed NumPy and CMA-ES with $s$; draw $\bm\theta_0 \sim \mathrm{Unif}[-\pi, \pi]^d$
		\State $\bm\theta^* \gets$ CMA-ES minimum of $\W(q_{\bm\theta}, p; C)$ from $\bm\theta_0$
		\State $q \gets q_{\bm\theta^*}$; score $\W$, $\MMD_k^2$ and $\TV$ against $p$ and $p_0$; Sinkhorn against $p$
		\State $J_{xk} \gets \tfrac12\bigl[q_{\bm\theta^* + \frac{\pi}{2}\bm e_k}(x) - q_{\bm\theta^* - \frac{\pi}{2}\bm e_k}(x)\bigr]$
		\State $F \gets J^{\top}\mathrm{diag}\bigl(1/\max(q, 10^{-10})\bigr)J$
		\State \Return $\bm\theta^*$, the scores, $\Tr F$ and $\#\{\lambda(F) > 10^{-6}\lambda_{\max}(F)\}$
		\Procedure{Block}{$a_1, \dots, a_m$}
		\If{$m = 1$} append $R_Y$ on $a_1$
		\Else
		\State \Call{Block}{$a_2, \dots, a_m$}; append CNOT$(a_1 \to a_m)$; \Call{Block}{$a_2, \dots, a_m$}
		\EndIf
		\EndProcedure
	\end{algorithmic}
\end{algorithm}

The perturbed full-support benchmark of Sec.~\ref{sec:b3p} runs Algorithm~\ref{alg:run} at cost $C_1$ with $p_0$ replaced by the B3 target of the same $n$ and the background $\nu$ drawn with the B1 seeds. The stress test of Sec.~\ref{sec:rank_results} runs it at cost $C_1$, $n = 3, 4, 5$ and seeds $1000$--$1009$ on three families: B1 at $\epsilon \in \{0, 0.05, 0.15\}$; a correlation shift of the uniform distribution, in which the joint probabilities of qubits $0$ and $1$ are $\tfrac14 + \tfrac{\zeta}{2}$ for equal and $\tfrac14 - \tfrac{\zeta}{2}$ for unequal bits, with $\zeta \in \{0, 0.10, 0.25\}$ (Family~2 of Sec.~\ref{sec:dominance} with $\beta = 2\zeta$); and a collapse $(1 - \alpha)\,p_0 + \alpha\,\delta_0$ of the two-peak target with $\alpha \in \{0, 0.10, 0.30\}$.

The optimizer replication of Table~\ref{tab:optimizers} replaces the CMA-ES step by Powell's method (SciPy, at most $20{,}000$ evaluations, tolerances $10^{-6}$ and $10^{-8}$) or by Adam (step size $0.05$, $400$ steps), the latter on $\MMD_k^2$ or on $\W$ with the optimal dual potential as gradient.

The trained-generator test of Sec.~\ref{sec:blindspot_demo} runs Algorithm~\ref{alg:run} at cost $C_1$ with twelve seeds on the defective target of each family, placed at the midpoint $\tau_{\W} - \eta/2$ of the window, where $\eta$ is the certification-grade $\eta_{\max}$ if it is at least $0.1\,\tau_{\W}$ and $\tau_{\W}/2$ otherwise, under the tightened $\tau_{\MMD}$ of Sec.~\ref{app:blindspot_detail}. The trained output is then certified against the uniform target with exact values.

\begin{algorithm}[t]
	\caption{AUROC with a seed-cluster interval}\label{alg:analysis}
	\begin{algorithmic}[1]
		\Require scores $s_r$ of the $100$ runs of a cell $(n, C)$, labels $y_r = \mathbf{1}[\epsilon_r > 0]$, seed indices $i_r$
		\State optional convergence cut: drop the runs with $y_r = 0$ and $\W(q_r, p_r; C) > 0.02$
		\State $A \gets \Pr(s_+ > s_-) + \tfrac12\Pr(s_+ = s_-)$ over positive--negative pairs
		\For{$B = 1, \dots, 2000$}
		\State draw $20$ seed indices with replacement and keep every run of each drawn seed
		\State $A_B \gets$ AUROC of the drawn runs; for a paired comparison, the difference of two scores on the same draw
		\EndFor
		\State \Return $A$ and the $2.5$ and $97.5$ percentiles of $\{A_B\}$
	\end{algorithmic}
\end{algorithm}

\begin{algorithm}[t]
	\caption{Finite-shot false-pass test}\label{alg:finite_shot}
	\begin{algorithmic}[1]
		\Require exact target $p$, defective generator $q$, shots $S$, $\tau_{\W} = 0.05$, $\tau_{\MMD}$, $\eta$, repetitions $R$
		\For{$r = 1, \dots, R$}
		\State draw independent counts $c, c' \sim \mathrm{Mult}(S, q)$
		\State $\hat w \gets \W(p, c/S; C_1)$; $D_1$ passes if $\hat w \leq \tau_{\W}$
		\If{$\hat w \leq \tau_{\W} - \eta$}
		\State $D_4$ passes
		\ElsIf{$\hat w \leq \tau_{\W}$}
		\State $U \gets (c'^{\top}Kc' - S)/[S(S - 1)] - 2\,c'^{\top}Kp/S + p^{\top}Kp$
		\State $D_4$ passes if $U \leq \tau_{\MMD}$
		\EndIf
		\EndFor
		\State \Return the pass fractions of $D_1$ and $D_4$
	\end{algorithmic}
\end{algorithm}

In the finite-shot test the defective pairs at each $n = 3, 4, 5$ are the mode-collapse witnesses $q_\lambda$ of Theorem~\ref{thm:dominance} with $\lambda = 2(\tau_{\W} + \Delta)/n$ and $\Delta \in \{0.005, 0.02\}$, and three campaign fits whose clean-reference $\W$ exceeds $\tau_{\W}$, each run at $\tau_{\MMD} = 10^{-2}$ and $10^{-4}$. The fixed-threshold test of Sec.~\ref{sec:b3p_validation} applies Algorithm~\ref{alg:finite_shot} to the $600$ fits of the full-support benchmark against the clean B3 target, with $200$ repetitions at $S = 10^3$, $10^4$ and $10^5$, and recalibrates the single threshold as in Sec.~\ref{app:b3p_detail}.

\section{Fisher trace: estimator, coordinates, finite shots and noise}\label{sm:trace}

This section supports Sec.~\ref{sec:cap}: the convention on the zero-amplitude set, the trace in other coordinates, the clamped estimator, the finite-shot cost of the identity, the explicit zero-amplitude locus, and the identity under depolarizing noise.

\subsection{The convention on the zero-amplitude set}\label{sm:z_convention}

The strict inequality on $\mathcal{Z}$ is a property of the convention that sums Eq.~\eqref{eq:fim} over outcomes with $p_{\bm{\theta}}(x) > 0$.
Wherever $\psi_{\bm{\theta}}(x) \neq 0$ the classical summand equals $(\partial_k p_{\bm{\theta}}(x))^2/p_{\bm{\theta}}(x) = 4(\partial_k\psi_{\bm{\theta}}(x))^2$, which has a finite limit as the amplitude vanishes. Hence the continuous extension of $F_{\mathrm{cl}}$ equals $F_{\mathrm{Q}}$ at every $\bm{\theta}$, and $F_{\mathrm{cl}}$ as defined is discontinuous on $\mathcal{Z}$.
The clamped estimator of Sec.~\ref{sec:design} does not take this limit, which is why it registers a deficit near $\mathcal{Z}$ (Sec.~\ref{sec:cap_saturation}).

\subsection{The trace in other coordinates}\label{sm:coordinates}

The value of the trace depends on the coordinates.
The matrix identity $F_{\mathrm{cl}} = F_{\mathrm{Q}}$ of part~(b) is covariant under reparameterization, since both matrices transform as $\Lambda^{-\top} F \Lambda^{-1}$ under $\bm{\vartheta} = \Lambda\bm{\theta}$. The value $\Tr F_{\mathrm{Q}} = d$, however, belongs to the circuit angles, in which the half-angle convention fixes every diagonal entry at one. A measure that normalizes by the trace, such as the effective dimension of Abbas \emph{et al.}~\cite{Abbas2021}, inherits this coordinate choice.

For the M\"ott\"onen node angles of Sec.~\ref{app:locus} the trace is pinned as well. On the $2^{k-1}$ angles of stage $k = 1, \dots, n$, $\Lambda$ is a $\pm 1$ matrix with $\Lambda_k\Lambda_k^{\top} = 2^{k-1} I$, so that $\Lambda^{-\top}\Lambda^{-1}$ equals $2^{-(k-1)} I$ there. The $2^{k-1}$ unit diagonal entries of each stage, weighted by $2^{-(k-1)}$, contribute one per stage, so that $\Tr F_{\bm{\vartheta}} = n$, which we confirmed numerically at $n = 3, 4$.
Because the trace is constant in both coordinate systems, the trace normalization of the effective dimension is the identity map for this ansatz, and the effective dimension reduces to a function of the shape of the Fisher spectrum alone.

\subsection{Empirical saturation and the clamped estimator}\label{sec:cap_saturation}

\begin{table}[t]
	\caption{Clamped Fisher-trace estimator across the campaign ($420$ runs per $n$). $\CFI_w^{\max} = 2^n - 1$ is the bound of Eq.~\eqref{eq:qfi_cap_mottonen}, attained off $\mathcal{Z}$; Saturated (\%) counts runs within $0.01$ of it. Off the zero-amplitude set the saturation is an identity, so the table validates the estimator; the sub-cap entries are the clamp, Eq.~\eqref{eq:clamp_deficit}.}\label{tab:qfi}
	\begin{ruledtabular}
		\begin{tabular}{ccccc}
			$n$ & $\CFI_w^{\max}$ & Mean   & Min    & Saturated (\%) \\
			\hline
			3   & 7               & 6.980   & 5.109   & 97.4         \\
			4   & 15              & 14.838  & 3.456   & 94.5         \\
			5   & 31              & 30.861  & 21.508  & 92.4         \\
			6   & 63              & 62.981  & 59.860  & 95.5         \\
			7   & 127             & 126.966 & 114.508 & 96.2         \\
			8   & 255             & 254.914 & 222.228 & 94.5         \\
		\end{tabular}
	\end{ruledtabular}
\end{table}

When we recompute the campaign's converged parameter points with the estimator's probability floor removed, we obtain $\Tr F_{\mathrm{cl}}(\bm{\theta}) = d$ in every run at every qubit count, to within $10^{-8}$, as Proposition~\ref{prop:cap_general}(b) requires, so that Table~\ref{tab:qfi} validates the estimator.
With the floor in place, the clamped trace lands within $0.03\%$ to $1.1\%$ of the cap in the mean, and within $0.01$ of it in $92\%$ to $97\%$ of runs at each scale.

The sub-cap entries originate in the clamp alone, since neither the zero-amplitude set nor the optimizer produces them.
Estimation near $\mathcal{Z}$ is the operational regime, because a converged fit to the two-peak B1 target drives $2^n - 2$ amplitudes toward zero (even at random parameters that saturate the identity, the median minimum basis probability is $2.8 \times 10^{-8}$ at $n = 6$; Sec.~\ref{sec:finite_shot}).
The exact parameter-shift Jacobian of Sec.~\ref{sec:design} removes finite-difference cancellation there. What remains is the inverse-probability weighting of rare outcomes.
Writing $p_{\mathrm{floor}} = 10^{-10}$ for the probability floor, the clamped estimator replaces $1/p_{\bm{\theta}}(x)$ by $1/p_{\mathrm{floor}}$ on outcomes below the floor. Off $\mathcal{Z}$ the estimate therefore falls short of $d$ by
\begin{equation}\label{eq:clamp_deficit}
	d - \Tr\widehat{F}_{\mathrm{cl}} = 4\!\!\!\sum_{x:\,0 < p_{\bm{\theta}}(x) < p_{\mathrm{floor}}}\!\!\!\Bigl(1 - \frac{p_{\bm{\theta}}(x)}{p_{\mathrm{floor}}}\Bigr)\sum_k\bigl(\partial_k\psi_{\bm{\theta}}(x)\bigr)^2,
\end{equation}
which is strictly positive exactly when some outcome falls below the floor.
We verify Eq.~\eqref{eq:clamp_deficit} run by run across the campaign, and we find that every sub-cap run carries at least one outcome below the floor, no run without such an outcome falls below the cap, and the observed deficit matches the prediction to a maximum relative residual of $1.2 \times 10^{-8}$.
The smallest $|\psi_{\bm{\theta}}(x)|$ carrying non-negligible $|\partial_k\psi_{\bm{\theta}}(x)|$ separates the two populations: its median is $\sim\!10^{-6}$ on the sub-cap runs against $10^{-5}$ to $10^{-2}$ on the saturated ones.

For the M\"ott\"onen ansatz the set $\mathcal{Z}$ is a union of hyperplane families in the node angles of the circuit's binary tree, and exact sparse fits lie on it (Sec.~\ref{app:locus}). How $\mathcal{Z}$ relates to the set of parameters the optimizer reaches remains open.
We quantify the cost of observing the identity from measurement shots in Sec.~\ref{app:finite_shot}.

\subsection{The cost of observing the identity at finite shots}\label{app:finite_shot}\label{sec:finite_shot}

The identity and its saturation are statements about the exact output distribution $p_{\bm{\theta}}$.
A hardware experimenter, however, estimates $p_{\bm{\theta}}$ from a finite number of measurement shots $S$, and the plug-in classical Fisher trace inherits the resulting estimation error through the inverse-probability weighting in $F_{\mathrm{cl}} = J^\top \mathrm{diag}(1/p)\, J$.

\begin{figure}[t]
	\centering
	\includegraphics[width=\columnwidth]{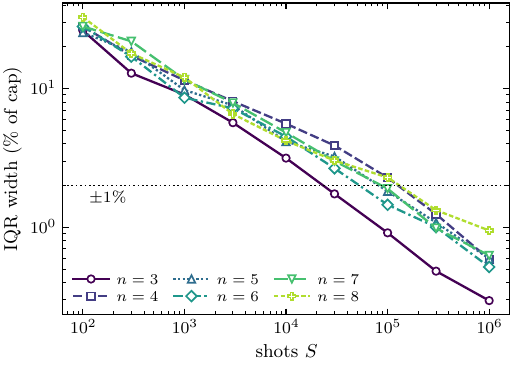}
	\caption{Interquartile width of the finite-shot Fisher trace under the Krichevsky--Trofimov estimator, normalized by the cap, over $8$ random saturated parameter points and $40$ shot replicates (at $n = 7$ one of the eight draws fell more than $0.01$ below the cap under the probability floor and was excluded, leaving $7$); the dotted line marks a $\pm 1\%$ band. The budget required to resolve the cap grows from $S \approx 3 \times 10^4$ at $n = 3$ to $3 \times 10^5$ at $n = 8$.}\label{fig:finite_shot}
\end{figure}

The saturation is supported in part on rare outcomes, whose probability decreases with $n$: the median minimum basis probability at random saturated parameters falls from $4.8 \times 10^{-4}$ at $n = 3$ to $2.8 \times 10^{-8}$ at $n = 6$, and further to $1.6 \times 10^{-11}$ at $n = 7$ and $2.4 \times 10^{-13}$ at $n = 8$.
The naive plug-in estimator $1/\hat{p}$ is dominated by zero-count outcomes and overshoots the cap by several orders of magnitude even at $S = 10^6$.
A smoothed estimator removes this pathology: with the Krichevsky--Trofimov estimate $\hat{p}_x = (c_x + \tfrac{1}{2})/(S + 2^{n-1})$, where $c_x$ is the shot count of outcome $x$, the plug-in trace concentrates onto the cap, as shown in Fig.~\ref{fig:finite_shot}.

The run-to-run spread sets the budget: the interquartile width reaches $\pm 1\%$ only at $S \approx 3 \times 10^4$ for $n = 3$ and $S \approx 10^5$--$3 \times 10^5$ beyond, and at $S = 10^2$ the median overshoots the cap by $78\%$ at $n = 8$.
A formal finite-shot bound accounting for the rare-outcome tail is left to future work.

\subsection{The zero-amplitude locus of the M\"ott\"onen ansatz}\label{app:locus}

For the M\"ott\"onen ansatz the set $\mathcal{Z}$ of Proposition~\ref{prop:cap_general}(b) is explicit. Each stage of the circuit is a uniformly controlled $R_Y$ rotation on qubit $k$ conditioned on qubits $0, \dots, k-1$~\cite{Mottonen2004,Mottonen2005}, so every basis amplitude factorizes along a binary tree into $n$ half-angle trigonometric factors~\cite[Lemma~3.1]{Falco2026GroverRudolph},
\begin{equation}\label{eq:tree_amplitude}
	\psi_{\bm{\theta}}(x) = \prod_{k=0}^{n-1} \kappa_{x_k}\!\bigl(\tfrac12\vartheta_{k,\,x_{<k}}\bigr), \qquad \kappa_0 = \cos,\; \kappa_1 = \sin,
\end{equation}
where $x_{<k} = (x_0, \dots, x_{k-1})$ labels the node of the tree at depth $k$ and the $2^n - 1$ node angles are $\bm{\vartheta} = \Lambda\bm{\theta}$, with $\Lambda$ an invertible matrix of entries $\pm 1$ fixed by the Gray-code compilation of each stage. Our compilation omits the closing CNOT of each Gray-code ladder, which only relabels the computational basis and leaves Eq.~\eqref{eq:tree_amplitude} intact up to that relabeling; stage $k = 1, \dots, n$, which rotates qubit $k - 1$, then uses $2^{k-1}$ rotations and $2^{k-1} - 1$ CNOTs, so the circuit has $\sum_{k=1}^{n} 2^{k-1} = 2^n - 1$ rotations and $\sum_{k=2}^{n} (2^{k-1} - 1) = 2^n - n - 1$ CNOTs, the counts of Sec.~\ref{sec:mottonen}. We verified the factorization and the invertibility of $\Lambda$ numerically for $n = 2$ to $7$.

If two or more factors of $\psi_{\bm{\theta}}(x)$ vanish, every first derivative of the product vanishes, whereas if exactly one vanishes its $\vartheta$-derivative is nonzero and, $\Lambda$ being invertible, so is some $\partial_k\psi_{\bm{\theta}}(x)$. Hence $\mathcal{Z}$ is the set of $\bm{\theta}$ at which, for some outcome $x$, exactly one factor in Eq.~\eqref{eq:tree_amplitude} vanishes. Its closure is the union over the $2^n - 1$ nodes of the hyperplane families $\{\vartheta_j \in \pi\mathbb{Z}\}$, flat in the node angles and oblique in the circuit angles, and $\mathcal{Z}$ omits from this union only the codimension-two intersections on which a factor further up the same branch also vanishes.

The same structure places exact sparse fits on $\mathcal{Z}$. An exact representation of a target with vanishing probabilities prunes whole subtrees of Eq.~\eqref{eq:tree_amplitude}, and since the angles inside a pruned subtree are unconstrained, each pruned subtree is generically cut by a single vanishing factor.

\subsection{The identity under depolarizing noise}\label{sec:noise}

The proof of Proposition~\ref{prop:cap_general} uses reality of the state vector to remove the phase term and the involutive generator to fix the tangent norm (Sec.~\ref{app:cap_proof}). Neither survives a mixed state, so $\Tr F_{\mathrm{Q}}$ is no longer pinned to $d$ and the trace regains information about the fit.

The pinning fails already at small noise rates. A depolarizing channel of rate $p_{\mathrm{dep}}$ maps the output distribution to $(1 - p_{\mathrm{dep}})p_{\bm\theta} + p_{\mathrm{dep}}2^{-n}$ and its Jacobian rows $J_x = \nabla_{\bm\theta}\,p_{\bm\theta}(x)$ to $(1 - p_{\mathrm{dep}})J_x$, so that
\begin{equation}\label{eq:depol_trace}
	\Tr F_{\mathrm{cl}} = \sum_x \frac{(1 - p_{\mathrm{dep}})^2\,\|J_x\|^2}{(1 - p_{\mathrm{dep}})\,p_{\bm\theta}(x) + p_{\mathrm{dep}}\,2^{-n}}.
\end{equation}
In the noiseless case each outcome contributes $\|J_x\|^2/p_{\bm\theta}(x) = 4\sum_k(\partial_k\psi_{\bm\theta}(x))^2$ however small $p_{\bm\theta}(x)$ is, which is why the identity holds. Under the channel an outcome below the noise floor $p_{\mathrm{dep}}2^{-n}$ is suppressed, and sparse fits put nearly all of their outcomes below that floor at once. Post-processing the stored Jacobians of the campaign with Eq.~\eqref{eq:depol_trace} gives the values of Table~\ref{tab:depol}.

\begin{table}[t]
	\caption{Classical Fisher trace relative to the parameter count under a depolarizing channel of rate $p_{\mathrm{dep}} = 10^{-3}$, obtained by post-processing the parameter-shift Jacobians of the converged unperturbed $C_1$ fits, evaluated at the warm-restarted parameters of Sec.~\ref{app:restart}; this is the one table not computed on the campaign as run.}\label{tab:depol}
	\begin{ruledtabular}
		\begin{tabular}{lcccc}
			$\Tr F_{\mathrm{cl}}/d$ & $n=3$ & $4$ & $5$ & $6$ \\
			\hline
			sparse two-peak (B1)   & 0.14 & 0.067 & 0.033 & 0.042 \\
			near-uniform (B3)      & 0.998 & 0.998 & 0.998 & 0.998 \\
		\end{tabular}
	\end{ruledtabular}
\end{table}

A depolarizing rate of one part in a thousand thus removes the identity on the sparse targets of the deployment regime and leaves diffuse targets untouched. In this regime, the trace regains a dependence on the entropy of the target, which is unrelated to fit quality. Table~\ref{tab:depol} therefore measures the fragility of the identity and provides no route to certification under noise. This sweep applies a distribution-level noise model to stored Jacobians; without circuit-level simulation it cannot show how gate and readout error propagate into the two protocol metrics.

\section{Fisher rank: estimators, resolution and robustness}\label{sm:rank}\label{app:rank_estimators}

This section supports Sec.~\ref{sec:rank_results}: the gap separation of the rank classes, the clamped and quantum ranks, a per-run certificate for their agreement, the node count that predicts the rank, the eigenvalue-threshold sweep, the optimizer replication, spectral refinements and the fragility experiment.

\subsection{Gap separation and post hoc cuts}\label{sm:gap}

Through $n = 7$ at $C_1$ and $C_2$ the classes are gap-separated outright, every perturbed run outranking every unperturbed one (Table~\ref{tab:rank_structure}). The rank interval is therefore degenerate there, whereas the distributional intervals are too wide to order the metrics at all.
Doubling the unperturbed class to $40$ seeds at every scale leaves all $360$ added runs rank-deficient and, at $n \leq 7$, below the smallest perturbed rank of their cells.

We identified the separating cuts, including the amplitude-tree count $2n - 1$, after inspecting the campaign, so the perfect scores carry no distribution-free lower confidence bound. Such a bound requires a cut fixed in advance, and the $80$ perturbed runs of a cell, four levels of the same $20$ seeds, would enter it as $20$ clusters. The degenerate bootstrap interval $[1.000, 1.000]$ of a gap-separated cell describes the observed sample.

\subsection{Severity split}\label{sm:severity}

First, a median split on $\epsilon$ among the perturbed runs, asking each diagnostic to grade defect severity, reduces the mean rank AUROC at $n = 3$--$6$ to $0.54$--$0.56$ while $\W$ and MMD keep $0.89$--$0.95$ against the clean reference, the appropriate convention since severity is defined against the clean target. Against the trained-on target the distributional means fall to $0.67$--$0.68$; the rank uses no reference and is unmoved.

Pooling the severity-split runs of $n = 3$--$6$ at $C_1$, the rank AUROC is $0.510\,[0.450, 0.570]$, at chance, and $0.511\,[0.447, 0.572]$ on the restarted campaign. The rank AUROC is also the only score in this section that convergence conditioning leaves unchanged and that the restart policy moves by at most $+0.02$, at $n = 8$.

\subsection{The clamped estimator and the quantum Fisher rank}\label{app:rank_clamp}

No run of the campaign has an amplitude that vanishes to machine precision, so the support-restricted matrix coincides with $F_{\mathrm{Q}}$ on every run, and the clamped estimator of Sec.~\ref{sec:design} departs from both on the $599$ runs that carry an outcome below the floor.

We recomputed the rank at the stored parameters of all $2{,}520$ runs from the exact amplitude Jacobian, with and without the clamp. The clamped rank equals the rank of $F_{\mathrm{Q}}$ on $2{,}431$ runs, including every run at $n \leq 5$ and every B2 and B3 run. On the $89$ sub-floor B1 runs at $n = 6$ to $8$ where the two differ, the clamped rank is smaller by one or two in a dimension of at least $63$.

Proposition~\ref{prop:rank_certificate} (Sec.~\ref{app:rank_certificate}) turns the trace deficit of Eq.~\eqref{eq:clamp_deficit} into a per-run certificate of agreement. It certifies $2{,}024$ of the $2{,}520$ runs, every run without a sub-floor outcome and $103$ of the $599$ with one. All $89$ disagreements lie among the $496$ uncertified runs, of which $407$ agree nonetheless. The smallest observed decision margin in a certified run is $10^{-9}\lambda_{\max}$, about four orders of magnitude above the nominal double-precision eigenvalue-error scale; this is a numerical robustness check without interval arithmetic.
The fraction at $2n - 1$ in Table~\ref{tab:rank_structure} is the same for both matrices at every $n$, and the rank AUROC on B1 at $C_1$ agrees to three decimals through $n = 7$ and reads $0.966$ against $0.964$ at $n = 8$.

The reported clamped ranks therefore track the rank of $F_{\mathrm{Q}}$, and through it the support at the resolution of the eigenvalue threshold. The common trace cap of Eq.~\eqref{eq:qfi_cap_mottonen} does not by itself make the spectra or the eigendirections of the two matrices identical.

\subsection{A node count that predicts the rank}\label{app:rank_nodes}

To make the resolution quantitative, we let $L_c(q)$ be the number of internal nodes of the amplitude tree, at depths $0$ to $n - 1$, that are prefixes of some outcome with $q(x) > c$. For an exact two-peak fit $L_c = 2n - 1$, and for full support $L_c = d$.
The cutoff $c$ is an absolute mass on the leaves, distinct from the relative eigenvalue cutoff of the rank, and a node whose subtree mass exceeds $c$ only through several sub-cutoff leaves is not counted. At $c = 10^{-6}$, $L_c$ and the count of nodes whose subtree mass exceeds $c$ agree on $81\%$ of campaign runs and on every B3 run, and differ by up to $20$ nodes on B1 runs at $n = 7$ and $8$.

At $c = 10^{-6}$, numerically equal to the relative eigenvalue threshold, $L_c$ predicts $\mathrm{rank}_{10^{-6}}$ across all $2{,}520$ runs with Pearson $r = 0.9996$, equality in $75\%$ of runs, a median deviation of zero and a $90$th-percentile deviation of $0.02\,d$. On the converged unperturbed B1 runs at $n = 6$ and $7$, pooled over the three costs, its median is $19.5$ and $54$ against median ranks $22$ and $56$.

The converged B2 fits are full rank for the same reason, apart from $7$ of the $59$ at $n = 3$: at $n = 3$ the B2 target has entropy $0.19$ nats but carries mass above $10^{-6}$ on every outcome, so $L_{10^{-6}} = d$ in $90\%$ of converged B2 fits and in every converged B3 fit.
The unclamped trace still equals the cap of Eq.~\eqref{eq:qfi_cap_mottonen} on every run, so rank deficiency concentrates the eigenvalue mass on the live subspace.

\subsection{A per-run certificate for the clamped rank}\label{app:rank_certificate}

The clamp of Sec.~\ref{sec:design} changes the Fisher matrix wherever an outcome lies below the floor, and Sec.~\ref{app:rank_clamp} reports how often the numerical rank survives the change. We now turn the trace deficit of Eq.~\eqref{eq:clamp_deficit} into a sufficient condition, checkable run by run, for the clamped and the quantum Fisher ranks to agree.

\begin{proposition}[Rank certificate for the clamped estimator]\label{prop:rank_certificate}
Let $A = \widehat{F}_{\mathrm{cl}}(\bm{\theta})$ be the clamped estimator with floor $p_{\mathrm{floor}}$ and $B = F_{\mathrm{Q}}(\bm{\theta})$ the quantum Fisher matrix at the same real-amplitude parameters, and write $g_x = \nabla_{\bm{\theta}}\psi_{\bm{\theta}}(x)$. Then
\begin{equation}\label{eq:clamp_difference}
	B - A = 4\!\!\sum_{x:\,p_{\bm{\theta}}(x) < p_{\mathrm{floor}}}\!\!\Bigl(1 - \frac{p_{\bm{\theta}}(x)}{p_{\mathrm{floor}}}\Bigr) g_x g_x^{\top} \succeq 0,
\end{equation}
so that with $\varepsilon = \Tr(B - A)$ one has $\|B - A\|_{\mathrm{op}} \leq \varepsilon$ and $0 \leq \lambda_i(B) - \lambda_i(A) \leq \varepsilon$ for the eigenvalues in decreasing order. For a relative cutoff $\kappa \geq 0$ let $m_i = \lambda_i(A) - \kappa\,\lambda_{\max}(A)$. Then
\begin{equation}\label{eq:margin_bound}
	-\kappa\,\varepsilon \;\leq\; \bigl[\lambda_i(B) - \kappa\,\lambda_{\max}(B)\bigr] - m_i \;\leq\; \varepsilon ,
\end{equation}
every eigenvalue with $m_i > \kappa\,\varepsilon$ lies above the cutoff in both matrices, every eigenvalue with $m_i < -\varepsilon$ lies below it in both, and if each eigenvalue satisfies one of the two alternatives then $\mathrm{rank}_{\kappa}(A) = \mathrm{rank}_{\kappa}(B)$.
\end{proposition}

\begin{proof}
Wherever $\psi_{\bm{\theta}}(x) \neq 0$ the classical summand is
\begin{equation}\label{eq:clamp_summand}
	\frac{(\partial_k p)(\partial_l p)}{p} = 4\,\partial_k\psi\,\partial_l\psi,
\end{equation}
which is also its continuous extension at $\psi_{\bm{\theta}}(x) = 0$ (Proposition~\ref{prop:cap_general}).
Summing over outcomes gives $B = 4\sum_x g_x g_x^{\top}$, whereas the clamp multiplies the summand of a sub-floor outcome by $p_{\bm{\theta}}(x)/p_{\mathrm{floor}}$. The difference is Eq.~\eqref{eq:clamp_difference}, a sum of positive semidefinite rank-one terms, whose trace $\varepsilon$ bounds its operator norm.

Weyl's inequality for a positive semidefinite perturbation gives $0 \leq \lambda_i(B) - \lambda_i(A) \leq \varepsilon$ for every $i$, in particular for the largest eigenvalue, and subtracting $\kappa$ times the bound on $\lambda_{\max}$ from the bound on $\lambda_i$ gives Eq.~\eqref{eq:margin_bound}. An eigenvalue counts toward $\mathrm{rank}_{\kappa}$ when it exceeds $\kappa\,\lambda_{\max}$, so $m_i > \kappa\,\varepsilon \geq 0$ places it above the cutoff in $A$ and, by the lower bound of Eq.~\eqref{eq:margin_bound}, in $B$, while $m_i < -\varepsilon \leq 0$ places it below in $A$ and, by the upper bound, in $B$; when no margin $m_i$ falls in $[-\varepsilon, \kappa\,\varepsilon]$ the two counts agree.
\end{proof}

The certificate says nothing about eigenvectors, which require spectral-gap information in addition, and it is silent on a run with a margin in the ambiguous band, where the two ranks may or may not agree.
It is evaluated in floating point, with $\varepsilon$ taken as the trace difference and the margins from the computed spectra, and is not a rounding-rigorous certificate; Sec.~\ref{app:rank_clamp} reports how far the margins of the certified runs stay from the two decision boundaries.

\subsection{Eigenvalue threshold and optimizer replication}\label{sm:sweep}

Third, every run behind these numbers uses one optimizer, so we retrained B1 at $n = 3, 4, 5$ on the same stored targets with Powell's method on the $\W$ objective, Adam on an MMD objective with exact gradients, and Adam on the $\W$ objective itself, with the exact gradient given by the optimal dual potential of the Kantorovich problem (Table~\ref{tab:optimizers}).

Adam on $\W$ reproduces the CMA-ES result exactly, a rank AUROC of $1.000$ with no unperturbed fit at full rank at any of the three scales. Adam on MMD leaves every unperturbed fit at full rank at $n = 3$ and scores at chance there. The count the rank reaches is thus set by the objective: a $\W$ fit drives the background of a sparse target to zero and an MMD fit does not.
Powell and Adam on $\W$ converge poorly at $n = 5$, with median unperturbed residuals of $0.29$ and $0.41$, so their values there partly re-express the convergence confound.

Once the parameter count is large enough, every optimizer we tried leaves unperturbed fits rank-deficient and perturbed fits substantially higher-rank. The scale at which this sets in depends on the optimizer and on the objective (Table~\ref{tab:optimizers}).

Every rank verdict depends on a single hyperparameter, so we vary it (Table~\ref{tab:rank_sweep}).
Recomputing the diagnostic at relative eigenvalue thresholds of $10^{-5}$ and $10^{-7}$ alongside the $10^{-6}$ used throughout leaves the pattern unchanged; the only weakening is at $n = 3$ for $10^{-7}$, where the parameter count is small enough that a single admitted eigenvalue changes the score, and the $n = 8$ breakdown is present at every threshold.

\begin{table}[t]
	\caption{Rank AUROC on B1 at three relative eigenvalue thresholds, cost $C_1$, with the range over the three ground costs in brackets; and the optimizer replication of the rank confound at $n = 3$, $4$ and $5$, listed in that order within each cell (rank AUROC, fraction of unperturbed fits at full rank, median unperturbed training residual). CMA-ES is the campaign optimizer; Powell keeps the $\W$ objective; Adam is run on an MMD objective and on the $\W$ objective, both with exact gradients.}\label{tab:rank_sweep}\label{tab:optimizers}
	\footnotesize
	\begin{ruledtabular}
		\begin{tabular}{cccc}
			$n$ & $10^{-5}$ & $10^{-6}$ & $10^{-7}$ \\
			\hline
			3 & $1.000$ & $1.000$ & $0.900\ [0.875, 0.925]$ \\
			4 & $1.000$ & $1.000$ & $1.000$ \\
			5 & $1.000$ & $1.000$ & $0.999\ [0.999, 1.000]$ \\
			6 & $1.000$ & $1.000$ & $1.000\ [0.999, 1.000]$ \\
			7 & $1.000$ & $1.000\ [0.999, 1.000]$ & $0.999\ [0.997, 0.999]$ \\
			8 & $0.976\ [0.976, 0.987]$ & $0.966\ [0.959, 0.966]$ & $0.923\ [0.923, 0.947]$ \\
		\end{tabular}
	\end{ruledtabular}
	\begin{ruledtabular}
		\begin{tabular}{lccc}
			optimizer & rank AUROC & full rank & residual \\
			\hline
			CMA-ES & $1.000/1.000/1.000$ & $0/0/0$ & $0.000/0.000/0.000$ \\
			Powell & $0.775/0.970/0.976$ & $0.45/0.05/0$ & $0.000/0.000/0.290$ \\
			Adam (MMD) & $0.500/0.744/0.973$ & $1.00/0.50/0$ & $0.002/0.001/0.001$ \\
			Adam ($\W$) & $1.000/1.000/1.000$ & $0/0/0$ & $0.005/0.014/0.409$ \\
		\end{tabular}
	\end{ruledtabular}
\end{table}
In its present formulation the rank is a simulation-time diagnostic, because it requires resolving the CFIM spectrum down to a small relative eigenvalue, a strictly harder estimation task than the scalar trace, whose shot cost already grows steeply with $n$ (Fig.~\ref{fig:finite_shot}).
Two non-saturating Fisher-spectrum diagnostics refine the rank, namely the entropy-based effective rank of Roy and Vetterli~\cite{RoyVetterli2007} and the participation ratio of the spectrum. Recomputed on the same converged CFIM spectra, they inherit the same design-recovery confound. On the B1 perturbation labels the two reach $\mathrm{AUROC} = 0.96$--$1.00$ across $n = 3$--$6$, against $1.000$ for the numerical rank.

\subsection{Rank deficiency among certified runs}\label{sec:rank_fragility}

Among the runs the bi-metric protocol certifies as $\W$-PASS and MMD-PASS at the operating point of Sec.~\ref{sec:disagreement}, the count with $\mathrm{rank}_{10^{-6}}(F_{\mathrm{cl}}) < d$ is $67$ of $399$, $52$ of $370$, $58$ of $341$ and $107$ of $332$ at $n = 3, 4, 5, 6$ respectively, and $171$ of $249$ and $101$ of $132$ at $n = 7, 8$.

A plausible hypothesis is that they are \emph{fragile certifications}, runs whose verdict a small parameter perturbation could break.
We perturb the converged parameters of certified runs ($30$ rank-deficient and $30$ full-rank per scale, $24$ random unit directions per run at $\|\delta\bm\theta\| \in \{0.02, 0.05, 0.1, 0.2\}$) and find the opposite of generic fragility: at $n = 4$--$6$ and $\|\delta\bm\theta\| = 0.1$, a rank-deficient certified verdict breaks in $24$--$50\%$ of random directions against $73$--$80\%$ for full-rank certified runs.

With $\Tr F_{\mathrm{cl}} = d$ fixed by the identity of Sec.~\ref{sec:cap}, the isotropic mean-square response is the same for every run, so rank deficiency concentrates the sensitivity in few directions (median $\lambda_{\max} = 19.2$ against $1.1$ at $n = 6$) and a random perturbation lands mostly in the insensitive subspace.
In contrast, adversarial fragility holds: along its top Fisher eigendirection the same rank-deficient run breaks essentially always ($100\%$ at $n = 6$, $\|\delta\bm\theta\| = 0.1$, against $24\%$ in random directions).
The rank is thus a flag for anisotropic verdict robustness: a flagged run resists generic drift but breaks along a few sensitive directions, and the flag complements the bi-metric verdict without overriding it.

The rank flags certified verdicts with anisotropic robustness (Sec.~\ref{sec:rank_fragility}), and we use it to trigger manual inspection where structured drift is plausible, without overriding the bi-metric decision. Every full-rank certified run is a fit to a full-support target, except $52$ fits to the five-outcome B2 target at $n = 3$ that leave mass on its empty outcomes, whereas every certified fit to the sparse B1 target is rank-deficient, as are, from $n = 6$, a growing share of certified fits to the perturbed targets ($62$ of $208$, $128$ of $163$ and $55$ of $55$ at $n = 6, 7, 8$), whose low-mass background falls below the eigenvalue threshold; the flag therefore reads the resolved support of the certified output (Sec.~\ref{sec:rank_results}).

\section{Pooled distributional scores}\label{sm:pooled}

This section supports Sec.~\ref{sec:pooled}: the AUROC tables with their intervals, the training residual by perturbation level, the cross-metric cut, the warm-restarted campaign, the reference and cost effects, the Sinkhorn baseline, the kernel bandwidth and the full table of the full-support benchmark.

\subsection{Benchmark AUROCs}\label{sm:auroc_tables}\label{app:auroc_ci}

\begin{figure}[t]
	\centering
	\includegraphics[width=\columnwidth]{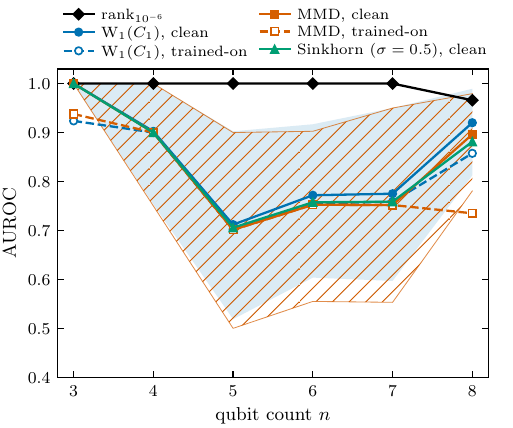}
	\caption{AUROC vs.\ qubit count on B1 at cost $C_1$ for the rank diagnostic, $\W$ and MMD under both references, and the debiased Sinkhorn divergence against the trained-on target (regularization $\sigma = 0.5$). The rank reaches $1.000$ through $n = 7$ and $0.966$ at $n = 8$ (per-cost values and intervals in Table~\ref{tab:rank_auroc}); the distributional depression over $n = 5$--$7$ is the labeling effect removed in Fig.~\ref{fig:conditioned_auroc}, and the two conventions differ by more than $0.06$ only at $n = 3$ and $n = 8$. Shading: $95\%$ seed-cluster bootstrap bands of the clean-reference $\W$ (blue, filled) and MMD (vermilion, hatched) curves, which span $0.18$--$0.40$ at $n \geq 5$ (Table~\ref{tab:auroc_ci} in Sec.~\ref{app:auroc_ci}).}\label{fig:roc}
\end{figure}

In Fig.~\ref{fig:roc} we compare the rank-diagnostic AUROC on the unperturbed-vs-perturbed labels of benchmark B1 with the corresponding $\W$ and MMD AUROCs at cost $C_1$.
The rank diagnostic separates the two classes perfectly through $n = 7$ at costs $C_1$ and $C_2$ and reaches $0.999$ at $C_3$. The distributional scores $\W(C_1)$ and MMD track each other closely and fall to $\mathrm{AUROC} \approx 0.70$--$0.78$ over $n = 5$--$7$, with interval widths of $0.31$ to $0.40$ (Table~\ref{tab:rank_auroc}). At $n = 8$ the rank falls to $0.959$--$0.966$ at every cost.
The $\W$ and MMD decline in Table~\ref{tab:rank_auroc} is the convergence labeling artifact removed in Sec.~\ref{sec:auroc}.

\begin{table*}[t]
	\caption{AUROC on B1 perturbation labels ($100$ runs per cell) for the rank diagnostic, $\W$ and MMD under both references, and the debiased Sinkhorn divergence against the trained-on target ($\sigma = 0.5$). The last column gives the range of the widths of the seed-cluster bootstrap $95\%$ CIs ($2{,}000$ resamples of the $20$ seeds of a cell) over the five distributional scores of the row; Table~\ref{tab:auroc_ci} in Sec.~\ref{app:auroc_ci} lists every interval. The rank uses no reference. Its interval is shown where it is not degenerate and is $[1.000, 1.000]$ in every other row, where the classes are gap-separated. Metrics within a row are compared through the paired differences of Sec.~\ref{sec:pooled}.}\label{tab:auroc}\label{tab:rank_auroc}
	\begin{ruledtabular}
		\begin{tabular}{ccccccccc}
			\multirow{2}{*}{$n$} & \multirow{2}{*}{Cost} & \multirow{2}{*}{rank} & \multicolumn{2}{c}{$\W$} & \multicolumn{2}{c}{MMD} & \multirow{2}{*}{Sinkhorn} & \multirow{2}{*}{CI width} \\
			 &  &  & trained-on & clean & trained-on & clean &  &  \\
			\hline
			3 & $C_1$ & 1.000 & 0.924 & 1.000 & 0.938 & 1.000 & 1.000 & 0.00--0.13 \\
			3 & $C_2$ & 1.000 & 0.878 & 1.000 & 0.917 & 1.000 & 1.000 & 0.00--0.13 \\
			3 & $C_3$ & 1.000 & 0.797 & 0.950 & 0.780 & 0.950 & 0.950 & 0.15--0.26 \\[2pt]
			4 & $C_1$ & 1.000 & 0.900 & 0.902 & 0.900 & 0.900 & 0.900 & 0.25 \\
			4 & $C_2$ & 1.000 & 0.848 & 0.854 & 0.848 & 0.851 & 0.851 & 0.30--0.31 \\
			4 & $C_3$ & 1.000 & 0.791 & 0.801 & 0.791 & 0.800 & 0.800 & 0.35--0.36 \\[2pt]
			5 & $C_1$ & 1.000 & 0.702 & 0.712 & 0.701 & 0.701 & 0.706 & 0.38--0.40 \\
			5 & $C_2$ & 1.000 & 0.952 & 0.954 & 0.952 & 0.952 & 0.953 & 0.13--0.15 \\
			5 & $C_3$ & 1.000 & 0.716 & 0.729 & 0.713 & 0.715 & 0.721 & 0.37--0.39 \\[2pt]
			6 & $C_1$ & 1.000 & 0.755 & 0.772 & 0.752 & 0.752 & 0.757 & 0.31--0.35 \\
			6 & $C_2$ & 1.000 & 0.802 & 0.814 & 0.801 & 0.801 & 0.804 & 0.31--0.34 \\
			6 & $C_3$ & 1.000 & 0.674 & 0.734 & 0.667 & 0.667 & 0.678 & 0.29--0.38 \\[2pt]
			7 & $C_1$ & 1.000 & 0.759 & 0.775 & 0.752 & 0.752 & 0.759 & 0.35--0.40 \\
			7 & $C_2$ & 1.000 & 0.764 & 0.794 & 0.757 & 0.757 & 0.767 & 0.33--0.39 \\
			7 & $C_3$ & $0.999\,[0.995, 1.000]$ & 0.686 & 0.711 & 0.657 & 0.657 & 0.691 & 0.30--0.39 \\[2pt]
			8 & $C_1$ & $0.966\,[0.923, 0.994]$ & 0.858 & 0.920 & 0.735 & 0.896 & 0.881 & 0.18--0.28 \\
			8 & $C_2$ & $0.959\,[0.921, 0.990]$ & 0.823 & 0.901 & 0.718 & 0.862 & 0.875 & 0.22--0.26 \\
			8 & $C_3$ & $0.962\,[0.933, 0.987]$ & 0.766 & 0.812 & 0.664 & 0.769 & 0.770 & 0.28--0.33 \\
		\end{tabular}
	\end{ruledtabular}
\end{table*}

In Table~\ref{tab:auroc} we report the AUROC with the perturbation level as a proxy quality label, together with the range of the seed-cluster bootstrap $95\%$ interval widths in each row; Table~\ref{tab:auroc_ci} in Sec.~\ref{app:auroc_ci} lists every interval.
Each cell has $20$ negatives and $80$ positives, so a single unperturbed run misranked against the whole positive class moves the AUROC by $1/20 = 0.05$; the quantization visible in the recurring values $0.950$ and $0.900$ is exactly this granularity, and the cluster intervals span $0.13$ to $0.40$ at $n \geq 5$.
Metrics within a scale are compared only through their paired seed-cluster differences, reported below, since the marginal intervals carry the between-seed variation that a paired comparison removes.

Within those intervals the pooled scores do not decay monotonically with $n$: the clean-reference $\W(C_1)$ AUROC falls to $0.712$ at $n = 5$ and recovers to $0.920$ at $n = 8$ (Table~\ref{tab:auroc}). They are, however, depressed over $n = 5$--$7$ by an amount the intervals resolve.
The depression is a labeling effect: the label calls every $\epsilon = 0$ run ``good'', the optimizer does not always fit the unperturbed target, and a run labeled good while distributionally wrong penalizes a metric for correctly flagging it.

Table~\ref{tab:auroc_ci} lists the seed-cluster bootstrap $95\%$ interval of every AUROC in Table~\ref{tab:auroc}. The clean-reference $\W$ and MMD intervals at cost $C_1$ are the bands of Fig.~\ref{fig:roc}. Table~\ref{tab:b3p_ci} gives the intervals of the pooled clean-reference scores on the full-support benchmark of Table~\ref{tab:b3p}.

\begin{table*}[t]
	\caption{AUROC on B1 perturbation labels ($100$ runs per cell) with seed-cluster bootstrap $95\%$ CIs over $2{,}000$ resamples of the $20$ seeds of a cell. The rank uses no reference and its intervals are degenerate wherever the classes are gap-separated.}\label{tab:auroc_ci}
	\begin{ruledtabular}
		\begin{tabular}{cccccccc}
			\multirow{2}{*}{$n$} & \multirow{2}{*}{Cost} & \multirow{2}{*}{rank} & \multicolumn{2}{c}{$\W$} & \multicolumn{2}{c}{MMD} & \multirow{2}{*}{Sinkhorn} \\
			 &  &  & trained-on & clean & trained-on & clean &  \\
			\hline
			3 & $C_1$ & $1.000\,[1.000, 1.000]$ & $0.924\,[0.849, 0.976]$ & $1.000\,[1.000, 1.000]$ & $0.938\,[0.876, 0.981]$ & $1.000\,[1.000, 1.000]$ & $1.000\,[1.000, 1.000]$ \\
			3 & $C_2$ & $1.000\,[1.000, 1.000]$ & $0.878\,[0.806, 0.939]$ & $1.000\,[1.000, 1.000]$ & $0.917\,[0.846, 0.970]$ & $1.000\,[1.000, 1.000]$ & $1.000\,[1.000, 1.000]$ \\
			3 & $C_3$ & $1.000\,[1.000, 1.000]$ & $0.797\,[0.659, 0.913]$ & $0.950\,[0.850, 1.000]$ & $0.780\,[0.640, 0.895]$ & $0.950\,[0.850, 1.000]$ & $0.950\,[0.850, 1.000]$ \\
			4 & $C_1$ & $1.000\,[1.000, 1.000]$ & $0.900\,[0.750, 1.000]$ & $0.902\,[0.753, 1.000]$ & $0.900\,[0.750, 1.000]$ & $0.900\,[0.750, 1.000]$ & $0.900\,[0.750, 1.000]$ \\
			4 & $C_2$ & $1.000\,[1.000, 1.000]$ & $0.848\,[0.692, 1.000]$ & $0.854\,[0.703, 1.000]$ & $0.848\,[0.693, 1.000]$ & $0.851\,[0.700, 1.000]$ & $0.851\,[0.700, 1.000]$ \\
			4 & $C_3$ & $1.000\,[1.000, 1.000]$ & $0.791\,[0.594, 0.950]$ & $0.801\,[0.601, 0.950]$ & $0.791\,[0.594, 0.950]$ & $0.800\,[0.600, 0.950]$ & $0.800\,[0.600, 0.950]$ \\
			5 & $C_1$ & $1.000\,[1.000, 1.000]$ & $0.702\,[0.502, 0.900]$ & $0.712\,[0.519, 0.902]$ & $0.701\,[0.500, 0.900]$ & $0.701\,[0.500, 0.900]$ & $0.706\,[0.508, 0.901]$ \\
			5 & $C_2$ & $1.000\,[1.000, 1.000]$ & $0.952\,[0.852, 1.000]$ & $0.954\,[0.871, 1.000]$ & $0.952\,[0.852, 1.000]$ & $0.952\,[0.852, 1.000]$ & $0.953\,[0.858, 1.000]$ \\
			5 & $C_3$ & $1.000\,[1.000, 1.000]$ & $0.716\,[0.517, 0.903]$ & $0.729\,[0.537, 0.908]$ & $0.713\,[0.510, 0.903]$ & $0.715\,[0.515, 0.903]$ & $0.721\,[0.524, 0.905]$ \\
			6 & $C_1$ & $1.000\,[1.000, 1.000]$ & $0.755\,[0.566, 0.903]$ & $0.772\,[0.604, 0.917]$ & $0.752\,[0.555, 0.903]$ & $0.752\,[0.555, 0.903]$ & $0.757\,[0.569, 0.906]$ \\
			6 & $C_2$ & $1.000\,[1.000, 1.000]$ & $0.802\,[0.612, 0.951]$ & $0.814\,[0.644, 0.951]$ & $0.801\,[0.605, 0.950]$ & $0.801\,[0.605, 0.950]$ & $0.804\,[0.615, 0.951]$ \\
			6 & $C_3$ & $1.000\,[1.000, 1.000]$ & $0.674\,[0.486, 0.854]$ & $0.734\,[0.584, 0.877]$ & $0.667\,[0.472, 0.854]$ & $0.667\,[0.472, 0.854]$ & $0.678\,[0.496, 0.856]$ \\
			7 & $C_1$ & $1.000\,[1.000, 1.000]$ & $0.759\,[0.568, 0.950]$ & $0.775\,[0.597, 0.950]$ & $0.752\,[0.553, 0.950]$ & $0.752\,[0.553, 0.950]$ & $0.759\,[0.568, 0.950]$ \\
			7 & $C_2$ & $1.000\,[1.000, 1.000]$ & $0.764\,[0.570, 0.952]$ & $0.794\,[0.626, 0.953]$ & $0.757\,[0.559, 0.950]$ & $0.757\,[0.559, 0.950]$ & $0.767\,[0.574, 0.952]$ \\
			7 & $C_3$ & $0.999\,[0.995, 1.000]$ & $0.686\,[0.516, 0.859]$ & $0.711\,[0.569, 0.865]$ & $0.657\,[0.460, 0.850]$ & $0.657\,[0.460, 0.850]$ & $0.691\,[0.529, 0.859]$ \\
			8 & $C_1$ & $0.966\,[0.923, 0.994]$ & $0.858\,[0.731, 0.964]$ & $0.920\,[0.812, 0.989]$ & $0.735\,[0.590, 0.869]$ & $0.896\,[0.781, 0.979]$ & $0.881\,[0.757, 0.977]$ \\
			8 & $C_2$ & $0.959\,[0.921, 0.990]$ & $0.823\,[0.702, 0.926]$ & $0.901\,[0.767, 0.998]$ & $0.718\,[0.586, 0.841]$ & $0.862\,[0.722, 0.970]$ & $0.875\,[0.744, 0.980]$ \\
			8 & $C_3$ & $0.962\,[0.933, 0.987]$ & $0.766\,[0.617, 0.913]$ & $0.812\,[0.675, 0.951]$ & $0.664\,[0.513, 0.811]$ & $0.769\,[0.595, 0.925]$ & $0.770\,[0.623, 0.918]$ \\
		\end{tabular}
	\end{ruledtabular}
\end{table*}

\begin{table*}[t]
	\caption{Pooled clean-reference AUROC on the perturbed full-support benchmark of Table~\ref{tab:b3p} (B3 blends, cost $C_1$, $100$ runs per $n$) with seed-cluster bootstrap $95\%$ CIs.}\label{tab:b3p_ci}
	\footnotesize
	\begin{ruledtabular}
		\begin{tabular}{ccccccc}
			 & $n = 3$ & $n = 4$ & $n = 5$ & $n = 6$ & $n = 7$ & $n = 8$ \\
			\hline
			$\W$ & $0.981\,[0.959, 1.000]$ & $0.991\,[0.978, 1.000]$ & $0.890\,[0.791, 0.974]$ & $0.811\,[0.690, 0.904]$ & $0.696\,[0.595, 0.789]$ & $0.734\,[0.617, 0.833]$ \\
			MMD & $0.968\,[0.927, 1.000]$ & $0.973\,[0.944, 0.995]$ & $0.816\,[0.692, 0.921]$ & $0.706\,[0.561, 0.838]$ & $0.529\,[0.414, 0.641]$ & $0.604\,[0.456, 0.743]$ \\
		\end{tabular}
	\end{ruledtabular}
\end{table*}

\subsection{Convergence cut, restart, reference and cost}\label{sm:conditioning}\label{app:robustness}

\begin{figure}[t]
	\centering
	\includegraphics[width=\columnwidth]{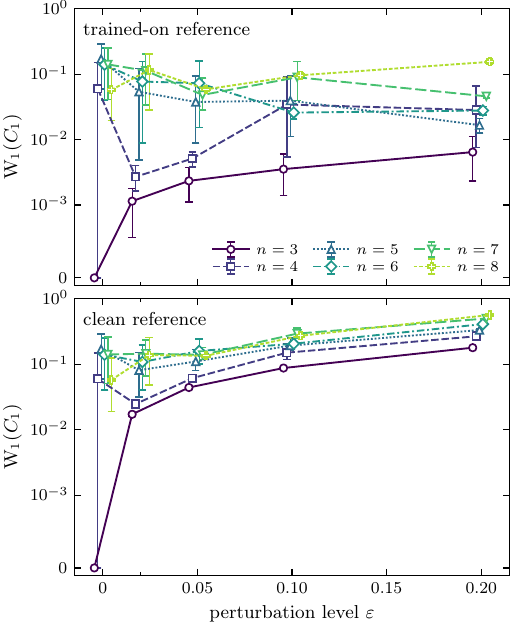}
	\caption{$\W(C_1)$ vs.\ perturbation level on B1, both references, with $95\%$ bootstrap CIs over $20$ seeds (symmetric-log axes). The trained-on residual (top) is non-monotone in $\epsilon$ and ordered by fit difficulty; the clean-reference distance (bottom) rises with $\epsilon$ across the perturbed levels.}\label{fig:response}
\end{figure}

We report the training residual $\W(C_1)$ on B1 in Fig.~\ref{fig:response}; Table~\ref{tab:response} in Sec.~\ref{app:robustness} lists the values.
At $n \geq 4$ the unperturbed entries are elevated ($0.058$--$0.170$) because CMA-ES does not always fit the clean target within budget, so the residual is non-monotone in $\epsilon$ and reflects fit quality. In contrast, defect magnitude is the distance to the clean target, which rises with $\epsilon$ across the perturbed levels.

Because the $\W$ residual at $\epsilon = 0$ is itself the score being evaluated against the clean reference, we also cut the negatives on the other channel, $\MMD_k^2 \leq 10^{-4}$; the $\W$ AUROC is again $1.000$ at every $n$ (Sec.~\ref{app:crosscut}).

The cut does, however, spend negatives, leaving $20, 18, 14, 15, 15$ and $13$ converged negatives per $C_1$ cell at $n = 3$ through $8$. As in Sec.~\ref{sec:rank_results}, the cut lies in an observed gap and selects the negatives by their residual after the fact, so the degenerate intervals describe the surviving sample. At $n = 8$ under cost $C_3$ only four negatives survive, too few to support a claim.

A post hoc warm restart of CMA-ES, which we keep out of the primary analysis because its coverage differs between the label classes ($99\%$ of the eligible unperturbed B1 runs against $48\%$ of the perturbed ones), leads to the same conclusion: it raises the pooled distributional AUROC by $+0.05$ to $+0.34$ and leaves the convergence-conditioned AUROC at $1.000$ everywhere (Sec.~\ref{app:robustness}).

In addition, the choice of reference distribution moves the pooled score independently of both.
At $n = 8$ the trained-on and clean scores separate for MMD, $0.735\,[0.590, 0.869]$ against $0.896\,[0.781, 0.979]$ at $C_1$.
The marginal intervals overlap, whereas a paired seed-cluster bootstrap of the AUROC difference on the same runs gives $+0.161\,[0.084, 0.243]$ at $C_1$, $+0.144\,[0.082, 0.214]$ at $C_2$ and $+0.104\,[0.052, 0.173]$ at $C_3$ for MMD, and $+0.062\,[0.018, 0.121]$ at $C_1$ for $\W$, each excluding zero.
At $n \leq 7$ the two conventions differ by at most $0.06$ outside $n = 3$.
As noted in Sec.~\ref{sec:design}, a training residual measures fit quality, and at $n = 8$ the perturbed targets are the harder ones to fit, so scoring against the trained-on target increasingly rewards the wrong runs.

The best-scoring $\W$ ground cost also changes with scale (Table~\ref{tab:auroc}). The paired seed-cluster difference between $C_1$ and $C_2$ on the clean-reference $\W$ AUROC excludes zero at $n = 5$ alone, $-0.24\,[-0.46, -0.04]$, and covers zero at every other scale, so no cost is best at every scale.

The same paired statistic also orders the two metrics: $\W$ scores at or above MMD against the clean reference in all $18$ B1 cells, by up to $0.07$, and the difference excludes zero in $8$ of them, at every cost for $n = 7, 8$ and at $C_3$ for $n = 5, 6$.
The gap is an order of magnitude below the convergence effect of Fig.~\ref{fig:conditioned_auroc}, and certification depends on calibration at a fixed operating point (Sec.~\ref{sec:protocol_sec}), which this ordering leaves open.

The debiased Sinkhorn divergence~\cite{Feydy2019} interpolates between $\W$ and MMD and dominates pure MMD on QCBM training in Coyle \emph{et al.}~\cite{Coyle2020}, so it is the obvious candidate to replace the pair.
Scored against the trained-on target at entropic regularization $\sigma = 0.5$, its AUROC lies within $0.02$ of the clean-reference $\W$ or MMD value in every cell (Table~\ref{tab:auroc} and Fig.~\ref{fig:roc}). The kernel bandwidth, in turn, explains neither the scale-dependent dip nor the $n = 8$ MMD shortfall; Sec.~\ref{app:robustness} gives both checks.

\subsection{Training residual by perturbation level}\label{app:response}

Table~\ref{tab:response} lists the values of Fig.~\ref{fig:response} at three perturbation levels.

\begin{table}[t]
	\caption{$\W(C_1)$ on B1 against both references (mean and $95\%$ CI over $20$ seeds), abridged to $\epsilon \in \{0, 0.05, 0.20\}$; the full response is in Fig.~\ref{fig:response}. At $\epsilon = 0$ the two references coincide and the columns are identical.}\label{tab:response}
	\begin{ruledtabular}
		\begin{tabular}{cccc}
			$n$ & $\epsilon$ & trained-on & clean            \\
			\hline
			3   & 0.00       & $0.000\,[0.000, 0.000]$ & $0.000\,[0.000, 0.000]$ \\
			3   & 0.05       & $0.002\,[0.001, 0.004]$ & $0.044\,[0.041, 0.048]$ \\
			3   & 0.20       & $0.006\,[0.002, 0.012]$ & $0.179\,[0.166, 0.193]$ \\
			\hline
			4   & 0.00       & $0.060\,[0.000, 0.150]$ & $0.060\,[0.000, 0.150]$ \\
			4   & 0.05       & $0.005\,[0.004, 0.007]$ & $0.061\,[0.059, 0.064]$ \\
			4   & 0.20       & $0.028\,[0.008, 0.065]$ & $0.267\,[0.241, 0.308]$ \\
			\hline
			5   & 0.00       & $0.170\,[0.060, 0.290]$ & $0.170\,[0.060, 0.290]$ \\
			5   & 0.05       & $0.038\,[0.009, 0.094]$ & $0.110\,[0.080, 0.167]$ \\
			5   & 0.20       & $0.017\,[0.013, 0.021]$ & $0.333\,[0.323, 0.343]$ \\
			\hline
			6   & 0.00       & $0.140\,[0.050, 0.260]$ & $0.140\,[0.050, 0.260]$ \\
			6   & 0.05       & $0.073\,[0.015, 0.158]$ & $0.159\,[0.101, 0.246]$ \\
			6   & 0.20       & $0.028\,[0.024, 0.032]$ & $0.411\,[0.403, 0.417]$ \\
			\hline
			7   & 0.00       & $0.141\,[0.031, 0.261]$ & $0.141\,[0.031, 0.261]$ \\
			7   & 0.05       & $0.048\,[0.028, 0.086]$ & $0.139\,[0.118, 0.179]$ \\
			7   & 0.20       & $0.046\,[0.043, 0.049]$ & $0.500\,[0.495, 0.506]$ \\
			\hline
			8   & 0.00       & $0.058\,[0.019, 0.122]$ & $0.058\,[0.019, 0.122]$ \\
			8   & 0.05       & $0.059\,[0.055, 0.064]$ & $0.135\,[0.130, 0.140]$ \\
			8   & 0.20       & $0.154\,[0.140, 0.168]$ & $0.566\,[0.554, 0.579]$ \\
		\end{tabular}
	\end{ruledtabular}
\end{table}

\subsection{The cross-metric convergence cut}\label{app:crosscut}

Restricting to unperturbed runs with $\MMD_k^2 \leq 10^{-4}$ and scoring $\W$ against the clean reference gives $1.000$ at every $n$ (negatives $20, 18, 14, 15, 15, 16$), and the MMD AUROC under the $\W$ cut is likewise $1.000$ everywhere. At the looser $\MMD_k^2 \leq 10^{-3}$ cut the $\W$ AUROC is $1.000$ through $n = 7$ and $0.966$ at $n = 8$ with $19$ negatives.

\subsection{The warm-restarted campaign}\label{app:restart}

We re-optimized part of the campaign with a warm restart of CMA-ES.
The restart targets CMA-ES under-convergence, which elevates unperturbed residuals and can mimic a degradation of discrimination with scale. A run was eligible when its training residual exceeded $\tau_{\W} = 0.05$, and the restart started from the run's incumbent parameters under the campaign's stopping rule. The as-run parameters analyzed here are byte-preserved pre-restart snapshots, and we verified per run that the stored metrics reproduce from the stored parameters, so the two campaign variants are exactly reconstructible from the stored data.

We applied the restart driver to both label classes and to all three benchmarks, but its coverage was not uniform. Of the runs meeting the restart criterion it reached $99\%$ of unperturbed B1 runs against $48\%$ of perturbed ones, and the shortfall concentrates at $n = 7$ and $8$, where perturbed coverage falls to $34\%$ and $36\%$ while unperturbed coverage stays at $100\%$ and $93\%$.
Through $n \leq 5$ coverage is complete in both classes; beyond it an intervention that improves one class's convergence more than the other's acts directly on the label structure.

We therefore report the campaign as run and treat the restarted one as a robustness check. Where the restart raises the pooled distributional AUROC, it does so by $+0.05$ to $+0.34$, with the largest shifts where its coverage is most asymmetric. It moves the rank AUROC by at most $+0.02$, and by nothing through $n = 7$, while the convergence-conditioned AUROC stays at $1.000$ everywhere.

\subsection{The Sinkhorn baseline}\label{app:sinkhorn}

The Sinkhorn divergence, computed against the trained-on target, sits between the clean-reference $\W$ and MMD values at every scale through $n = 7$ and slightly below both at $n = 8$ and $C_1$ ($0.881$ against $0.896$ and $0.920$).
The $\W$/Sinkhorn verdict disagreement is $0.0\%$ at $n = 3, 4, 5$, rising to $1.0\%$, $5.0\%$ and $17.0\%$ at $n = 6, 7, 8$ as the fits degrade.
The regularizations $\sigma \in \{0.1, 1.0\}$ leave this unchanged, and we ran no paired comparison between Sinkhorn and either metric.

\subsection{Kernel bandwidth}\label{app:bandwidth}

The MMD scores are insensitive to the kernel bandwidth within the cluster intervals: recomputing the B1 $C_1$ MMD AUROC at $\gamma \in \{1/(2n),\ 1/n,\ 2/n\}$ and at the median-heuristic bandwidth changes it by at most $0.02$ at every $n \leq 7$ and by at most $0.05$ at $n = 8$, so the $\gamma = 1/n$ choice explains neither the scale-dependent dip nor the $n = 8$ MMD shortfall.

\subsection{Full table of the full-support benchmark}\label{sm:b3p_full}

The runs at $n = 7$ and $8$ were launched after the outcome at $n \leq 6$ was known, under a dated addendum, written before their launch, that keeps the design, P1 and P3, and restates P2 for these two scales as a residual criterion: the $\epsilon = 0.20$ per-level AUROC stays at or above $0.90$, and so does every level whose clean-reference defect exceeds the median unperturbed residual by a factor of two.

At $n = 7$ and $8$ the clean near-uniform target does not converge within the publication budget (median residual $0.043$ and $0.046$), and at most one unperturbed fit survives the $\W$ cut, so P2 is untestable there under that cut. We report the restated criterion without a verdict, since the addendum names no metric and its conditioned form leaves at most one negative: at $\epsilon = 0.20$ the per-level AUROC is $0.978$ and $0.975$ for $\W$ and $0.777$ and $0.807$ for MMD, and no level reaches twice the residual.

\begin{table*}[t]
	\caption{Perturbed full-support benchmark (B3 blends, cost $C_1$, $100$ runs per $n$), AUROC on the $\epsilon = 0$ vs $\epsilon > 0$ labels. Pooled: the rank and the clean-reference $\W$ and MMD; their seed-cluster bootstrap $95\%$ CIs are in Table~\ref{tab:b3p_ci} in Sec.~\ref{app:auroc_ci}. Conditioned: the same scores after the convergence cut on the negatives named in the column, the $\W$ cut (training residual $\leq 0.02$) or the MMD cut ($\MMD_k^2 \leq 10^{-4}$), with surviving negatives in parentheses; a cell is dashed when fewer than two negatives survive. Defect resolution: the median training residual of the unperturbed fits, the exact clean-reference $\W$ of the injected defect at $\epsilon = 0.02$ (median over seeds), and the per-level AUROC ($\W$/MMD) of the $\epsilon = 0.02$ and $\epsilon = 0.20$ levels against $\epsilon = 0$.}\label{tab:b3p}
	\footnotesize
	\begin{ruledtabular}
		\begin{tabular}{ccccccccccc}
			& \multicolumn{3}{c}{Pooled} & \multicolumn{3}{c}{Conditioned} & \multicolumn{4}{c}{Defect resolution} \\
			\cline{2-4} \cline{5-7} \cline{8-11}
			$n$ & rank & $\W$ & MMD & $\W$, $\W$ cut & MMD, $\W$ cut & $\W$, MMD cut & residual$_{\epsilon = 0}$ & defect$_{0.02}$ & AUROC$_{0.02}$ & AUROC$_{0.20}$ \\
			\hline
			3 & $0.500$ & $0.981$ & $0.968$ & $0.981\ (20)$ & $0.968$ & $0.981\ (20)$ & $0.000$ & $0.0069$ & $0.930/0.910$ & $1.000/1.000$ \\
			4 & $0.500$ & $0.991$ & $0.973$ & $0.991\ (20)$ & $0.973$ & $0.991\ (20)$ & $0.002$ & $0.0063$ & $0.965/0.897$ & $1.000/1.000$ \\
			5 & $0.500$ & $0.890$ & $0.816$ & $0.996\ (16)$ & $0.939$ & $0.935\ (18)$ & $0.004$ & $0.0062$ & $0.823/0.657$ & $1.000/0.953$ \\
			6 & $0.494$ & $0.811$ & $0.706$ & $0.937\ (13)$ & $0.863$ & $0.822\ (19)$ & $0.014$ & $0.0061$ & $0.598/0.575$ & $1.000/0.965$ \\
			7 & $0.525$ & $0.696$ & $0.529$ & --\ (1) & -- & $0.738\ (17)$ & $0.043$ & $0.0057$ & $0.495/0.425$ & $0.978/0.777$ \\
			8 & $0.494$ & $0.734$ & $0.604$ & --\ (0) & -- & $0.734\ (20)$ & $0.046$ & $0.0055$ & $0.453/0.410$ & $0.975/0.807$ \\
		\end{tabular}
	\end{ruledtabular}
\end{table*}

P1 and P3 hold at every $n$.
The rank AUROC lies between $0.494$ and $0.525$ at every $n$, $597$ of the $600$ fits are full rank, and the median rank is $d$ in both classes, so the rank carries no label information once the support is fixed, as the support reading of Sec.~\ref{sec:rank_results} requires.

The pooled clean-reference scores stay below the conditioned B1 values even at $n = 3$--$6$, where the unperturbed fits converge (median residual at most $0.014$); there they range from $0.811$ to $0.991$ for $\W$ and from $0.706$ to $0.973$ for MMD.

The per-level columns show why.
On the near-uniform target the injected defect is small: at $\epsilon = 0.02$ its clean-reference $\W$ is $0.005$--$0.007$ at every $n$, against $0.016$--$0.058$ on B1 at $n = 3$--$8$ (median over seeds, cost $C_1$), and from $n = 6$ on it lies below the unperturbed fit residual, where the per-level $\W$ AUROC falls to $0.598$, $0.495$ and $0.453$.

At $\epsilon = 0.20$ the defect ($0.055$--$0.069$) exceeds the median residual at every $n$, and the $\W$ AUROC is $1.000$ at $n = 3$--$6$ and $0.978$ and $0.975$ at $n = 7$ and $8$, where the defect exceeds the residual by only $20$--$30\%$.

The trained-on reference scores at $0.39$--$0.62$ throughout, near chance, as a fit-quality measure should when both classes fit comparably well.

\section{Bounds and proofs for the protocol}\label{sm:bounds}\label{app:proofs}

This section gives the concentration bounds and the proofs behind Sec.~\ref{sec:protocol_sec}, the baselines and the same-data form of the false-pass bound, the scale of the correlation signal and its kernel repair, the comparison of Theorem~\ref{thm:parity}(c) with Ref.~\cite{Kurkin2025}, the relation of Theorem~\ref{thm:redundancy} to earlier bounds, and the full analysis of matched tolerances.

\subsection{Concentration of the two metrics}\label{app:concentration}

\begin{proposition}[W$_1$ concentration]\label{prop:w1}
	Let $p, q \in \simplex$, let $p$ be known exactly, and let $\hat{q}_m$ be the empirical distribution from $m$ i.i.d.\ samples from the sampled distribution $q$.
	Let $C$ be a ground-cost matrix with $D_{\max} = \max_{x,y} C(x,y)$.
	Then, writing $\bar{W}_m = \mathbb{E}\,\W(p, \hat{q}_m; C)$ for the mean of the empirical estimate,
	\begin{equation}\label{eq:w1_conc}
		\Pr\bigl(|\W(p, \hat{q}_m; C) - \bar{W}_m| > t\bigr) \leq 2\exp\!\left(-\frac{2mt^2}{D_{\max}^2}\right),
	\end{equation}
	and the bias of the estimate is bounded by
	\begin{equation}\label{eq:w1_bias}
		\bigl|\bar{W}_m - \W(p, q; C)\bigr| \;\leq\; \frac{D_{\max}}{2\sqrt{m}}\sum_{x}\sqrt{q(x)} \;\leq\; \frac{D_{\max}\sqrt{N}}{2\sqrt{m}},
	\end{equation}
	in which the sum runs over the distribution that is sampled.
	If both distributions are sampled, with $m_p$ and $m_q$ i.i.d.\ samples, the same bounded-differences argument, in which one sample of $p$ moves the statistic by at most $D_{\max}/m_p$ and one sample of $q$ by at most $D_{\max}/m_q$, gives
	\begin{multline}\label{eq:w1_conc_two}
		\Pr\bigl(|\W(\hat{p}_{m_p}, \hat{q}_{m_q}; C) - \mathbb{E}\,\W(\hat{p}_{m_p}, \hat{q}_{m_q}; C)| > t\bigr)\\
		\leq 2\exp\!\left(-\frac{2t^2}{D_{\max}^2\,(1/m_p + 1/m_q)}\right),
	\end{multline}
	whose exponent at $m_p = m_q = m$ is $-mt^2/D_{\max}^2$, half the one-sample rate; the bias is at most $\tfrac12 D_{\max}\bigl(m_p^{-1/2}\sum_x\sqrt{p(x)} + m_q^{-1/2}\sum_x\sqrt{q(x)}\bigr)$, the sum of the two one-sample terms, by the triangle inequality, and since $\W$ is jointly convex in $(p, q)$, Jensen's inequality still gives $\mathbb{E}\,\W(\hat{p}_{m_p}, \hat{q}_{m_q}; C) \geq \W(p, q; C)$.
\end{proposition}

\begin{corollary}[W$_1$ sample complexity]\label{cor:w1_samples}
	To achieve concentration about the mean, $|\W(p, \hat{q}_m) - \bar{W}_m| \leq t$ with probability $\geq 1 - \delta$:
	\begin{equation}
		m \geq \frac{D_{\max}^2 \ln(2/\delta)}{2t^2}.
	\end{equation}
	For the total error $|\W(p, \hat{q}_m) - \W(p, q)| \leq t$ the bias bound of Eq.~\eqref{eq:w1_bias} must be added: splitting $t$ evenly between fluctuation and bias gives
	\begin{equation}\label{eq:w1_total_samples}
		m \geq \max\!\left\{ \frac{2 D_{\max}^2 \ln(2/\delta)}{t^2},\; \frac{D_{\max}^2 \bigl(\sum_x \sqrt{q(x)}\bigr)^{2}}{t^2} \right\}.
	\end{equation}
	For our ground costs $D_{\max} = O(n)$, so the fluctuation complexity is $O(n^2/t^2)$; the bias term carries $\sum_x \sqrt{q(x)}$ of the sampled distribution, which is $O(1)$ when the generator's output is sparse, as for a near-exact fit of the two-peak B1 target, and at most $2^{n/2}$ for a full-support generator~\cite{Weed2019}; the sparsity of an exact target does not enter, because the target is not sampled.
\end{corollary}

\begin{proposition}[MMD concentration]\label{prop:mmd}
	Let $k : \mathcal{X} \times \mathcal{X} \to \mathbb{R}$ be a bounded kernel whose values lie in an interval of width $K_{\max}$; for the RBF--Hamming kernel $K_{\max} = 1$.
	Let $\widehat{\MMD}_u^2$ be the unbiased $U$-statistic estimator of $\MMD_k^2(p,q)$, computed either from $m$ i.i.d.\ samples of each distribution (two-sample case) or from $m$ i.i.d.\ samples of one distribution with the other known exactly (one-sample case).
	In both cases, for every $t > 0$, each of the two tails satisfies
	\begin{equation}\label{eq:mmd_conc}
		\Pr\bigl(\pm\bigl[\widehat{\MMD}_u^2 - \MMD_k^2(p,q)\bigr] > t\bigr) \leq \exp\!\left(-\frac{mt^2}{16K_{\max}^2}\right).
	\end{equation}
\end{proposition}

The MMD sample complexity is $m \geq 16K_{\max}^2 \ln(1/\delta)/t^2$.
With $K_{\max} = 1$ (RBF kernel), this is $O(\ln(1/\delta)/t^2)$, independent of the qubit count $n$ and of the support size $N$.

\subsection{Equal-budget comparison and window sizing}\label{app:budget}

Fix a total budget of $M$ samples per sampled distribution, counted across both stages. The $\W$-only rule spends all of it on one statistic; $D_4$ splits it disjointly and gives each metric $M/2$ (Table~\ref{tab:allocation}). Substituting these allocations into Propositions~\ref{prop:w1} and~\ref{prop:mmd}, we obtain the three exponential rates, meaning the coefficient of $-M$ in each exponent,
\begin{equation}\label{eq:equal_budget_rates}
	\begin{split}
	r_{D_1} &= \frac{2t^2}{D_{\max}^2}, \qquad
	r_{\text{miss}} = \frac{\eta^2}{D_{\max}^2}, \\
	r_{\text{joint}} &= \frac{t^2}{D_{\max}^2} + \frac{\xi^2}{32 K_{\max}^2}
	\end{split}
\end{equation}
in the one-sample model (exact target, sampled generator).
Halving the sample size reduces the transport part of the joint rate to half of $r_{D_1}$, and the MMD margin must compensate for this loss. Hence, the joint term has the larger rate at equal total budget if and only if
\begin{equation}\label{eq:equal_budget_breakeven}
	\xi \;>\; \frac{4\sqrt{2}\,K_{\max}\,t}{D_{\max}},
\end{equation}
and the rate of the full bound of Eq.~\eqref{eq:combined_bound} exceeds the single-metric rate when in addition the escalation-miss term is subordinate, $r_{\text{miss}} \geq 2 r_{D_1}$, that is $\eta \geq 2t$; the joint term retains the full quadratic rate, $r_{\text{joint}} \geq 2 r_{D_1}$, when $\xi \geq 4\sqrt{6}\,K_{\max}t/D_{\max}$.

The corresponding per-metric-budget requirement is $\xi \geq 4\sqrt{2}\,K_{\max}t/D_{\max}$, so the fair normalization stiffens the requirement on the MMD margin by exactly $\sqrt{3}$ and the requirement on the window from $\sqrt{2}\,t$ to $2t$; nothing else changes.
At $K_{\max} = 1$ and the Hamming cost $D_{\max} = n$, Eq.~\eqref{eq:equal_budget_breakeven} reads $\xi > 5.66\,t/n$ for break-even and $\xi \geq 9.80\,t/n$ for the full quadratic, both of which relax as $n$ grows.
When both distributions are sampled, with $M$ draws per distribution, the two transport rates halve by Eq.~\eqref{eq:w1_conc_two} while the discrepancy rate does not, since Proposition~\ref{prop:mmd} holds at the same rate in the two-sample case:
\begin{equation}\label{eq:equal_budget_rates_two}
	\begin{split}
	r_{D_1} &= \frac{t^2}{D_{\max}^2}, \qquad
	r_{\text{miss}} = \frac{\eta^2}{2D_{\max}^2}, \\
	r_{\text{joint}} &= \frac{t^2}{2D_{\max}^2} + \frac{\xi^2}{32 K_{\max}^2},
	\end{split}
\end{equation}
so the conditions become $\xi > 4K_{\max}t/D_{\max}$ for break-even and $\xi \geq 4\sqrt{3}\,K_{\max}t/D_{\max}$ for the full quadratic rate ($4.00\,t/n$ and $6.93\,t/n$ at the Hamming cost), with $\eta \geq 2t$ unchanged; if $M$ instead counts the target and generator draws together, every rate in Eq.~\eqref{eq:equal_budget_rates_two} is halved and the three conditions are the same.

These are comparisons of exponential rates, which order the guarantees asymptotically in $M$: at finite $M$ the prefactors and the additive miss term remain, so a larger rate does not by itself make the sum of the two $D_4$ terms smaller than the $D_1$ bound at every budget, and Sec.~\ref{sec:finite_shot_falsepass} measures the realized rates.
At equal cost, then, the quadratic advantage holds only for defects whose discrepancy margin is large relative to the transport accuracy demanded. Sec.~\ref{sec:shot_budget} measures what that margin costs to resolve.

\begin{table}[t]
	\caption{Sample allocation behind Eqs.~\eqref{eq:equal_budget_rates} and~\eqref{eq:equal_budget_rates_two}, per sampled distribution and across both stages. In the one-sample model only the generator is sampled; in the two-sample model the target is sampled as well, with the same counts.}\label{tab:allocation}
	\footnotesize
	\begin{ruledtabular}
		\begin{tabular}{lcc}
			Rule & transport stage & discrepancy stage \\
			\hline
			$D_1(\tau_{\W})$, $D_1(\tau_{\W} - \eta)$ & $M$ & -- \\
			$D_4$ (bound) & $M/2$ & $M/2$ \\
			$D_4$ (adaptive, expected) & $M$ & $M \times$ escalation rate \\
		\end{tabular}
	\end{ruledtabular}
\end{table}

The escalation-miss term $\delta_{\W}(\eta)$ decays in $\eta$ at the same Gaussian rate at which $\delta_{\W}(t)$ decays in $t$, so a fixed multiplicative widening of the window suppresses the miss term below any fixed multiple of the product term.
Taking $\eta \geq \sqrt{2}\,t$ already gives $\delta_{\W}(\eta) \leq \delta_{\W}(t)^2/2$, so the additive term is at most the square of the single-metric error and the quadratic character of Eq.~\eqref{eq:combined_bound} is preserved independently of $\delta_{\MMD}$.

Unlike the sample size, the window can therefore be set generously, at one to two concentration margins, at no cost in the bound and with no effect on the in-zone false-fail bound of Sec.~\ref{sec:combined}; the realized false-fail rate does grow with the window, because a wider window escalates more good pairs and subjects them to the AND rule (Sec.~\ref{sec:finite_shot_falsepass}).

\subsection{Baselines and the false-fail bound}\label{sm:baselines}

This compares upper bounds at equal sample size per metric, and both conventions flatter $D_4$: the realized rates of both rules sit far below either bound at the budgets we test (Sec.~\ref{sec:finite_shot_falsepass}), and Sec.~\ref{app:budget} restates the comparison at equal total budget, where the advantage survives with a stiffened precondition.

A second baseline is the single-metric rule at the shifted threshold, $D_1(\tau_{\W} - \eta)$, whose PASS region is contained in that of $D_4$: it has the lower false-pass probability, at most $\delta_{\W}(\eta + \Delta)$, and $D_4$ differs from it only by admitting, inside the window, the runs on which the discrepancy vote also passes.
The escalation therefore sits between the two single-metric rules $D_1(\tau_{\W})$ and $D_1(\tau_{\W} - \eta)$, trading false passes against false fails, and Sec.~\ref{sec:finite_shot_falsepass} reports both sides of the operating characteristic against both.

In return, the AND rule incurs a false-fail probability on a jointly-good pair. For good-side margins $\W(p,q;C) \leq \tau_{\W} - t'$ and $\MMD_k^2(p,q) \leq \tau_{\MMD} - \xi'$, with $t' > b_m$, it is at most $\delta_{\W}(t' - b_m) + \delta_{\MMD}(\xi')$, where $b_m$ is the upward bias of Eq.~\eqref{eq:w1_bias}. The bias pushes toward FAIL, so on this side it must be subtracted from the margin.
The composition itself is the intersection--union test of Berger and Hsu~\cite{BergerHsu1996}; we contribute its instantiation with explicit constants on $\{0,1\}^n$, the separation of the window from the analysis margins, and the cross-run agreement statistic of Sec.~\ref{sec:cross_run}.

\subsection{Same-data estimation}\label{sec:dependent_alpha}

The independence assumption holds by construction for the disjoint sample sets of Sec.~\ref{sec:protocol}.
When both metrics are computed from the same data, or exactly as in the present statevector campaign, independence can fail, and the joint-misvote term is replaced by a bounded conditional error: if $\Pr(v_{\MMD}=1 \mid v_{\W}=1) \leq \alpha_{CE} < 1$ on the defective pair, then
\begin{equation}
	\Pr(D_4 \text{ false-pass}) \leq \delta_{\W}(\eta) + \alpha_{CE}\,\delta_{\W}(\Delta).
\end{equation}
The exact statevector campaign cannot estimate $\alpha_{CE}$ for a fixed pair, and Sec.~\ref{sec:cross_run} reports the cross-run statistic it measures in its place.

\subsection{Scale of the correlation signal and kernel repair}\label{sm:kernel_repair}

At the operational bandwidth $\gamma = 1/n$ the correlation signal of Eq.~\eqref{eq:family2_mmd} decays with scale.
At $\gamma = 1/n$, $b^2 = (4n^2)^{-1}[1 + O(n^{-1})]$ and $c^{\,n-2} \to e^{-1/2}$, so
\begin{equation}\label{eq:family2_asymptotic}
	\MMD_k^2(p, q) \;=\; \frac{e^{-1/2}}{4}\,\frac{\beta^2}{n^2}\,\bigl(1 + O(n^{-1})\bigr),
\end{equation}
with $e^{-1/2}/4 \approx 0.1516$.
Detecting a fixed-$\beta$ correlation defect therefore requires an MMD tolerance that tightens as $n^{-2}$, and since the sample complexity of Proposition~\ref{prop:mmd} scales as $t^{-2}$ at additive accuracy $t$, taking $t = \Theta(\beta^2/n^2)$ gives a sufficient shot count $m = O(n^4/\beta^4)$ under the Hoeffding-type certificate.
This is a range-based upper bound on what the certificate requires, which a variance-adaptive bound could tighten (Sec.~\ref{sec:finite_shot_falsepass}).

One response to the $n^{-2}$ decay is to repair the kernel, and the repair is exactly computable.
We augment the RBF--Hamming kernel with the parity feature of the suspected pair, $k'(x,y) = k(x,y) + \varphi(x)\varphi(y)$ with $\varphi(x) = (-1)^{x_0 \oplus x_1}$.
For the Family 2 witness, $\mathbb{E}_q[\varphi] - \mathbb{E}_p[\varphi] = \beta$ exactly, so the augmentation adds $\beta^2$ to $\MMD_{k'}^2$ at every $n$, with no $2^{-n}$ suppression, and the correlation-shift window reopens at every scale at the cost of widening the kernel range from $[0, 1]$ to $[-1, 2]$.
If the defect axis is unknown, the same construction summed over all $\binom{n}{2}$ pairs still adds exactly $\beta^2$, because only the true pair's feature pairs with the discrepancy (any other parity is uncorrelated with it under $p$ and $q$ alike; we verified both statements numerically at $n = 3, 4, 6$).

The cost, however, does not vanish: the kernel now ranges from about $-n/2$ to $1 + \binom{n}{2}$, so the range width $K_{\max}$ of Proposition~\ref{prop:mmd} grows as $\binom{n}{2} + O(n)$, and the certificate scales as $K_{\max}^2/t^2 = \Theta(n^4/\beta^4)$, the same scaling as Eq.~\eqref{eq:family2_asymptotic} gives for the un-augmented kernel. Normalizing the sum instead returns the $n^{-2}$ decay to the signal.
Certification against a specified defect family thus costs $O(1)$ in the kernel range; against an unspecified one the certified shot cost is $O(n^4/\beta^4)$ under every repair of this kind, and Sec.~\ref{sec:parity} extends the limitation to every normalized kernel.

A third mechanism, ground-cost mismatch, has neither an exact witness nor a trained-generator demonstration: the transport acceptance regions $\{q : \W(p, q; C) \leq \tau\}$ for two ground costs are generically different polytopes in $\simplex$, and the MMD acceptance region is an RKHS ball, sensitive to the kernel bandwidth where the polytope is sensitive to the cost.

\subsection{Comparison with Kurkin et al.}\label{sm:kurkin}

Two qualifications bound the statement.
It concerns the population signal, and it fixes the kernel before the defect direction is chosen. A test that knows the plausible correlations can add their parity features, as in Sec.~\ref{sec:blindspots}; a test restricted to the $\binom{n}{2}$ pairs needs only $O(\beta^{-2}\log(n/\delta))$ samples by a union of Hoeffding bounds; and adaptive kernel selection~\cite{Biggs2023} is the general tool.

Kurkin \emph{et al.}~\cite{Kurkin2025} write the discrepancy of any stationary bounded kernel on $\{0,1\}^n$ as an average over the kernel's spectral measure of squared parity-moment mismatches, of which Eq.~\eqref{eq:kernel_fourier} is the RBF--Hamming instance. They show that the Gaussian-kernel discrepancy of a particular pair is $O(2^{-n})$ at total variation of order one. They also show that for every sufficiently large $n$ some pair at total variation one, built from a random split of the cube, has discrepancy at most $2^{-bn}$, with an unspecified constant $b > 0$, for every stationary characteristic kernel whose spectral measure is a probability measure. This last condition is the unit-diagonal normalization, under which the bound is meaningful since $\MMD_{Ck}^2 = C\,\MMD_k^2$.

Part~(c) differs in three respects.
The kernel class is every positive-semidefinite kernel with unit diagonal, stationary or not, so that nonstationary kernels and the normalization of any sum of kernels are covered.
The alternatives form the explicit family $q_{S,\beta}$, each member at the same transport and total-variation distance $\beta/2$ from the target and prepared by a shallow circuit.
The bound is an exact value with an attaining kernel.
Its quantifier order is, in addition, the one certification needs: the kernel is fixed before the defect is chosen. In Ref.~\cite{Kurkin2025} the exponentially small pair is fixed first and then defeats every kernel in its class, so neither statement contains the other.

\subsection{The redundancy boundary in context}\label{sm:redundancy}

The inequality is an instance of the general bound $\MMD_k(p,q) \leq \W(p,q;\rho_k)$ of Theorem~21(i) of Sriperumbudur \emph{et al.}~\cite{Sriperumbudur2010}, in which the ground metric is the kernel-induced distance $\rho_k(x,y) = \|k(\cdot,x) - k(\cdot,y)\|_{\mathcal{H}} = \sqrt{2(1 - e^{-\gamma d_H(x,y)})}$: for integer $d \geq 1$, $1 - e^{-\gamma d} \leq d\,(1 - e^{-\gamma})$ and $\sqrt{d} \leq d$ give $\rho_k \leq \sqrt{2(1 - e^{-\gamma})}\,d_H$, and Eq.~\eqref{eq:redundancy} follows.
The proof in Sec.~\ref{app:parity_proofs} takes the same route, a shortest Hamming path in feature space, where adjacent strings are $\sqrt{2(1 - e^{-\gamma})}$ apart, through an optimal coupling.
The theorem adds the sharpness, since the adjacent pair attains the constant, and the decision statement, the equivalence on the tolerance ratio. It is the elementary converse, on the Hamming cube, of the results that control transport distances by kernel norms on $\mathbb{R}^d$~\cite{Vayer2023,Modeste2024}.

The theorem concerns population values, whereas the discrepancy vote of the sampled protocol comes from an independent draw. It guarantees every pair whose population values lie in the window a population margin $\tau_{\MMD} - 2(1 - e^{-\gamma})\tau_{\W}^2 \geq 0.0086$ below the loose tolerance, but a draw that lands in the window does not certify that premise for the pair it came from. For a pair that does meet it, the Hoeffding tail of Proposition~\ref{prop:mmd} at that margin is $0.63$ at $S = 10^5$ and falls to $0.05$ only near $S = 6 \times 10^5$.
The finite-shot agreement is therefore a measurement, which the fixed-threshold test of Sec.~\ref{sec:b3p_validation} reports.

The adjacent pair is also the extremal defect for the window analysis: its discrepancy per unit transport distance is the largest any pair attains, so its window, $\eta^{\mathrm{adj}}_{\max} = \tau_{\W} - \sqrt{\tau_{\MMD}/(2(1 - e^{-\gamma}))} = 0.037$ to $0.029$ at $n = 3$--$8$ for the certification-grade point, contains the window of every defect family.
Both statements are for the Hamming cost; under $C_2$ and $C_3$ the constant in Eq.~\eqref{eq:redundancy} changes with the cost of an adjacent flip.

\subsection{Matched tolerances and the equivalent transport threshold}\label{sm:matched}

Both windows are statements about a fixed pair $(\tau_{\W}, \tau_{\MMD})$, so we ask which single transport threshold reproduces the escalation verdict on each family.
On each family the escalation acts as a transport-only rule with an effective threshold: by Theorem~\ref{thm:dominance} it rejects exactly the $\lambda$ with $\lambda n/2 > T_1$, and by Eq.~\eqref{eq:family2_mmd} exactly the $\beta$ with $\beta/2 > T_2$, where
\begin{align}\label{eq:equiv_threshold}
	T_i &= \min\bigl\{\tau_{\W},\ \max\{\tau_{\W} - \eta,\ \tau^{*}_i\}\bigr\}, \quad i = 1, 2, \nonumber\\
	\tau^{*}_1 &= \tfrac{n}{2}\sqrt{\tau_{\MMD}/\Phi_n}, \qquad \tau^{*}_2 = \tfrac12\sqrt{\tau_{\MMD}/A_n},
\end{align}
and $\tau^{*}_i$ is the transport distance at which the family's discrepancy crosses $\tau_{\MMD}$, the MMD-implied transport threshold.
The outer minimum records that a clear transport FAIL receives no discrepancy vote, and the inner maximum that a clear transport PASS receives none either.
Since $\eta_{\max,i} = \tau_{\W} - \tau^{*}_i$, the window of family $i$ is open at $\eta$ exactly when $\tau^{*}_i \leq \tau_{\W} - \eta$, and then $T_i = \tau_{\W} - \eta$: inside its own window the escalation is, on that family, the shifted single-metric rule $D_1(\tau_{\W} - \eta)$, and inside the deployment window of Sec.~\ref{sec:blindspots}, $\eta \leq \min_i \eta_{\max,i}$, one transport threshold reproduces the two-metric verdict on both families.
At $n = 3$ and $\eta = 0.005$, for instance, $\tau^{*}_1 = 0.025$ and $\tau^{*}_2 = 0.038$ both lie below $\tau_{\W} - \eta = 0.045$, and $T_1 = T_2 = 0.045$.

We keep two notions apart here. The window of family $i$ is \emph{open} when $\tau^{*}_i \leq \tau_{\W} - \eta$, so that every in-window member of the family is rejected and $T_i = \tau_{\W} - \eta$. The mismatch interval $(\max\{\tau_{\W} - \eta, \tau^{*}_i\}, \tau_{\W}]$ is non-empty under the weaker condition $\tau^{*}_i < \tau_{\W}$, which also holds for windows wider than $\eta_{\max,i}$. We use ``open'' in the first sense throughout.
In contrast, when both MMD-implied thresholds exceed the transport threshold, $\tau^{*}_i > \tau_{\W}$ (equivalently $\eta_{\max,i} < 0$), we have $T_1 = T_2 = \tau_{\W}$, and the escalation is $D_1(\tau_{\W})$ on both.
The two effective thresholds differ only for windows wider than the narrower family's, $\eta > \eta_{\max,2}$, and then at most by the ratio of the MMD-implied thresholds,
\begin{equation}\label{eq:threshold_mismatch}
	\frac{\tau^{*}_2}{\tau^{*}_1} = \sqrt{\frac{\Phi_n}{n^2 A_n}} = \frac{\rho^{*}_1(n)}{\rho^{*}_2(n)},
\end{equation}
which equals $1.54$, $1.55$, $1.56$, $1.57$, $1.58$ and $1.58$ at $n = 3$--$8$, with limit $2\sqrt{e^{1/2} - 1} \approx 1.61$: relative to $\W$, the discrepancy is more sensitive to mode collapse than to correlation shift by this factor at every $n$, and one discrepancy tolerance is equivalent to transport tolerances that differ between the families by it.

The factor enters through the two transport responses, $\W$ per unit $\lambda$ on Family~1 and per unit $\beta$ on Family~2 ($n/2$ and $1/2$ under $C_1$), and so depends on the ground cost. Under $C_2$ with the campaign weights it is $1.52$--$1.58$. Under $C_3$ with parity weight $0.5$, where the responses are $n/2 + (n-1)/4$ and $3/4$ (verified by exact LP), it is $1.73$ at $n = 3$ falling to $1.65$ at $n = 8$.

Matching the tolerances on Family~1, so that both metrics reject the same $\lambda$, means $\tau_{\MMD} = (2\tau_{\W}/n)^2\Phi_n$, the closing value of Eq.~\eqref{eq:eta_max}; the Family-1 window then has zero width by construction, and the Family-2 window is closed at every $n$, with relative width $-0.54$ to $-0.58$ for $n = 3$--$8$.
Matching on Family~2 instead, $\tau_{\MMD} = 4\tau_{\W}^2 A_n$, closes that window and leaves the Family-1 window open at every $n$ with the nearly $n$-independent relative width $1 - \sqrt{n^2 A_n/\Phi_n} = 0.35$--$0.37$.

The factor of Eq.~\eqref{eq:threshold_mismatch} between the two MMD-implied thresholds does not depend on the calibration, so that a certifier who wants a given transport tolerance enforced on both must name the family.

\subsection{Proofs of the concentration and false-pass bounds and of Theorem~\ref{thm:dominance}}

\begin{proof}[Proof of Proposition~\ref{prop:w1}]
	We first show that $\W(\cdot, q; C)$ is $D_{\max}$-Lipschitz with respect to total variation.
	For any $r, r' \in \simplex$,
	\begin{equation}\label{eq:w1_lipschitz}
		|\W(r, q; C) - \W(r', q; C)| \leq D_{\max} \cdot \TV(r, r').
	\end{equation}
	This follows from the dual formulation of Eq.~\eqref{eq:w1_dual}.
	Any $C$-Lipschitz potential $f$ has oscillation $\max_x f(x) - \min_x f(x) \leq D_{\max}$, and because the signed measure $r - r'$ has zero total mass, $\langle f, r - r' \rangle$ is unchanged by shifting $f$ by a constant; centering $f$ at $c = \tfrac12(\max_x f + \min_x f)$ gives $\|f - c\|_\infty \leq D_{\max}/2$, so
	\begin{align}
		|\langle f, r - q \rangle - \langle f, r' - q \rangle| &= |\langle f, r - r' \rangle| \nonumber\\
		&= |\langle f - c, r - r' \rangle| \nonumber\\
		&\leq \tfrac{D_{\max}}{2} \cdot \|r - r'\|_1 \nonumber\\
		&= D_{\max} \cdot \TV(r, r'). \label{eq:w1_lipschitz_chain}
	\end{align}
	Taking the supremum over $C$-Lipschitz $f$ gives the result, and by the symmetry of $\W$ in its two arguments the same bound holds in the second argument, which is the one the observation model uses.

	We now apply McDiarmid's inequality~\cite{McDiarmid1989}.
	Let $X_1, \ldots, X_m$ be i.i.d.\ draws from $q$, and let $\hat{q}_m = \frac{1}{m}\sum_i \delta_{X_i}$.
	Replacing $X_i$ by $X_i'$ moves mass $1/m$ from one outcome to another, so $\hat{q}_m$ changes by $1/m$ in total variation under the convention $\TV = \tfrac12\|\cdot\|_1$ fixed above (a change of $2/m$ in $L^1$), and by the Lipschitz bound it changes $\W(p, \hat{q}_m; C)$ by at most $\frac{D_{\max}}{m}$.
	By McDiarmid's inequality with bounded differences $c_i = \frac{D_{\max}}{m}$:
	\begin{multline}\label{eq:w1_mcdiarmid}
		\Pr\bigl(|\W(p, \hat{q}_m; C) - \mathbb{E}[\W(p, \hat{q}_m; C)]| > t\bigr)\\
		\leq 2\exp\!\left(-\frac{2t^2}{\sum_i c_i^2}\right) = 2\exp\!\left(-\frac{2m t^2}{D_{\max}^2}\right).
	\end{multline}
	This establishes Eq.~\eqref{eq:w1_conc} (concentration about the mean).
	For the bias of Eq.~\eqref{eq:w1_bias}, the Lipschitz bound of Eq.~\eqref{eq:w1_lipschitz} gives $|\bar{W}_m - \W(p,q;C)| \leq D_{\max}\,\mathbb{E}[\TV(\hat{q}_m, q)]$, and
	\begin{equation}\label{eq:tv_expectation}
		\mathbb{E}[\TV(\hat{q}_m, q)] = \tfrac12 \sum_x \mathbb{E}|\hat{q}_m(x) - q(x)| \leq \frac{1}{2\sqrt{m}}\sum_x \sqrt{q(x)},
	\end{equation}
	where the inequality bounds each mean absolute deviation by the binomial standard deviation $\sqrt{q(x)(1 - q(x))/m} \leq \sqrt{q(x)/m}$. Cauchy--Schwarz then gives the worst-case $\sqrt{N}$ factor (cf.\ the $m^{-1/2}$ rate of empirical measures on a finite support~\cite{Weed2019}).
\end{proof}

\begin{proof}[Proof of Proposition~\ref{prop:mmd}]
	For the two-sample case with $0 \leq k \leq K_{\max}$ and even $m$, this is Theorem~10 of Gretton \emph{et al.}~\cite{Gretton2012}, whose exponent $t^2\lfloor m/2 \rfloor/8K_{\max}^2$ equals $mt^2/16K_{\max}^2$; the argument below holds for every $m$, for both cases and for any kernel whose values lie in an interval of width $K_{\max}$.
	The unbiased estimator $\widehat{\MMD}_u^2$ is a one-sample $U$-statistic of the pooled sample of size $2m$.
	Each observation enters $O(m)$ terms, and each term varies by at most $K_{\max}$.
	Replacing one observation changes $\widehat{\MMD}_u^2$ by at most $4K_{\max}/m$ (since each observation appears in at most $2(2m-1)$ of the $\binom{2m}{2}$ kernel evaluations, contributing $O(K_{\max}/m)$ each).
	McDiarmid's inequality with $2m$ variables and bounded differences $c_i = 4K_{\max}/m$ bounds the upward and the downward deviation separately by $\exp(-mt^2/16K_{\max}^2)$, which gives both tails of Eq.~\eqref{eq:mmd_conc}.

	In the one-sample case, used in Sec.~\ref{sec:finite_shot_falsepass}, one distribution is known exactly and $m$ samples are drawn from the other. Replacing one sample changes the $U$-statistic by at most $2K_{\max}/m$ through the pairwise kernel average and by at most $2K_{\max}/m$ through the cross term, so McDiarmid's inequality with $m$ variables and $c_i = 4K_{\max}/m$ gives $\exp(-mt^2/8K_{\max}^2)$ for each tail, which implies Eq.~\eqref{eq:mmd_conc}. The lower tail is the one a false pass in Eq.~\eqref{eq:combined_bound} requires.
\end{proof}

\begin{proof}[Proof of the false-pass bounds of Eqs.~\eqref{eq:combined_clear} and~\eqref{eq:combined_bound}]
	Throughout, a false pass requires $\hat{w}$ to fall below the true value, and since $\bar{W}_m \geq \W(p,q;C)$ by convexity of $\W(p,\cdot\,;C)$ and Jensen's inequality, such a deviation from the true value is at least as large a deviation from the mean, so Proposition~\ref{prop:w1} applies without the bias term of Eq.~\eqref{eq:w1_bias}.

	\textbf{Outside the margin zone.} Outside $\mathcal{M}_\eta$ the true $\W$ value differs from the threshold by at least $\eta$.
	A $D_4$ PASS always requires the $\W$ vote to pass, so a false pass requires the empirical $\W$ verdict to be wrong, an estimation error exceeding $\eta$; by Proposition~\ref{prop:w1} this has probability at most $\delta_{\W}(\eta)$, whether or not escalation fires, which is Eq.~\eqref{eq:combined_clear}.

	\textbf{Inside the margin zone.} A $D_4$ PASS falls into one of two disjoint events determined by $\hat{w}$: the clear case $P_1 = \{\hat{w} \leq \tau_{\W} - \eta\}$ (no escalation), and the escalated case $P_2 = \{\tau_{\W} - \eta < \hat{w} \leq \tau_{\W}\} \cap \{v_{\MMD} = 1\}$ (both votes PASS); empirical values $\hat{w} > \tau_{\W}$ return FAIL on either branch, since a PASS requires $v_{\W} = 1$.
	On $P_1$ the pair is defective with $\W(p,q;C) = \tau_{\W} + \Delta$, so $\hat{w} \leq \tau_{\W} - \eta$ forces an estimation error exceeding $\eta + \Delta > \eta$; by Proposition~\ref{prop:w1}, $\Pr(P_1) \leq \delta_{\W}(\eta)$, and the protocol certifies without evaluating MMD, so no product structure is available.

	On $P_2$ the $\W$ vote passes only if $\hat{w} \leq \tau_{\W}$, an error exceeding $\Delta$, giving $\Pr(v_{\W}=1) \leq \delta_{\W}(\Delta)$; the MMD vote passes only if its empirical statistic falls below $\tau_{\MMD}$ while the population value sits $\xi$ above threshold, giving $\Pr(v_{\MMD}=1) \leq \delta_{\MMD}(\xi)$ by Proposition~\ref{prop:mmd}.
	The two statistics use disjoint samples, so the misvotes are independent and $\Pr(P_2) \leq \delta_{\W}(\Delta)\,\delta_{\MMD}(\xi)$.
	This is the acceptance region of an intersection--union test~\cite{BergerHsu1996}. The union bound over $P_1 \cup P_2$ gives Eq.~\eqref{eq:combined_bound}.
\end{proof}

\begin{proof}[Proof of Theorem~\ref{thm:dominance}]
	We write $q_\lambda - p = \lambda(\delta_{x_0} - p)$ from Eq.~\eqref{eq:mixture_witness}.

	\textbf{Step 1: W$_1$ (claim (a)).}
	The signed measure $q_\lambda - p$ has an excess $\lambda(1 - 2^{-n})$ at $x_0$ and a deficit $\lambda\,2^{-n}$ at each other point. It has a single source, $x_0$, so the plan is forced: it moves mass $\lambda\,2^{-n}$ from $x_0$ to each $x$ at Hamming cost $d_H(x_0, x)$. Hence
	\begin{equation}
		\W(p, q_\lambda; C_1) = \lambda \sum_{x} 2^{-n} d_H(x_0, x) = \lambda\, n/2,
	\end{equation}
	using $\mathbb{E}_{x \sim \mathrm{Unif}}[d_H(x, x_0)] = \sum_{i=1}^n \Pr[x_i \neq (x_0)_i] = n/2$.

	\textbf{Step 2: Total variation (claim (b)).}
	Since $q_\lambda - p = \lambda(\delta_{x_0} - p)$ and $\|\delta_{x_0} - p\|_1 = 2(1 - 2^{-n})$, the total variation is $\TV(p, q_\lambda) = \lambda\,(1 - 2^{-n})$.

	\textbf{Step 3: MMD (claim (c)).}
	MMD is the RKHS norm of the mean-embedding difference, and the embedding is linear in the distribution, so
	\begin{align}
		\MMD_k(p, q_\lambda) &= \|\mu_p - \mu_{q_\lambda}\|_{\RKHS} \nonumber\\
		&= \lambda\,\|\mu_{\delta_{x_0}} - \mu_p\|_{\RKHS} = \lambda\,\MMD_k(p, \delta_{x_0}). \label{eq:witness_mmd_linear}
	\end{align}
	We set $a := \mathbb{E}_{x \sim p}[k(x, x_0)]$. By the binomial expansion of the Hamming-distance generating function,
	\begin{equation}\label{eq:mean_embed}
		a = 2^{-n}\sum_{d=0}^n \binom{n}{d} e^{-\gamma d} = \left(\frac{1 + e^{-\gamma}}{2}\right)^{\!n} = c^n.
	\end{equation}
	Translation invariance of $d_H$ gives $\mathbb{E}_{x' \sim p}[k(x,x')] = a$ for every fixed $x$, hence $\mathbb{E}_{x,x' \sim p}[k(x,x')] = a$; with $k(x_0,x_0) = 1$ the point-mass expansion yields $\MMD_k^2(p, \delta_{x_0}) = a + 1 - 2a = 1 - c^n = \Phi_n$. Squaring Eq.~\eqref{eq:witness_mmd_linear} gives claim~(c).

	\textbf{Step 4: Strict inclusion.}
	For any $\lambda$ in the interval of Eq.~\eqref{eq:witness_interval}, claim~(a) gives $\W(p, q_\lambda; C_1) = \lambda n/2 \leq \tau_{\W}$ with $\W > \tau_{\W} - \eta$, so $q_\lambda \notin \mathcal{R}_{\W}$ and the one-sided margin condition of $D_4$ holds, triggering the MMD escalation. Because $\MMD_k^2(p, q_\lambda)$ is increasing in $\lambda$ by claim~(c), its minimum over the interval is at the left endpoint; if the condition of Eq.~\eqref{eq:dominance_condition} holds there, then $\MMD_k^2(p, q_\lambda) > \tau_{\MMD}$ throughout, the MMD vote is FAIL, and the conservative AND rule returns FAIL. Thus $q_\lambda \in \mathcal{R}_{\text{bi}} \setminus \mathcal{R}_{\W}$, and since $\mathcal{R}_{\W} \subseteq \mathcal{R}_{\text{bi}}$ by construction, $\mathcal{R}_{\text{bi}} \supsetneq \mathcal{R}_{\W}$.

	The operational claim is a direct substitution: at the knife-edge endpoint $\lambda = 2\tau_{\W}/n$ (the limit $\eta \to 0$) the right-hand side of Eq.~\eqref{eq:dominance_condition} is $4.1 \times 10^{-4}, 2.3 \times 10^{-4}, 1.5 \times 10^{-4}, 1.1 \times 10^{-4}$ at $n = 3, 4, 5, 6$, each exceeding $\tau_{\MMD} = 10^{-4}$; the corresponding largest admissible windows are $\eta_{\max}(n) = 0.025, 0.017, 0.009, 0.001$ from Eq.~\eqref{eq:eta_max}.
\end{proof}

\begin{proof}[Transport value of the Family 2 correlation shift]
	Let $p = \mathrm{Unif}(\{0,1\}^n)$ and $q(x) = 2^{-n}[1 + \beta\,(-1)^{x_0 \oplus x_1}]$ with $\beta \in (0,1)$, and let $C_1$ be the Hamming cost. We show $\W(p, q; C_1) = \beta/2$.

	\textbf{Upper bound.} The signed measure $q - p$ carries $+2^{-n}\beta$ on each of the $2^{n-1}$ strings with $x_0 = x_1$ and $-2^{-n}\beta$ on each of the $2^{n-1}$ strings with $x_0 \neq x_1$, so the total excess is $\beta/2$. We pair each excess string $x$ with the deficit string obtained by flipping $x_0$, a bijection between the two sets with $d_H = 1$ on every pair. Transporting along it moves the whole excess at unit cost, so $\W(p, q; C_1) \leq \beta/2$.

	\textbf{Lower bound.} We take $f(x) = \tfrac12 (-1)^{x_0 \oplus x_1}$. Then $|f(x) - f(y)| \in \{0, 1\}$, and it equals $1$ only when $x$ and $y$ differ in $x_0$ or in $x_1$, in which case $d_H(x,y) \geq 1$; hence $f$ is $C_1$-Lipschitz. Its pairing with the discrepancy is
	\begin{equation}
		\langle f, q - p \rangle = \sum_x \tfrac12 (-1)^{x_0 \oplus x_1}\, 2^{-n} \beta\, (-1)^{x_0 \oplus x_1} = \frac{\beta}{2},
	\end{equation}
	since the squared sign is one at every $x$. By Kantorovich duality, Eq.~\eqref{eq:w1_dual}, $\W(p, q; C_1) \geq \beta/2$.

	The two bounds meet, so the value is exact. We have additionally confirmed it by exact linear programming at $n = 2$--$7$ and $\beta \in \{0.05, 0.3\}$, where the primal optimum, $\beta/2$, and the dual pairing agree to machine precision.
\end{proof}

\section{The finite-shot experiment}\label{sm:finite_shot}\label{app:finite_shot_exp}

This section gives the design and the full rates of the finite-shot test of Sec.~\ref{sec:finite_shot_falsepass} and the shot cost of certification-grade tolerances.

Across all $120$ configurations these realized rates stay below the bound of Eq.~\eqref{eq:combined_bound}, with no violations and by a wide margin: the bound is non-vacuous in only $22$ of the $120$ configurations, because $\delta_{\W}(\eta)$ with the window $\eta = \sqrt{2}\,t$ at the concentration margin $t = 0.01$ does not fall below unity until $S \approx 10^5$ at these defect sizes, while the realized rates are already at the percent level by $S = 10^3$.
The gap comes from the Hoeffding-type tails of Propositions~\ref{prop:w1} and~\ref{prop:mmd}, which use only the range of the statistic; an empirical-Bernstein tail at the measured variance is tighter by a factor of $2.6 \times 10^{3}$ to $2.0 \times 10^{5}$, so a variance-adaptive certificate would be non-vacuous where the realized rates already sit.

On the false-fail side (Sec.~\ref{app:finite_shot_exp}), the escalation keeps the false-fail behavior of $D_1(\tau_{\W})$ to within a factor of two and takes about half of the false-pass gap toward $D_1(\tau_{\W} - \eta)$; the shifted rule closes the rest of that gap with a false-fail rate forty times larger inside the window.

Pairs placed in the window are rejected by the shifted rule by construction, so the deployment cost is set by how often ordinary fits land there: of the $776$ unperturbed campaign fits that the transport threshold admits against the clean reference, $70$ ($9.0\%$; $8$, $17$, $15$ and $30$ at $n = 5$--$8$, none below $n = 5$) have $\W \in (\tau_{\W} - \eta, \tau_{\W}]$ and would be rejected outright by $D_1(\tau_{\W} - \eta)$. The escalation evaluates them and, on exact values, keeps $62$ and rejects $8$ ($1.0\%$ of the admitted fits). At smaller budgets both rules reject borderline good runs frequently, through the same $\W$ bias mechanism recorded below.

The false-pass rate is not monotone in $S$: it is small at $S = 10^2$, peaks near $S = 10^3$ or $10^4$, and then decays.
The small-$S$ suppression comes from bias: the $\W$ bias term of Eq.~\eqref{eq:w1_bias} inflates $\hat{w}$ at low shot counts and pushes borderline defective pairs across $\tau_{\W}$ into a FAIL verdict reached for the wrong reason.

Sec.~\ref{sec:shot_budget} quantifies what certification-grade tolerances cost in shots.

\subsection{Design}

The MMD vote uses the unbiased $U$-statistic of Proposition~\ref{prop:mmd}, in its one-sample form against the exact target. On the same draws we also evaluate the plug-in $V$-statistic, whose positive bias (Sec.~\ref{sec:shot_budget}) pushes the vote toward FAIL.
Defective pairs are of two kinds: members of the mode-collapse family of Theorem~\ref{thm:dominance} placed a controlled distance $\Delta \in \{0.005, 0.02\}$ above $\tau_{\W}$, and trained generators from the campaign scored against the clean benchmark target, retained when $\W$ exceeds $\tau_{\W}$ (nine fits, three of which are the warm-restarted versions of Sec.~\ref{app:restart}). Each pair is defined by its stored distribution, so the choice of version does not enter the rates.

\subsection{False-pass rates}

At the benchmark-comparison thresholds $D_4$ and $D_1$ differ in $5$ of the $60$ configurations. These exceptions are all campaign-fit pairs at $S = 100$, where the shot noise of the discrepancy estimate exceeds the margin to the loose tolerance and $D_4$ falls below $D_1$ by $0.002$ to $0.009$; none occurs at $S \geq 10^3$.

At certification grade the largest realized $D_1$ false-pass rate is $0.357$, on a configuration where $D_4$ realizes $0.20$, the largest $D_4$ rate in the regime.

Against the shifted single-metric rule $D_1(\tau_{\W} - \eta)$, evaluated on the same draws, the false-pass ordering reverses, as the inclusion of Sec.~\ref{sec:combined} requires. At $S = 10^{3}$ its mean false-pass rate over the certification-grade configurations is $0.006$, against $0.033$ for $D_4$ and $0.071$ for $D_1(\tau_{\W})$. Among the $29$ configurations with a nonzero $D_1(\tau_{\W})$ rate $D_4$ matches $D_1(\tau_{\W} - \eta)$ in $11$ and exceeds it in $18$, by a median factor of $2.3$. At $S = 10^{5}$ all three rates are zero in $2{,}000$ repetitions (upper limit $0.0015$ per configuration).

On the same draws the plug-in estimator reports $13$ configurations driven to zero, a median factor of $9.5$, and a largest $D_4$ rate of $0.062$. The difference is the plug-in bias, which also raises the false-fail rate below.

The Hoeffding-type tails of Propositions~\ref{prop:w1} and~\ref{prop:mmd} use only the range of the statistic, while the MMD estimator's variance at these budgets is smaller than its range by orders of magnitude. Measuring the realized standard deviation of $\MMD_k^2$ on the mode-collapse witness and comparing an empirical-Bernstein tail against the Hoeffding tail at matched confidence gives a ratio of $2.6 \times 10^{3}$ at $S = 10^{3}$, rising to $7.5 \times 10^{4}$ at $S = 10^{5}$ for $n = 3$ and to $2.0 \times 10^{5}$ at $n = 6$.
A variance-adaptive certificate requires a Bernstein bound for the degenerate $U$-statistic, with the empirical variance estimated from the same shots, which is a separate technical problem.

\subsection{False-fail rates and escalation frequency}

We measure the false-fail rate and the escalation frequency, which sets the adaptive scheme's extra shot cost, on pairs that each exact channel judges good at the regime's tolerances.

In the clear region neither rule incurs false fails at large budgets: they are zero in $2{,}000$ repetitions by $S = 10^5$ (upper limit $0.0015$), and the escalation fires in at most a few percent of repetitions, so the expected cost of $D_4$ is one sample set.

The cost of the conservative AND rule appears on jointly-good pairs placed in the window. At $S = 10^5$ and certification grade the escalation fires in $59$--$100\%$ of repetitions on these pairs. The realized false-fail rate of $D_4$ is at most $0.11$, against $0.053$ for $D_1(\tau_{\W})$, and where both are nonzero it exceeds $D_1$'s by a factor of $1.2$ to $2.0$ ($2.6$ with the plug-in estimator).

The shifted rule $D_1(\tau_{\W} - \eta)$ obtains its lower false-pass rate at this cost: on the same window pairs at $S = 10^5$ it rejects $59\%$ to $100\%$ of repetitions (mean $0.87$), against a mean of $0.02$ for $D_4$ and $0.01$ for $D_1(\tau_{\W})$, while in the clear region the three rules agree at $S = 10^5$.

In contrast, on defective pairs the escalation frequency is small (median $0.00$, maximum $0.32$ across the $120$ configurations), because most estimates fall on a clear side of the window.

Four in-window witness pairs are excluded from the good-pair accounting at certification grade because their exact $\MMD_k^2$ exceeds the tolerance: they are the mismatch-region members of Theorem~\ref{thm:dominance}, and $D_4$ rejects them correctly.

\subsection{Shot cost of certification-grade thresholds}\label{sec:shot_budget}

The advantage of the escalation appears only at certification-grade tolerances, $\tau_{\MMD} \leq 10^{-4}$. In this subsection, we estimate the number of shots required to resolve a tolerance that tight.

The certificate of Proposition~\ref{prop:mmd} demands $m \geq 16 K_{\max}^2 \ln(1/\delta)/t^2$, i.e.\ $m \geq 4.8 \times 10^{9}$ at $t = 10^{-4}$ and $\delta = 0.05$; conversely $S = 10^5$ shots, the largest we test, certify $\MMD_k^2$ only to $\pm 2.2 \times 10^{-2}$. The transport side (Corollary~\ref{cor:w1_samples}) is milder by three orders, so the discrepancy channel binds.

The certificate is loose, so we also measure the requirement directly, on the exact witnesses of both defect families placed at their window midpoints. Where the certification-grade window is narrower than $0.1\,\tau_{\W}$ or closed (mode collapse at $n = 6$--$8$, correlation shift at $n = 4$--$8$), the witness is placed in the window of the tightened $\tau_{\MMD}$ of Sec.~\ref{sec:blindspot_demo}, a quarter of the closing value, which pins $\eta_{\max}$ to $\tau_{\W}/2$.

A useful vote must return FAIL on the defect and PASS on a perfect generator. An estimator with a positive bias meets the first condition at any budget, so we require both. At power $0.95$ and size $0.05$ the unbiased $U$-statistic needs $S \approx 10^5$ shots for the mode-collapse witness at $n = 3, 4$ and $n = 6$--$8$ and $10^6$ at $n = 5$, and $10^5$ for the correlation-shift witness at every scale.

The measured requirement is four orders of magnitude below the certificate. It is still at or above the top of the range Sec.~\ref{sec:finite_shot_falsepass} explores, and a full order above it for mode collapse at $n = 5$, the narrowest of the windows kept at the certification-grade tolerance.

The second condition matters because the plug-in ($V$-statistic) estimator, which Sec.~\ref{sec:finite_shot_falsepass} reports for comparison, carries a positive bias. With the reference known exactly and $m$ samples $y_i$ drawn from $q$, its double sum $m^{-2}\sum_{i,j} k(y_i, y_j)$ contains the $m$ diagonal terms $k(y_i, y_i) = 1$, so that its expectation is $m^{-1} + (1 - m^{-1})\,\mathbb{E}_{y,y' \sim q}[k(y,y')]$, while the cross term is unbiased. The bias is therefore $\bigl(1 - \mathbb{E}_{y,y' \sim q}[k(y,y')]\bigr)/m$. For a uniform reference it equals $\Phi_n/m$ for a perfect generator and $\Phi_n(1 - \lambda^2)/m$ for the mode-collapse witness, with $\Phi_n$ the bracket of Theorem~\ref{thm:dominance}(c), the same quantity that carries the separation there.

Since the witness has $\MMD_k^2 = \lambda^2 \Phi_n$, plug-in bias equals signal at $m = (1 - \lambda^2)/\lambda^2 \approx \lambda^{-2}$, independently of $n$ and of the bandwidth, which is $1.6 \times 10^3$ shots at $n = 3$ rising to $1.1 \times 10^4$ at $n = 8$. Below that budget the estimator exceeds $\tau_{\MMD} = 10^{-4}$ on a defective and a perfect generator alike, its size reaching $1.00$ at $S = 10^2$ and still $0.92$ at $S = 10^3$ for $n = 3$. This is the discrepancy-side counterpart of the transport bias of Sec.~\ref{sec:finite_shot_falsepass}. The unbiased $U$-statistic removes the term at the cost of a larger variance and reaches a usable operating point at $S \approx 10^5$.

For correlation defects the budget carries the additional $\Theta(n^4/\beta^4)$ scaling of Sec.~\ref{sec:blindspots}.

\section{The fixed-threshold test}\label{sm:fixed_threshold}\label{app:b3p_detail}

This section gives the recalibration and the paired comparisons of the fixed-threshold test of Sec.~\ref{sec:b3p_validation}.

We classify each fit by its exact values as transport-defective ($\W > \tau_{\W}$), jointly good ($\W \leq \tau_{\W}$ and $\MMD_k^2 \leq \tau_{\MMD}$) or mismatch ($\W \leq \tau_{\W}$ and $\MMD_k^2 > \tau_{\MMD}$), and compare $D_4$ with $D_1(\tau_{\W})$, with $D_1(\tau_{\W} - \eta)$ and with a recalibrated single-metric rule $D_1(\tau^{*})$.

The recalibrated threshold $\tau^{*}$ is cross-fitted by seed cluster to match the false-fail rate of $D_4$ on held-out seeds, the choice least favorable to the escalation, and a cost-matched variant also grants $D_1(\tau^{*})$ the expected shot cost of the escalation; Sec.~\ref{app:b3p_detail} gives the procedure, the paired comparisons on the transport-defective fits and Table~\ref{tab:b3p_validation}, which reports the certification-grade regime.

Moreover, at $S \leq 10^4$ the upward bias of $\hat{w}$ rejects all but a few draws of the good fits under every rule (false-fail above $0.999$ at $n = 5$ and $6$ with $S = 10^3$, and at $n \geq 7$ with $S \leq 10^4$), the bias mechanism of the preceding subsection on targets whose $\W$ sits near the threshold.

The escalation's false-pass rate equals, to the third decimal, that of a single transport threshold calibrated on held-out seeds to its false-fail rate and given its shot cost in six of the nine cells with false passes; the recalibrated rule is lower in the two cells at $n = 5$, and the escalation is lower in one cell, at $n = 3$ and $S = 10^4$, by $0.003$.

\subsection{Cross-fitted recalibration and paired comparisons}

The recalibration is cross-fitted by seed cluster. For each of the $20$ training seeds of a scale, $\tau^{*}$ is the smallest threshold whose false-fail rate on the jointly-good fits of the other $19$ seeds does not exceed that of $D_4$ on the same fits. $D_1(\tau^{*})$ is then evaluated on the held-out seed's fits. Its realized false-fail rates differ from the calibration target by up to $0.009$ (Table~\ref{tab:b3p_validation}), and $\tau^{*}$ varies by less than $0.003$ across the folds in every cell.

A cost-matched variant repeats the calibration and the evaluation on draws of $S' = \lceil S(1 + \bar{e})\rceil$ shots, where $\bar{e}$ is the escalation rate over all fits of the scale. That rate is zero where every rule rejects every fit and otherwise $0.09$ to $0.36$ (over the jointly-good fits alone, $0.01$ to $0.57$).

\begin{table*}[t]
	\caption{Fixed-threshold certification of the $600$ perturbed full-support fits at certification grade ($\tau_{\MMD} = 10^{-4}$), $200$ repetitions per fit, all $18$ cells. Counts: jointly good / transport-defective / mismatch fits. FF: false-fail rate on the good fits, printed as $1.000$ when it rounds there (the smallest such value is $0.9996$); $\tau^{*}$: mean of the cross-fitted threshold over the $20$ folds; FF $D_1(\tau^{*})$: realized false-fail rate of the cross-fitted rule on the held-out fits; $S'/S$: shot budget of the cost-matched variant, $S' = \lceil S(1 + \bar{e})\rceil$, relative to $S$, and ``cost'': the realized false-fail rate of the rule recalibrated and evaluated at $S'$; FP: false-pass rate on the transport-defective fits, for $D_1(\tau^{*})$ cross-fitted at $S$ and, under ``cost'', at $S'$; MR: rejection rate on the mismatch fits. An entry $0$ is zero in every repetition (one-sided $95\%$ upper limits $3/(200\,N_{\mathrm{def}})$ for FP on this fixed panel, $0.0002$--$0.0009$ for the $67$ to $17$ defective fits per scale), and $<\!0.001$ is a nonzero rate below $0.0005$.}\label{tab:b3p_validation}
	\scriptsize
	\begin{ruledtabular}
		\begin{tabular}{ccccccccccccccc}
			 &  &  & \multicolumn{2}{c}{FF} &  & \multicolumn{3}{c}{FF, cross-fitted} & \multicolumn{4}{c}{FP} & \multicolumn{2}{c}{MR} \\
			$n$ & $S$ & good/def./mis. & $D_1(\tau_{\W})$ & $D_4$ & $\tau^{*}$ & $D_1(\tau^{*})$ & $S'/S$ & cost & $D_1(\tau_{\W})$ & $D_4$ & $D_1(\tau^{*})$ & cost & $D_4$ & $D_1(\tau^{*})$ \\
			\hline
			3 & $10^3$ & 60/18/22 & $0.346$ & $0.462$ & $0.045$ & $0.462$ & $1.30$ & $0.462$ & $0.052$ & $0.017$ & $0.032$ & $0.017$ & $0.75$ & $0.71$ \\
			3 & $10^4$ &          & $0$ & $0.002$ & $0.039$ & $0.002$ & $1.12$ & $0.002$ & $0.052$ & $0.001$ & $0.004$ & $0.004$ & $0.49$ & $0.42$ \\
			3 & $10^5$ &          & $0$ & $0$ & $0.034$ & $0.002$ & $1.09$ & $0.009$ & $0.017$ & $0$ & $0$ & $0$ & $0.43$ & $0.55$ \\
			4 & $10^3$ & 72/17/11 & $0.880$ & $0.912$ & $0.048$ & $0.912$ & $1.09$ & $0.912$ & $0.005$ & $0.003$ & $0.004$ & $0.003$ & $0.98$ & $0.97$ \\
			4 & $10^4$ &          & $0.005$ & $0.025$ & $0.044$ & $0.025$ & $1.13$ & $0.025$ & $0.019$ & $0.001$ & $0.003$ & $0.001$ & $0.73$ & $0.60$ \\
			4 & $10^5$ &          & $0$ & $0.003$ & $0.040$ & $0.006$ & $1.09$ & $0.006$ & $0.007$ & $0$ & $0$ & $0$ & $0.72$ & $0.60$ \\
			5 & $10^3$ & 69/26/5  & $1.000$ & $1.000$ & $0.050$ & $1.000$ & $1.00$ & $1.000$ & $0$ & $0$ & $0$ & $0$ & $1.00$ & $1.00$ \\
			5 & $10^4$ &          & $0.094$ & $0.114$ & $0.048$ & $0.114$ & $1.22$ & $0.114$ & $0.007$ & $0.004$ & $0.003$ & $0.001$ & $0.99$ & $0.98$ \\
			5 & $10^5$ &          & $<\!0.001$ & $0.019$ & $0.045$ & $0.023$ & $1.13$ & $0.025$ & $0.016$ & $0.015$ & $<\!0.001$ & $0$ & $0.97$ & $0.19$ \\
			6 & $10^3$ & 72/27/1  & $1.000$ & $1.000$ & $0.083$ & $1.000$ & $1.00$ & $1.000$ & $0$ & $0$ & $0$ & $0$ & $1.00$ & $1.00$ \\
			6 & $10^4$ &          & $0.468$ & $0.471$ & $0.050$ & $0.471$ & $1.35$ & $0.471$ & $0$ & $0$ & $0$ & $0$ & $1.00$ & $1.00$ \\
			6 & $10^5$ &          & $0.010$ & $0.012$ & $0.050$ & $0.014$ & $1.21$ & $0.013$ & $<\!0.001$ & $<\!0.001$ & $0$ & $0$ & $0.65$ & $0$ \\
			7 & $10^3$ & 44/56/0  & $1.000$ & $1.000$ & $0.132$ & $1.000$ & $1.00$ & $1.000$ & $0$ & $0$ & $0$ & $0$ &  &  \\
			7 & $10^4$ &          & $1.000$ & $1.000$ & $0.050$ & $1.000$ & $1.00$ & $0.999$ & $0$ & $0$ & $0$ & $0$ &  &  \\
			7 & $10^5$ &          & $0.266$ & $0.266$ & $0.050$ & $0.266$ & $1.25$ & $0.267$ & $0$ & $0$ & $0$ & $0$ &  &  \\
			8 & $10^3$ & 33/67/0  & $1.000$ & $1.000$ & $0.198$ & $1.000$ & $1.00$ & $1.000$ & $0$ & $0$ & $0$ & $0$ &  &  \\
			8 & $10^4$ &          & $1.000$ & $1.000$ & $0.076$ & $1.000$ & $1.00$ & $1.000$ & $0$ & $0$ & $0$ & $0$ &  &  \\
			8 & $10^5$ &          & $0.706$ & $0.706$ & $0.050$ & $0.706$ & $1.10$ & $0.706$ & $0$ & $0$ & $0$ & $0$ &  &  \\
		\end{tabular}
	\end{ruledtabular}
\end{table*}

On the transport-defective fits we compare the rules through the paired difference $\mathrm{FP}(D_4) - \mathrm{FP}(D_1(\tau^{*}))$, with a seed-cluster interval and a shot-repetition interval at fixed fits. At $n = 3$, $S = 10^3$ the difference is $-0.014$, with cluster interval $[-0.022, -0.008]$ over the $16$ seeds that carry the $18$ defective fits and repetition interval $[-0.020, -0.009]$. This is the only cell in which the cluster interval excludes zero in the escalation's favor, and the two rules agree in the cost-matched variant ($0.017$ against $0.017$; cost-matched difference $0.000$, cluster interval $[-0.004, 0.004]$).

At $n = 3$, $S = 10^4$ the cost-matched difference is $-0.003$, with cluster interval $[-0.007, 0.000]$ and repetition interval $[-0.005, -0.001]$. At $n = 4$ and at $n = 5$, $S = 10^4$ it is $0.000$ to $0.003$ with cluster intervals that reach zero. At $n = 5$, $S = 10^5$ the recalibrated rule attains the lower rate, $0$ against $0.015$ for $D_4$, with cluster interval $[0.000, 0.046]$ and repetition interval $[0.013, 0.019]$: the false fails that $D_4$ incurs on in-window good fits, when spent on a lower single threshold instead, yield a cleaner false-pass rate than the discrepancy vote returns. From $n = 6$ every rule's false-pass rate is zero or below $0.001$.

The cluster intervals describe variation across the fitted generators of this benchmark, and the repetition intervals the shot noise on a fixed panel of fits. Both are conditional on the fitted cross-validation decisions, since the calibration is not repeated inside the resamples and the folds overlap, and we pre-specified no equivalence margin.

\section{Campaign decomposition}\label{sm:campaign}

This section decomposes the campaign verdicts at the certification-grade operating point: the cross-run agreement statistic, the disagreement between the two votes, the tightened thresholds of the trained-generator test and the split of the campaign mismatch runs.

On the pre-registered stress families, trained-model $\W$ and MMD AUROCs are both low ($0.40$--$0.77$ across cells), and neither approaches its convergence-conditioned performance.

\subsection{Cross-run agreement and threshold calibration}\label{sec:cross_run}

Because the campaign scores are exact, the conditional error $\alpha_{CE}$ of Sec.~\ref{sec:dependent_alpha} cannot be estimated for a fixed pair, and we record instead a descriptive statistic $a_{\mathrm{run}}$: the fraction of the perturbed ($\epsilon > 0$) B1 runs whose $\W$ vote passes on which the MMD vote also passes.
It counts the co-occurrence of exact-value votes across different runs, some of them transport-good. A bound on the conditional misvote probability of a specified transport-defective pair would need repeated same-data sampling of that pair; the finite-shot tests of Secs.~\ref{sec:finite_shot_falsepass} and~\ref{sec:b3p_validation} use disjoint draws and measure the false-pass rate directly.

At the benchmark tolerance $\MMD_k^2 \leq 10^{-2}$, and still at $10^{-3}$, $a_{\mathrm{run}} = 1.000$ at every scale and under both references: the escalation never overturns a $\W$ verdict, and Theorem~\ref{thm:redundancy} makes this exact at $10^{-2}$ for the population values of every pair.

In contrast, at the certification-grade $\MMD_k^2 \leq 10^{-4}$ the statistic falls below one, by different amounts under the two references.
Under the trained-on convention it reads $0.932$, $0.931$, $0.972$, $0.977$, $1.000$, $1.000$ at $n = 3$ through $8$, an improvement factor of at most $1.07$, because a well-fitted defective run has a small residual against the target it was trained on and passes both votes.
Against the clean reference it reads $0.634$ at $n = 3$, $0.984$ at $n = 4$, and $1.000$ thereafter: the escalation's overturns concentrate at $n = 3$, the only scale at which the deployment window of Sec.~\ref{sec:blindspots} is open, with a residual at $n = 4$.

The statistic agrees in pattern with the finite-shot measurement and with the closed-form window.
Realizing the margin-zone bound of Eq.~\eqref{eq:combined_bound} thus requires calibrating $(\tau_{\W}, \tau_{\MMD})$ at the operating point in use.

\subsection{Bi-metric disagreement}\label{sec:disagreement}

\begin{figure}[t]
	\centering
	\includegraphics[width=\columnwidth]{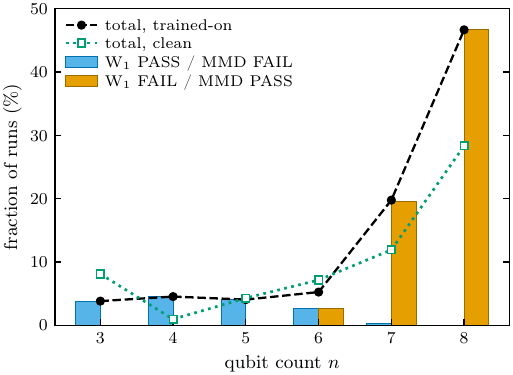}
	\caption{$\W$/MMD disagreement vs.\ qubit count at $\tau_{\W} = 0.05$ and $\MMD_k^2 \leq 10^{-4}$. Bars: the two patterns against the trained-on reference; lines: totals for both references. The rise at $n = 7, 8$ is almost entirely $\W$=FAIL / MMD=PASS.}\label{fig:disagreement}
\end{figure}

At $n \leq 6$ the disagreement rate is essentially constant at $3.8\%$--$5.2\%$ (Table~\ref{tab:disagreement}, Fig.~\ref{fig:disagreement}) and consists through $n = 5$ entirely of $\W$=PASS / MMD=FAIL, which is still half of it at $n = 6$.
For $n = 7, 8$ the total rises to $19.8\%$ and $46.7\%$, almost entirely in the opposite pattern ($82$ and $196$ $\W$F/MMDP runs), and the $\W$P/MMDF count falls to $1$ and $0$.
In contrast, against the clean reference the joint-failure population dominates at every scale, because a run trained on a perturbed target stays far from the clean one however well it fits.

We use the same transport threshold for benchmarks B1, B2, and B3, whose attainable fits differ, so that PASS/FAIL rates are comparable across them. A persistent flagged population at the certification-grade tolerance is what the window analysis of Sec.~\ref{sec:dominance} predicts. The counts pool all three benchmarks, including the synthetic near-uniform B3.

\begin{table}[t]
	\caption{$\W$/MMD disagreement at $\tau_{\W} = 0.05$ and the certification-grade $\MMD_k^2 \leq 10^{-4}$, trained-on reference. ``Dis.'' is the disagreement rate in percent of the $420$ runs at each $n$, the next three columns are run counts, and ``Dis.$^{\rm cl}$'' repeats the rate against the clean reference. $\W$P/MMDF dominates through $n = 5$, ties at $n = 6$, and inverts at $n = 7, 8$.}\label{tab:disagreement}
	\begin{ruledtabular}
		\begin{tabular}{cccccc}
			$n$ & Dis.\ & $\W$P/MMDF & $\W$F/MMDP & $\W$F/MMDF & Dis.$^{\rm cl}$ \\
			\hline
			3   & 3.81       & 16         & 0          & 5          & 8.10             \\
			4   & 4.52       & 19         & 0          & 31         & 0.95             \\
			5   & 4.05       & 17         & 0          & 62         & 4.29             \\
			6   & 5.24       & 11         & 11         & 66         & 7.14             \\
			7   & 19.76      & 1          & 82         & 88         & 11.90            \\
			8   & 46.67      & 0          & 196        & 92         & 28.33            \\
		\end{tabular}
	\end{ruledtabular}
\end{table}

The $\W$P/MMDF column holds $11$ to $19$ runs at each of $n = 3$--$6$; at $n = 3$ these are all B1 runs at the two largest perturbations. A per-run Fourier decomposition finds them no more correlation-dominated than agreeing runs, so they cannot be attributed to a single defect family of Sec.~\ref{sec:blindspots}. The complementary direction appears below $n = 7$ only at $n = 6$ ($11$ runs), and the joint-failure column grows with scale ($5$ to $92$ runs), tracking the loss of trainability.

\subsection{Trained generators}\label{app:blindspot_detail}

The cells of Table~\ref{tab:blindspot_demo} marked $\dagger$ use tightened discrepancy thresholds: $2.6 \times 10^{-5}$ for mode collapse at $n = 6$, and $2.4$, $1.5$, and $1.1 \times 10^{-5}$ for correlation shift at $n = 4, 5, 6$.

Four of the eight cells required tightening $\tau_{\MMD}$ below certification grade: the three correlation-shift cells at $n = 4$--$6$, where the window is closed at $\tau_{\MMD} = 10^{-4}$, and the mode-collapse cell at $n = 6$, where it is open but too narrow to train into. The correlation-shift window reopens once $\tau_{\MMD}$ drops below its closing value, $9.7 \times 10^{-5}$ to $2.4 \times 10^{-5}$ at $n = 4$ through $8$, at the shot cost of Sec.~\ref{sec:shot_budget}.

The $56$ mismatch runs among the campaign runs that $\W$-only certification admits number $34$, $4$, $11$, $6$ and $1$ at $n = 3$ through $7$, and the $37$ of them inside the escalation window $28$, $1$, $4$, $3$ and $1$.

Thirty-four of them are B1 runs at $n = 3$ and $\epsilon = 0.05$, where the injected defect itself has clean-reference $\W \approx 0.043 < \tau_{\W}$ and median $\MMD_k^2 = 2.0 \times 10^{-4}$; one more is a B1 run at $n = 4$, $\epsilon = 0.05$. These runs are the tolerance mismatch of Sec.~\ref{sec:dominance} on a trained generator and do not count toward the ordinary route.
The remaining $21$ are unperturbed B2 and B3 runs at $n = 4$--$7$. Across the $56$ the transport residuals sit at a median $\W = 0.039$ against a median $\MMD_k^2 = 1.7 \times 10^{-4}$, and $36$ of the $56$ meet the convergence criterion of Sec.~\ref{sec:pooled}.

The $56$ split into two populations. Thirty-five are B1 runs at $\epsilon = 0.05$, all but one at $n = 3$, whose injected defect lies inside the transport tolerance, so they are the tolerance mismatch of Sec.~\ref{sec:dominance} on a trained generator. The remaining $21$ are unperturbed B2 and B3 runs at $n = 4$--$7$, which are $21$ of the $776$ admitted unperturbed fits, or $2.7\%$; the $8$ of them inside the window are the rejections counted in Sec.~\ref{sec:finite_shot_falsepass} (Sec.~\ref{app:blindspot_detail}).

The population depends on the tolerance: at $\tau_{\MMD} = 10^{-3}$ and $10^{-2}$ it is empty at every scale, as Theorem~\ref{thm:redundancy} requires at $10^{-2}$, and at $10^{-5}$ it grows to $448$ of the $1{,}052$.
Whether rejecting these fits improves downstream utility depends on an external application criterion.

\end{document}